\documentclass[a4paper,11pt,fleqn]{article}
\usepackage{latexsym}
\usepackage{amsmath}
\usepackage{amssymb}
\usepackage[dvips]{graphicx}
\author{
\mbox{ { \footnotesize } } }
\title{\textbf{ Disk-averaged Spectra \& light-curves of Earth
 }}
\date{
 G.  Tinetti (NAI-NRC/Caltech), V. S. Meadows (Caltech),
D. Crisp (JPL), W. Fong (Caltech),  N. Kiang (GISS),  E. Fishbein (JPL), T. Velusamy (JPL),
E. Bosc (JPL) and M. Turnbull (Carnegie Washington) }
\makeindex
\begin{document}
\maketitle
\renewcommand{\theequation}{\arabic{equation}}
\renewcommand{\thefigure}{\arabic{figure}}
\newcommand{\ud}{\mathrm{d}}
\newcommand{\de}{\partial}
\section*{Abstract}
\begin{tabular}{|p{14.cm}|}
\hline
{\small
\emph{
We are using computer models to explore the observational sensitivity
to changes in atmospheric and surface properties, and the detectability of
biosignatures, in the globally averaged spectra and light-curves of the Earth. Using AIRS
(Atmospheric Infrared Sounder) data, as input for atmospheric and surface
properties, we have generated spatially resolved high-resolution synthetic
spectra using the SMART radiative transfer model,
for a variety of conditions, from the UV to the far-IR (beyond the range of
current Earth-based satellite data). We have then averaged over the visible
disk for a number of different viewing geometries to quantify the
sensitivity to surface types and atmospheric features as a function of
viewing geometry, and spatial and spectral resolution.
\newline
According to our model, in the case of Earth several atmospheric species   can be identified 
 in  disk-averaged spectra and
 potentially detected depending on the wavelength range and resolving power of the instrument.
O$_{3}$,  H$_{2}$O,
O$_{2}$, CO$_{2}$ (and probably oxygen dimer (O$_{2}$)$_{2}$) are clearly visible in the optical part of the spectrum, CO$_{2}$,  O$_{3}$,  H$_{2}$O in the MID-IR
(CH$_{4}$ and N$_{2}$O are hard to  discriminate because obscured by the strong water absorption). 
 \newline
The detectability of the red edge (a distinctive signature of vegetation) is affected by:  1) the extent of the vegetation over the planetary surface, 2) how much the vegetation is obscured by clouds, and 3) the strength of the signal for the particular vegetation type.
According to our model, the land surface cover of vegetation on Earth seems to be adequate to produce a  disk-averaged signal that is strong enough to show the red edge in disk-averaged spectra,
 even when the signal is averaged over the daily time scale.
To quantify how much the vegetation is obscured by clouds, we selected some specific
 bands  as indicators of photosynthetic activity.
The possibility of detecting phyto-plankton appears to be more challenging.
\newline
 These results have
been processed with an instrument simulator to improve our understanding of
the detectable characteristics of Earth-like planets as viewed by the first
generation  extrasolar terrestrial planet detection
and characterization missions (Terrestrial Planet Finder/Darwin and Life
finder). The wavelength range of our results are modelled over are applicable to both the proposed
visible coronograph and mid-infrared interferometer TPF architectures.
 We have validated this model against disk-averaged observations by the Mars Global Surveyor Thermal Emission Spectrometer (MGS TES).   
 This model was also used to analyze Earth-shine data for
detectability of planetary characteristics and biosignatures in disk-averaged spectra. 
   }} \\
\hline
\end{tabular}

\section{Introduction}

The principal goal of the NASA Terrestrial Planet Finder (TPF) and ESA Darwin mission concepts
   is to detect and  characterize extrasolar terrestrial planets.
   TPF-C (coronograph), TPF-I (interferometer) and Darwin are expected to survey nearby stars and directly detect planetary
   systems that include terrestrial-sized planets in their habitable zones (TPF website, TPF book, Darwin website).

This first generation of terrestrial planet exploration missions will 
only be able to  resolve the planets as point sources.
The distance and the relatively low brightness of the terrestrial sized planet compared to its
 parent star pose a daunting technological challenge:
 the spectral information provided will be necessarily an average over the visible disk and  also over the exposure time, which will probably be non negligible
(hours or days depending on the target).

The  task for the scientific community will be  to interpret this space-time-averaged single pixel, in which the information of the whole planet is irreversibly collapsed.
Due to the  nature of the problem, in the most general case  the interpretation of the recorded spectrum will not be unique. 
A family of solutions will provide an equally good explanation of the spectral features observed.
This expected degeneracy is  due in part to a lack of spectral, spatial and temporal 
sensitivity of these averaged spectra simulating the observations. 
Another source of degeneracy is given by space-time symmetries: those can be studied
and predicted  with our model as well.

Previous research on detectability of planetary characteristics in the disk-average, has focused on measurements of
earthshine on the unilluminated side of the moon (Woolf et al., 2002; Arnold et al., 2002; Goode, Monta\~n\'es Rodriguez, Pall\'e et al., 2001-2003), and models of diurnal photometric variability on an Earth-like planet  (Ford et al., 2001).

Ford et al. (2002) developed a model performing Monte Carlo integrations, with single scattering over a  planet with different surface types  specified by a map. They used this code to evaluate the photometric variability in some specific bands in the optical due to daily rotation and seasons. 
Although providing a good estimate of surface reflectivity in the optical,
  their model did not include an atmosphere or realistic clouds: both phenomena can significantly change the reflectivity of a planet.

Goode, Monta\~n\'es Rodriguez, Pall\'e et al., (2001, 2003) observed for several years earthshine
reflected by the moon to determine precisely the global and absolutely calibrated bond albedo of the Earth, and to study atmospheric phenomena
governing the Earth's radiation budget.
Woolf et al. (2002) examined earthshine data and speculated on the roles of different atmospheric and surface
features (clouds, ozone, water, oxygen, Rayleigh scattering,  detectability of the vegetation signal).
  Arnold et al. (2002) used a collection of earthshine data (400-800 nm) to derive normalized Earth albedo spectra and extract the signature of Earth vegetation.   
These observations, although extremely valuable, are necessary limited to views of the Earth available from astronomical observatories and over a relatively short time span. There are viewing geometries that can not be obtained
via earthshine (eg. South or North Pole views) and that might be of interest for missions like TPF and Darwin, which may observe planetary systems at significant inclinations.
The advantage of a comprehensive theoretical model  that realistically includes the contributions of surface atmosphere and clouds is the possibility  to generate the full range of scenarios (clouds, atmospheric gases, geometries) that are not available from a single observational data set. 

To better understand these issues, we  built a model which reproduces 
disk-averaged spectra 
as   faithfully as possible, starting from a description of a  spatially  resolved terrestrial planet on diurnal or seasonal time scales.
The model we  implemented, is a useful tool for exploring the sensitivity of the time-disk-averaged spectra to local properties and 
processes on time scales shorter than the integration time of the instrument. These results were then processed with an instrument simulator to improve
our understanding of the detectable characteristics as viewed by the first
generation of extrasolar terrestrial planet detection and characterization
missions.

We present here the first simulations of  disk-averaged
spectra and relative light curves (obtained by integrating the spectral signal over a specific band) of  Earth from the Infrared to the visible.
 This paper is the continuation of a previous one 
about Mars (a good example of a likely abiotic planet). We refer to that paper
for the details of the model and for a useful planetary comparison (Tinetti et al., 2004).
 

Disk-averaged spectra show sensitiveness  to seasons, viewing geometries, phases and surface signatures. 
The red edge, in particular, is a distinctive signature of plants.
 Astrobiologists are interested whether this biosignature could arise on an extrasolar planet and be detectable with TPF.  The question of the origin of the red edge is beyond the scope of this paper (see discussion section), but, assuming that it is plausible, our model and the earthshine data can help to assess the challenges for detecting it in a disk-averaged spectrum.  
  The primary challenges are: the spatial extent of vegetation  over the planetary surface, and how much they might be obscured by clouds.
Large part of the result and discussion sections is dedicated to the selection of the bands to be used as indicators of photosynthetic activity.
Some of them lose significantly their sensitivity when clouds are included in the simulations. 
The choice of the bands needs necessarily to take into
account  those effects.
Clouds in fact deeply affect the intensities and shape of the  Earth spectra, masking or reducing in some cases or
 enhancing (depending on cloud altitude, composition and optical depth) in some other cases some atmospheric and surface features. This consideration
 can probably be extended to the case of a general terrestrial planet, even if
  the specific results valid for the case of Earth and water clouds
 might not be applicable to the case of  exotic  condensate clouds and atmospheres.


The model has been validated against disk-averaged observations of
the Earth by the Mars Global Surveyor Thermal Emission Spectrometer (MGS
TES, fig. \ref{fig:validair}) (Christensen, Pearl, 1997) and earthshine spectra  (fig. \ref{fig:validavis}) (Woolf et al. 2002).
The model generates a variety of products  including disk averaged
synthetic spectra, light-curves and  the spectral variability at visible
and IR wavelengths as a function of the viewing angle.  These tools have been
used for  simulations of an increasingly 
cloudy/forested/oceanic Earth (fig. \ref{fig:surf} and \ref{fig:cloudre}) and analyzed to determine the detectability
of biosignatures on an Earth-like planet (e.g. red-edge signal, fig. \ref{fig:re}, \ref{fig:media} and \ref{fig:cloudre}; presence of plankton,
fig. \ref{fig:planktonx} and \ref{fig:planktonxx}). 
  We have therefore examined the sensitivity of the red-edge detectability to varying 1) planetary geometric views (varying visible land cover), 2) clouds (effect of both cloud type and amount of cover) over the current Earth's surface, and 3) the concentration of chlorophyll in phytoplankton.


\section{Modeling approach}

\newcommand{\degree}{\mbox{$^{\circ}$}}
The core of the model is a spectrum-resolving (line-by-line)
atmospheric/surface radiative transfer model, SMART,  (Meadows and Crisp, 1996; Crisp, 1997). AIRS (Atmospheric
Infrared Sounder) data were used as input for Earth  surface albedos and atmospheric properties   (Fishbein et al., 2003).

\begin{figure}[h!]
\begin{center}
    \includegraphics[width=11cm]{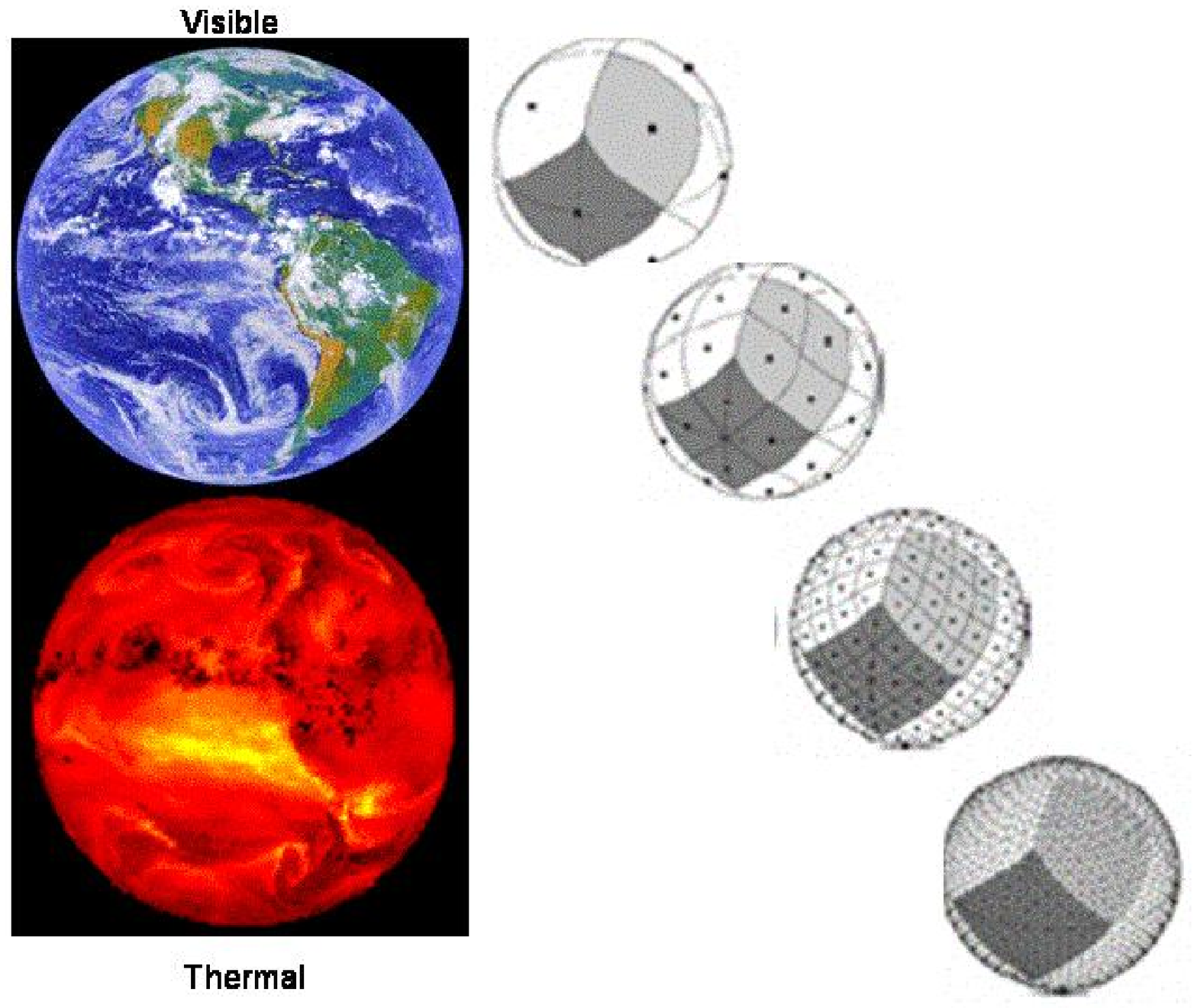}
\caption{ { \footnotesize \emph{ 
For producing disk-averages that could be viewed from any direction, spherical coordinates were not the best choice:
parallels and meridians do not divide the sphere into equal area
tiles, the pole is a singularity, the geometry is not invariant for a
general rotation.  
\newline
We chose to use Healpix (Hierarchical Equal
Area and iso-latitude Pixelization) grid as tiling program for our
model. Healpix, originally designed for the ESA Planck and
NASA WMAP missions (Gorski et al., 1998), is a curvilinear partition of the sphere into exactly
equal area quadrilaterals . This peculiarity is particularly desirable
since the geometry does not considerably change when we look at the planet
from different viewing angles, and the contribution of each region of
the sphere is accounted for with the same weight.  The ability to
easily degrade and upgrade the resolution of the map, is also
useful.  
\newline
We resolved the Earth with 48 pixels for mapping the atmospheric
properties and more than three thousands pixels for tiling the earth's
surface.
 } } } \label{fig:healpix}
	    \end{center} 
\end{figure}

These 
data  were processed and mapped on the sphere using a pixelization
scheme created for Planck and WMAP missions (Healpix) (Gorski et al., 1998)
 (fig. \ref{fig:healpix}). For each of the
pixels SMART was run to generate a database of synthetic spectra for a
variety of viewing angles, illuminations, surface types and cloud cover.
Finally, a model was created in which the user specifies spatial resolution
and observing position, and the program selects the appropriate synthetic
spectra  and creates the final view, which can then be
averaged on the disk (Tinetti et al, 2004). 

In the following paragraphs we give a 
fairly detailed description of the models, of the data we used as input to  model, 
and of the methods we have selected to estimate some specific parameters or
indexes.

\subsection{Input data}
\paragraph{Atmospheric  properties}
The atmospheric temperature, water vapor trace-gas and cloud
distributions needed to construct the model are extracted from the
NCEP Global Forecast System (NCEP, 1988; Kanamitsu et al, 1991;
Kanamitsu, 1989) and supplemented by the Upper Atmospheric Research
Satellite (UARS) and Harvard University tropospheric  ozone
climatologies by the AIRS Level~2 Simulation System (Fishbein et al.,
2003). The NCEP forecasts are produced
every three hours and have 1\degree times 1\degree horizontal
resolution and extend upto 10\,hPa.  The UARS climatologies 
are monthly zonal means with 10\degree latitude resolution and  extent
above 0.01\,hPa while the Harvard tropospheric ozone climatology
(Logan, 1998) is an annual mean with 4\degree times 5\degree resolution.

\begin{figure}[h!]
  \includegraphics[width=7cm]{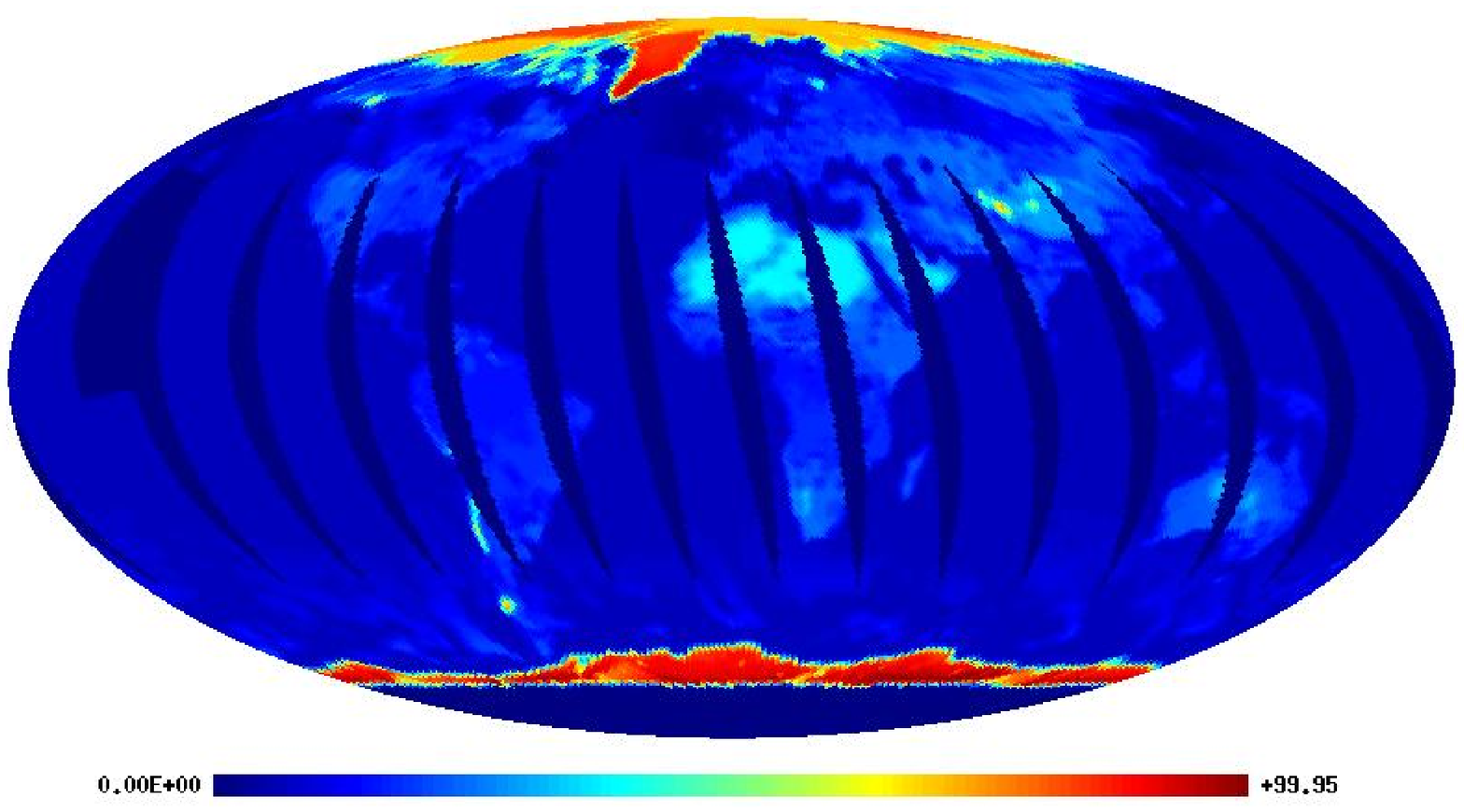} 
  \includegraphics[width=7cm]{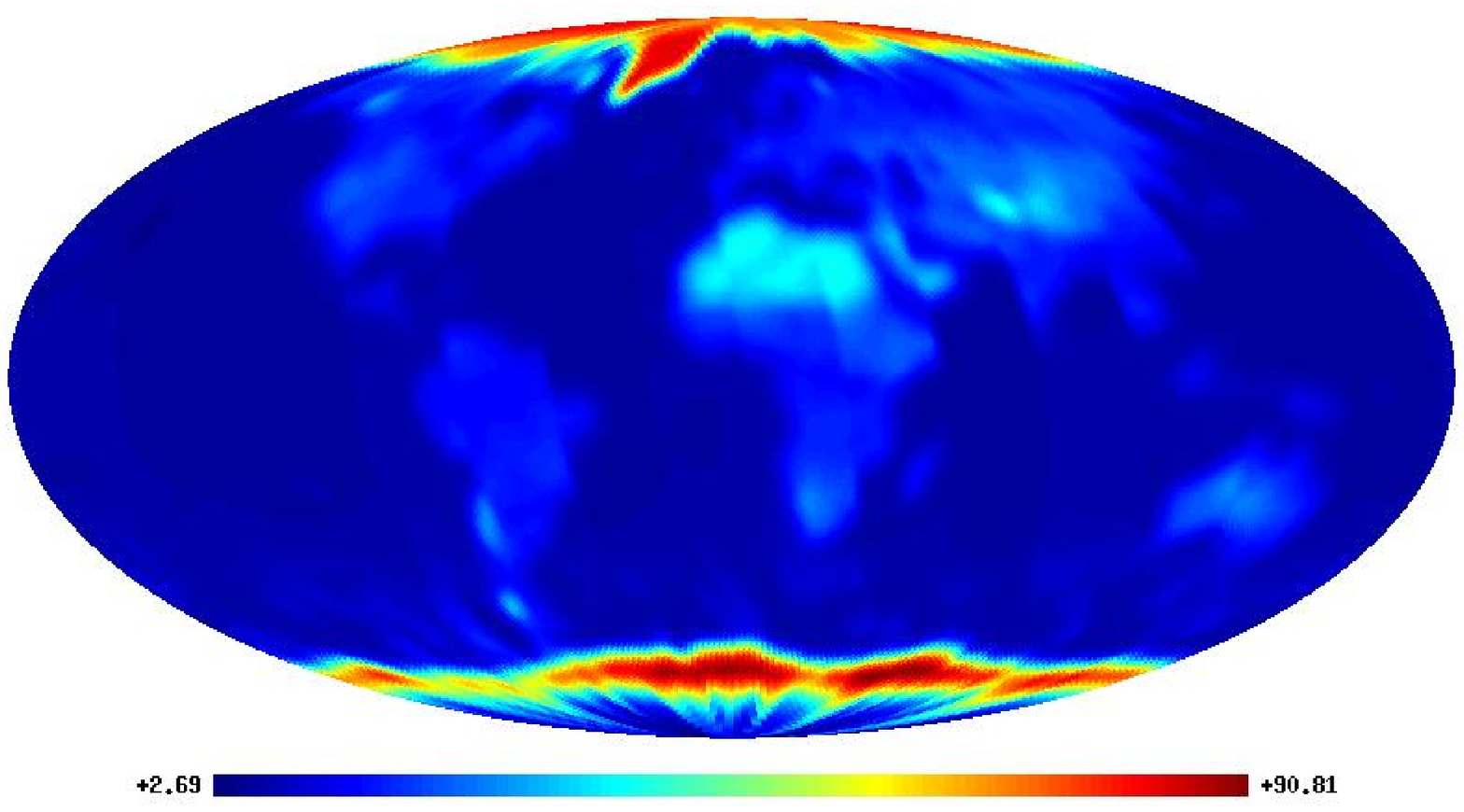} 
 \caption{ { \footnotesize \emph{ Left: This figure shows the actual
       AIRS solar albedo map. The dark blue strips are the regions
       where there are no data points.  Right: This figure shows the
       same data after using the two-dimensional Gaussian
       interpolation. The strips are no longer visible but the
       contours are less defined.  \newline
The spatial interpolation  method  assigned  each of the 3 million data
points (taken along the near polar orbit path) to a regular latitude-longitude grid by performing a weighted average, using   a two-dimensional
Gaussian distribution centered at the location of the data point.  The average value at each grid
point was normalized by summing the contributions of 3 million
Gaussians at each latitude/longitude grid point. This procedure was repeated for each parameter -
temperature and  trace gases.  } } } \label{fig:stempday}
\end{figure}

The simulation system samples the fields as the AIRS instrument would
view the earth from its 705\,km altitude sun-synchronous 1:30\,PM
local-time ascending equator-crossing polar orbit.  The satellite
orbits the earth 14.6 times per day collecting 3 million observations
daily.  The AIRS scan produces gores between the orbit passes near the
equator (Fig.  \ref{fig:stempday}) which have been filled using a
Gaussian interpolation.  The data have been remapped to a spherical
grid, 37 by 72 (latitude by longitude) (Fig.  \ref{fig:stempday}).

%

\begin{figure}[h!]
   \includegraphics[width=7cm,angle=0]{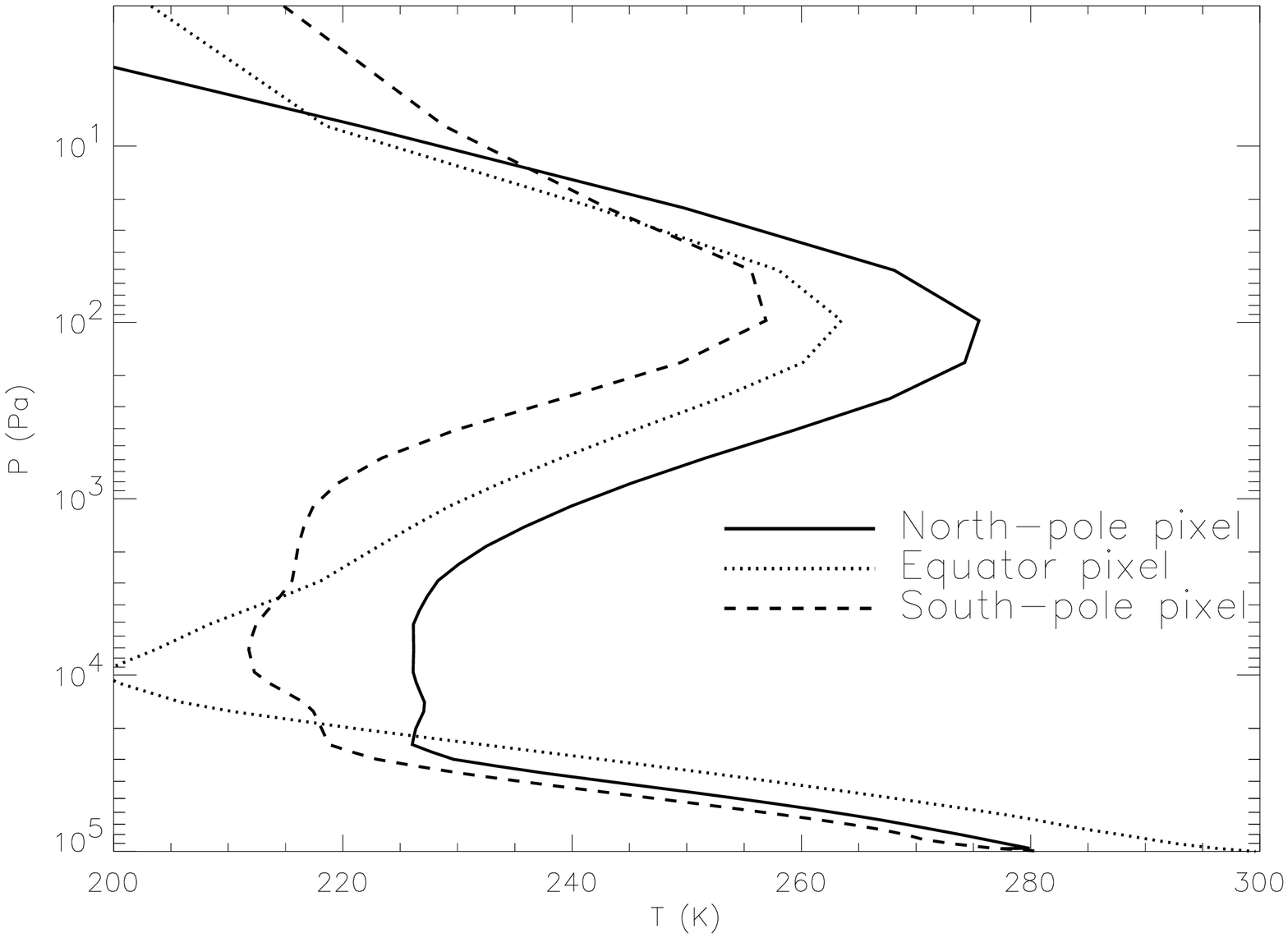}  
   \includegraphics[width=7cm,angle=0]{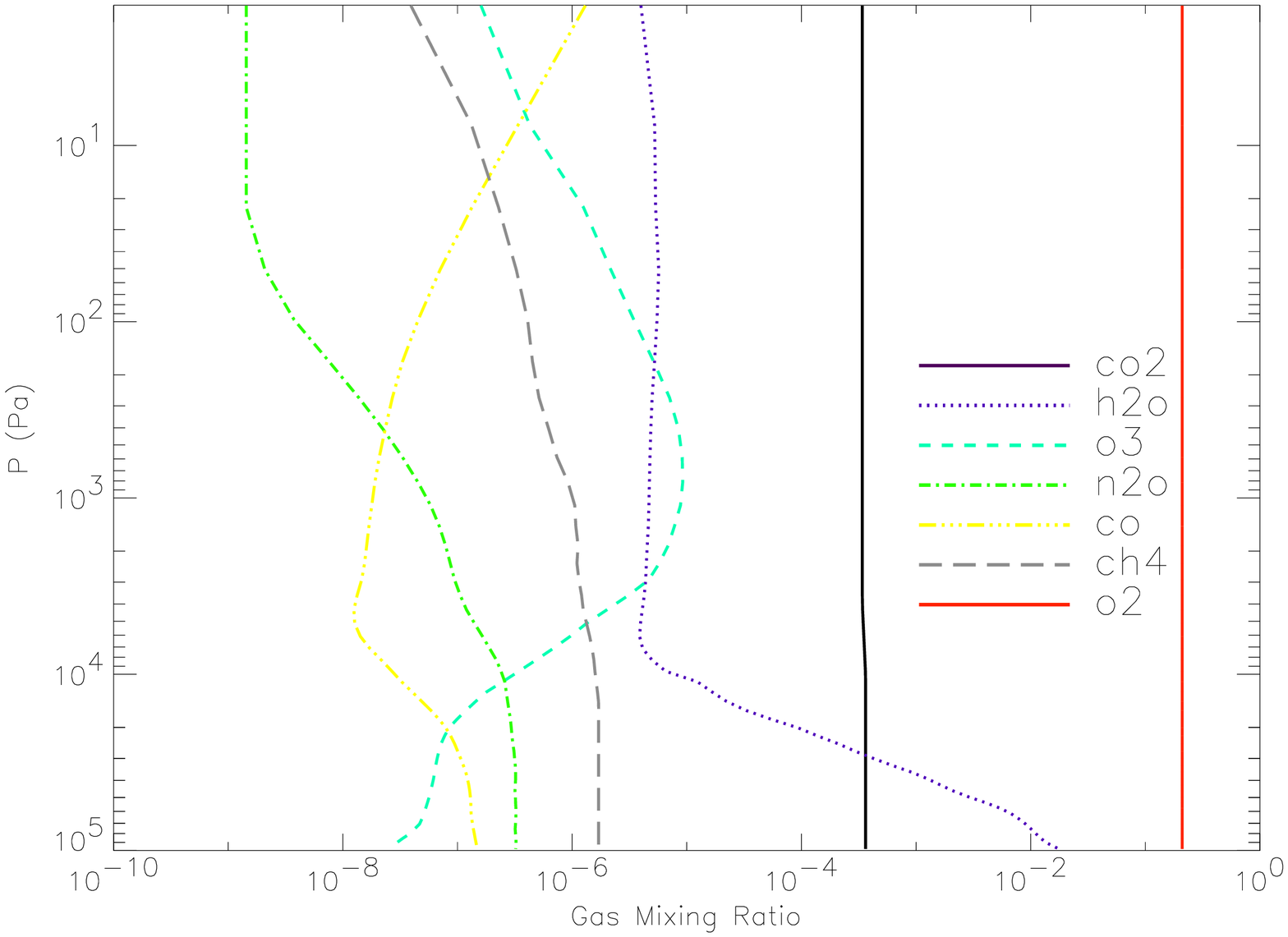} 
\caption{ { \footnotesize \emph{ 
Examples of temperature-pressure (fig. on the left) and gas-mixing ratios profiles 
(H$_{2}$O, O$_{2}$, O$_{3}$, CH$_{4}$,
CO$_{2}$, CO, N$_{2}$O, fig. on the right)   used for our model
 } } } \label{fig:atm}
\end{figure}


\paragraph{Solar spectrum}
                                                                                                                                                            
The upper boundary condition of the model is specified by the downward solar flux at the top of the atmosphere.  The solar spectrum used is a high-spectral-resolution spectrum compiled from satellite observations at UV wavelengths (Atlas-2 SUSIM) and stellar atmosphere models at longer wavelengths
(Kurucz, 1995). The position of the sun in our simulation was derived from the Jet Propulsion Laboratory Horizons Ephemeris System (Giorgini et al., 1997). \\

\paragraph{Molecular Line lists}
                                                                                                                                                            
The optical properties of the atmosphere were simulated using molecular line databases to generate the line-by-line description of infrared vibration-rotation bands of gases, and UV and visible cross-sections of gases associated with electronic and pre-dissociation transitions  (e.g. ozone).
                                                                                                                                                            
For our model, we have used  the HITRAN2K
(High resolution Transmission 2K) database [Rothman et al., 2003].

\begin{figure}[h!]
\begin{center}
  \includegraphics[width=14cm,bbllx=0bp,bblly=180bp,bburx=608bp,bbury=626bp,clip=]
  {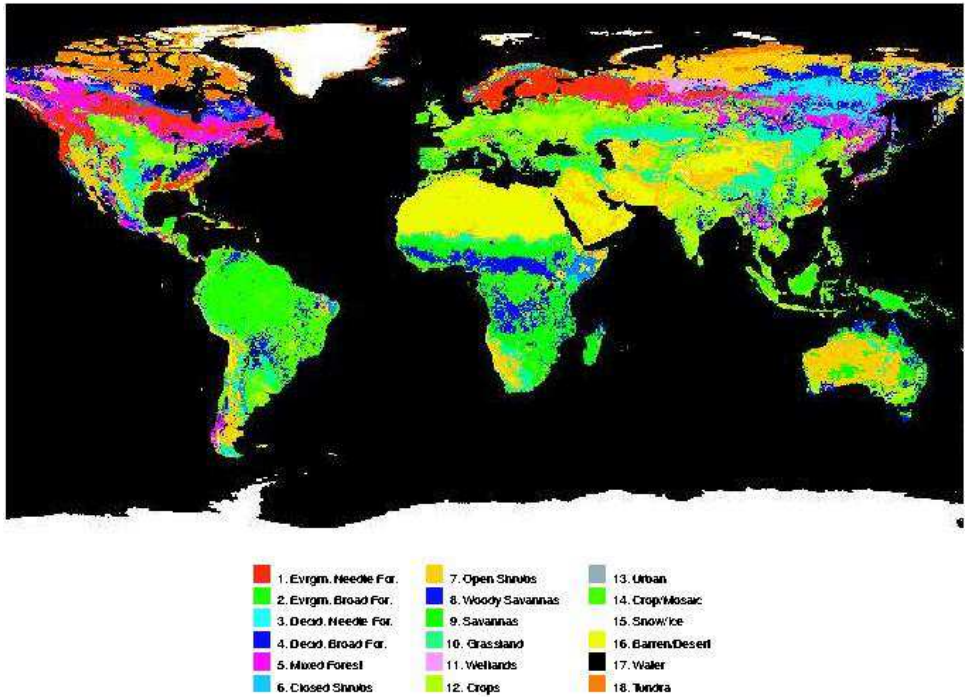}
  \caption{ { \footnotesize \emph{ The IGBP Global Cover Type map)
        \newline
      } } } \label{fig:ndvi}
\end{center} 
\end{figure}

\paragraph{Albedo/Surface Types}
The land cover map provided by AIRS assigns to each point one of
eighteen possible surface types (Fig. \ref{fig:ndvi}).  The
contribution of each material is determined by land fraction, the
amount of vegetation, and the types of vegetation defined by the
International Geosphere Biosphere Programme (IGBP) land use surface
classification (Loveland and Belward, 1997). For our purposes, these 18 divisions were
then reduced  to six general surface types: ocean,
forest, grass, tundra, ground, and ice.

These different surface types were simulated using as input to the
radiative transfer model six measured, wavelength dependent, surface
reflectance spectra, $\alpha_{k} (\lambda), \, k = 1 \ldots 6$.

\begin{figure}[h!]
\begin{center}
\includegraphics[width=8cm,angle=0]{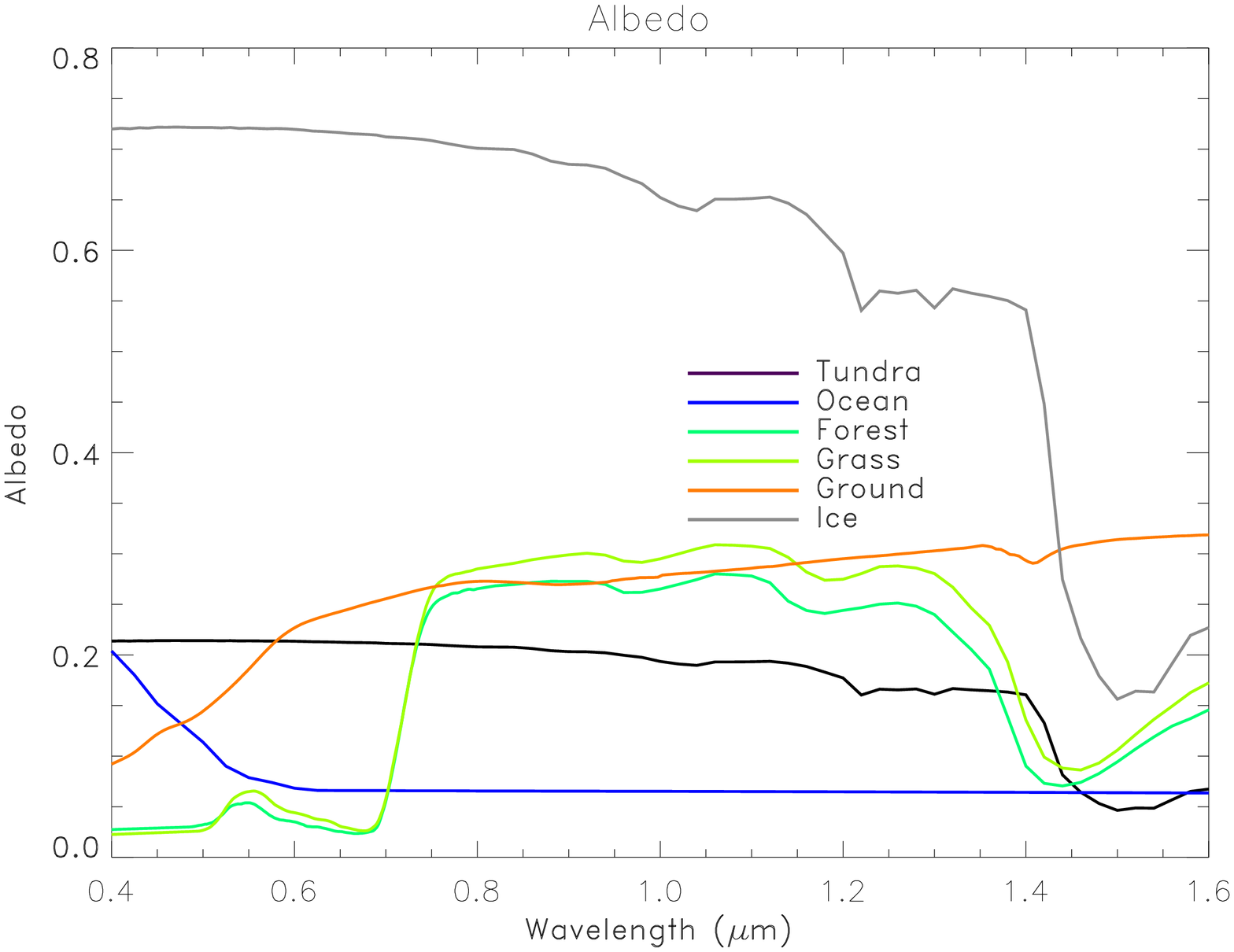} 
\caption{ { \footnotesize \emph{ Reflectance in the optical/NIR for
      the six surface types we used in our model } } }
\label{fig:albedo}
	    \end{center} 
\end{figure}

The IGBP map (Fig. \ref{fig:ndvi}) and the solar albedo map (fig.
\ref{fig:stempday}) must be consistent. The surface reflectance
spectra $\alpha_{k} (\lambda)$ must then  be recalibrated in order
to give finally the same global average albedo provided by the AIRS
albedo map. The scaling factor for each surface type can be calculated
by comparing the solar albedo value from the AIRS map (fig.
\ref{fig:stempday}) with the numerical integral in the 0.49-0.94
$\mu$m range (the wavelength range of the AIRS instrument which
provided the albedo map) of the $\alpha_{k} (\lambda)$.  Using the
simplified land cover map (6 surface types only), for each of the 48
atmospheric pixels, the fractional area covered by each of the six
surface types was determined. The radiative transfer code was run only
for the surface types with a non-negligible fractional area (greater
than 5\% of the entire pixel area).


\paragraph{Clouds}
Clouds vary considerably in amount and type over the globe. They have very important effects on longwave and solar energy transfer in the atmosphere, as we can see from Table 1  (Harrison et al. 1990),  (D. Hartmann, 1994), (Oort and Peixoto, 1992).
 


For radiative calculations optical and radiative properties must be specified. Normally, water clouds have  relatively weak solar absorption, but they effectively scatter solar radiation back toward space and so have high reflectivities. Thick clouds behave almost  like black bodies for long wave radiation: they absorb all incident long wave radiation and emit like a blackbody with the temperature of the atmosphere at that the same level of the cloud. A simple approach is to specify the properties of three types of clouds: high (cirrus), medium (cumulus) and low (stratus). \\ \\

\begin{tabular}[h!]{lccc}
Type & SW reflectivity & SW absorptivity & \% area \\
High (cirrus) & 0.21 & 0.005 & 0.228 \\
Medium (cumulus) & 0.48 & 0.02 & 0.09 \\
Low (stratus) & 0.69 & 0.035 & 0.313 \\
\end{tabular}
\\
{\par \footnotesize{ \textbf{Table 1 -} Values of cloud shortwave reflectivity and absorptivity and fractional area coverage assumed in (Manabe and Strickler, 1964)} } \\
 
In our model, the wavelength dependent optical properties of liquid water
droplets were computed  using Mie scattering theory  (Crisp, 1997), (Liou, 2002). 
Cirrus clouds were parameterized as polydispersions of hexagonal crystals, and their optical properties were derived using geometric optics 
(Muinonen et al., 1989). The liquid water and ice refractive indexes were obtained from (Segelstein,1981) and (Warren, 1984) respectively.

The top and bottom pressures and the optical depths were taken from AIRS Level~2 Simulation System (Fishbein et al.,
2003). Those sets of data are sometimes underestimating the presence of low clouds.
We have used both AIRS and MODIS data (MODIS website)  for the cloud maps.


\subsection{Models}
Most of the models listed here were already described  in a previous paper about Mars (Tinetti et al., 2004). We refer to that
work for a more detailed explanation and list of references.


\paragraph{LBLABC}
                                                                                                                                                            
  LBLABC is a line-by-line model (Meadows and Crisp, 1996)  that generates
 monochromatic gas absorption coefficients  from molecular line lists, for each of the gases present in the atmosphere at a specified temperature and pressure.                                                                                                This model was designed to evaluate absorption coefficients at all atmospheric levels over a very broad range of pressures (10$^{-5}$ to 100 bars),
temperatures (130 to 750 K), and line center distances (10$^{-3}$ to 10$^{3}$ \, cm$^{-1}$).

\paragraph{SMART}

     SMART  is a multi-stream, multi-level, spectrum-resolving (line-by-line) multiple-scattering algorithm designed to generate the high-resolution synthetic spectra of planetary atmospheres  (Meadows and Crisp, 1996). Using a high-resolution spectral grid, it completely resolves the wavelength dependence of all atmospheric constituents, including absorbing gases (infrared absorption bands, UV predissociation bands, and electronic bands) and airborne particles (clouds, aerosols) at all levels of the atmosphere, as well as the wavelength- dependent albedo of the planet's surface, and the spectrum of the incident stellar source.                                                                       
\begin{figure}[h!]
\begin{center}
   \includegraphics[width=8cm,bbllx=8bp,bblly=168bp,bburx=594bp,bbury=622bp,clip=]{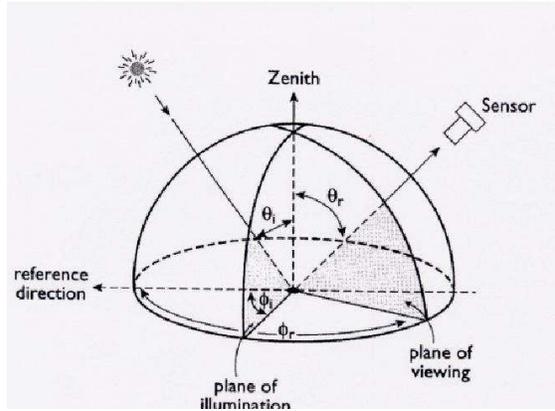} 
\caption{ { \footnotesize \emph{ This figure shows the three input angles for SMART: the solar zenith angle
($\theta_{i}$), the viewing angle zenith angle ($\theta_{r}$) and the viewing azimuth angle ($\phi_{r} -
\phi_{i}$). The solar azimuth angle ($\phi_{i}$) has been chosen as the reference direction.
SMART was run for each pixel for nine solar zenith angles (0, 15, 30, 45, 60, 75, 80, 85, and 90 degrees),
four viewing zenith angles (the four Gaussian angles: 21.48, 47.93, 70.73, and 86.01 degrees),
and seven viewing azimuth angles (0, 30, 60, 90, 120, 150, and 180 degrees), six surface types (tundra, ocean, ice, ground, grass and forest) and three cloud types (Cirrus, Strato-Cumulus and Alto-Stratus). 
This library was used for the creation of three-dimensional data cubes
 (two dimensions spatially and one dimension for the
spectra). In the spatial dimensions, only the pixels that are detectable by the
remote sensor will be considered, therefore the data cube will be comprised of pixels up to 90
degrees in each direction from the sub-sensor position on the planet. The resultant data cube will
display one half of the planet's surface with the corresponding spectra at each of the detectable pixels.
 } } } \label{fig:angles}
	    \end{center} 
\end{figure}

\paragraph{Averaging on the disk}

We used a very high resolution of the Healpix base (more than three thousands) 
 to assign to each of the 48 atmospheric pixels the fraction detectable by the sensor, the fration illuminated
and the fraction covered by different surface types and
 clouds (we took in consideration 3 cloud types: Strato-Cumulus, Alto-Stratus, Cirrus) according to the AIRS-IGBP land cover and cloud maps (AIRS -level 2, and MODIS).


SMART was run for a variety of  surface types (six in total), cloud types (3 main types), solar zenith angles (nine), viewing zenith angles (four), viewing azimuth
angles (seven) (fig. \ref{fig:angles}). 
The radiance values for arbitrary angles can be estimated  using interpolation (linear interpolation for the solar zenith angle
 and bicubic
spline interpolation  for the viewing angles)

Using the simplified land cover map and a cloud map, for each of the 48 atmospheric pixels, the fractional area covered by each
of the six surface types and by one of three cloud types was determined. We neglected the contribution of surface types covering
a fractional area of less than 5\%.

The radiance  for each pixel can be calculated 
using the method previously explained for the viewing/solar angles
 and summing the contributions of the all the radiances corresponding to different surface and cloud types weighted  by their
 fractional area.

 The disk-averaged synthetic spectrum can be easily calculated
by summing the radiance values for all of the detectable pixels at a given wavenumber and dividing
by the number of pixels.

\section{Earth: Validation with observed spectra}  
We have validated our model against two types of observations. We  used  MGS-TES data taken from Mars Global Surveyor en route to Mars
 for the IR (Christensen, Pearl, 1997) and earthshine data (Woolf et al., 2002) for the optical. 

Earthshine is sunlight reflected by the Earth that can be seen lighting up the dark portion of the crescent Moon.  
  To
obtain the earthshine, the spectrum of the dark Moon is divided by the spectrum of the bright Moon, thereby removing the reflection spectrum of the Moon itself and the extra pass through the Earth's atmosphere.
The result is the disk-averaged outgoing radiation at the top of the Earth's atmosphere. This can be directly compared with our model. 

\begin{figure}[h!]
\includegraphics[width=11. cm]{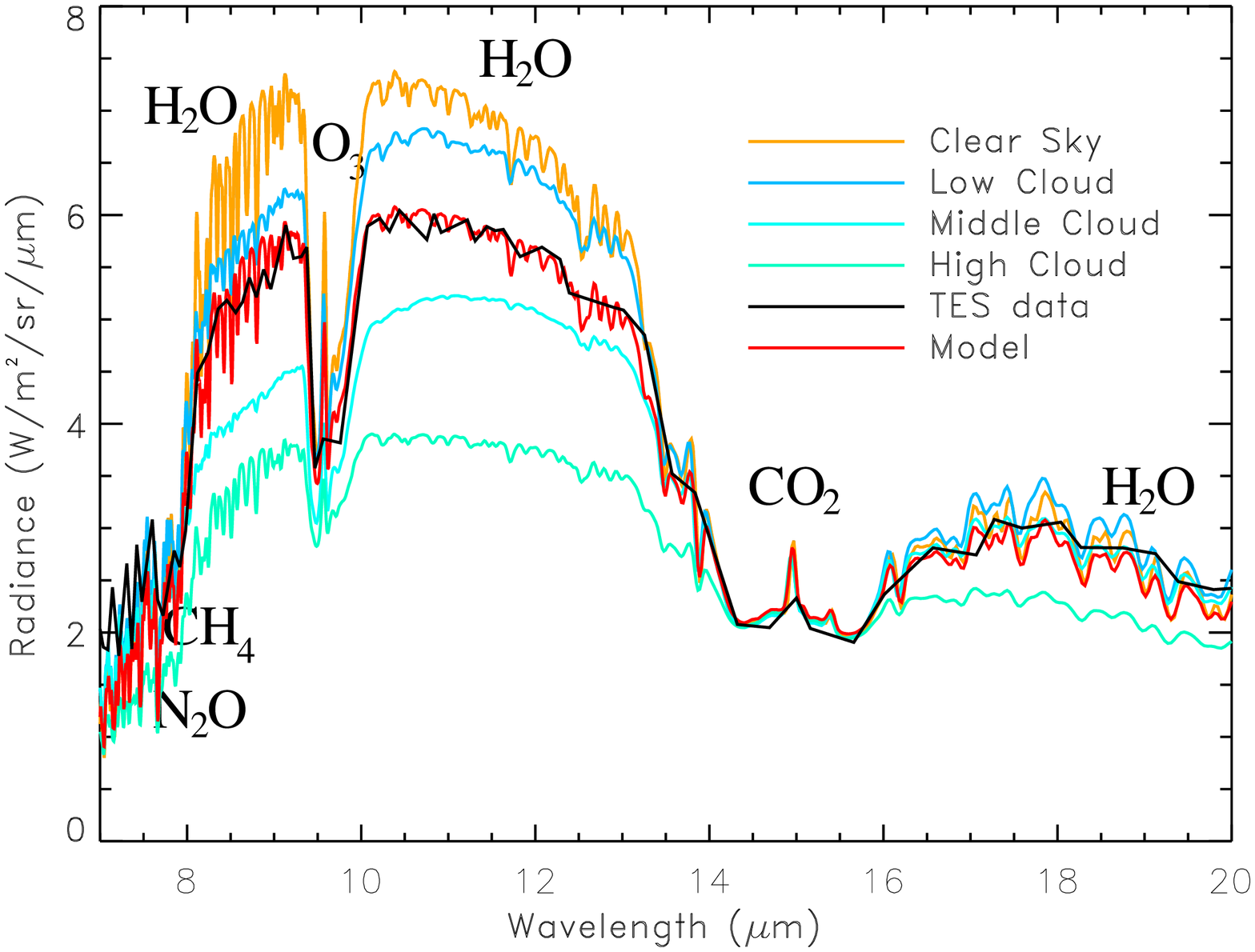}
\includegraphics[width=2.5 cm]{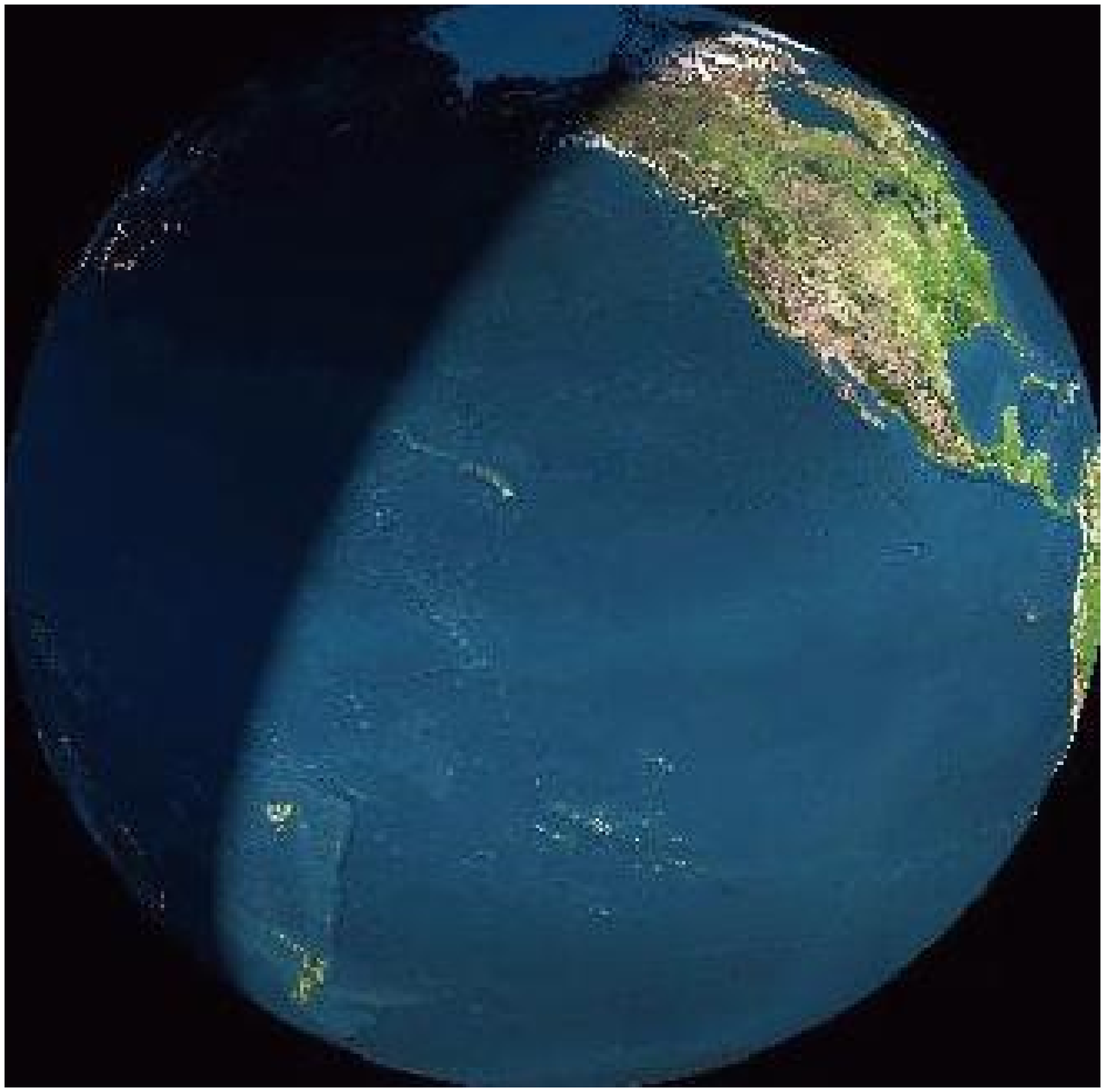} \\
 \caption{ { \footnotesize \emph{   Fig. on the left: disk-averaged spectrum of Earth observed by Mars Thermal Emission Spectrometer (TES) (black) (Christensen, Pearl, 1997) compared with the synthetic one produced in almost the same conditions with almost 50\% of clouds (red) and without
clouds  (yellow). Plots in light blue show the contribution of different cloud-types. Fig. on the right: we show the view of the Earth at 17:30 UT on November 24, 1996,
from the Mars Global Surveyor (Walker, 1994).   } } }  \label{fig:validair}
\end{figure}
\begin{figure}[h!]
\includegraphics[width=11. cm]{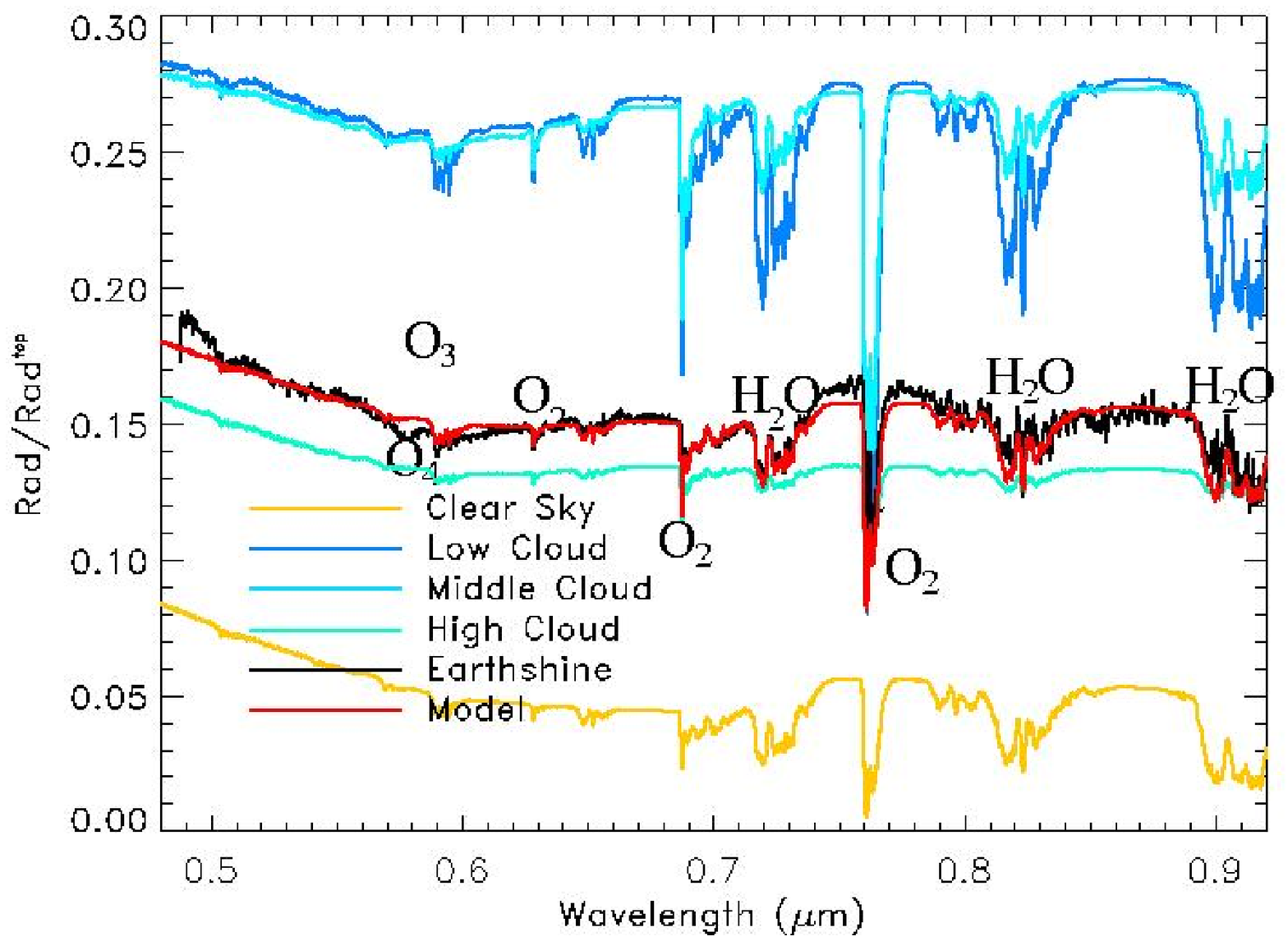}
\includegraphics[width=2.5 cm]{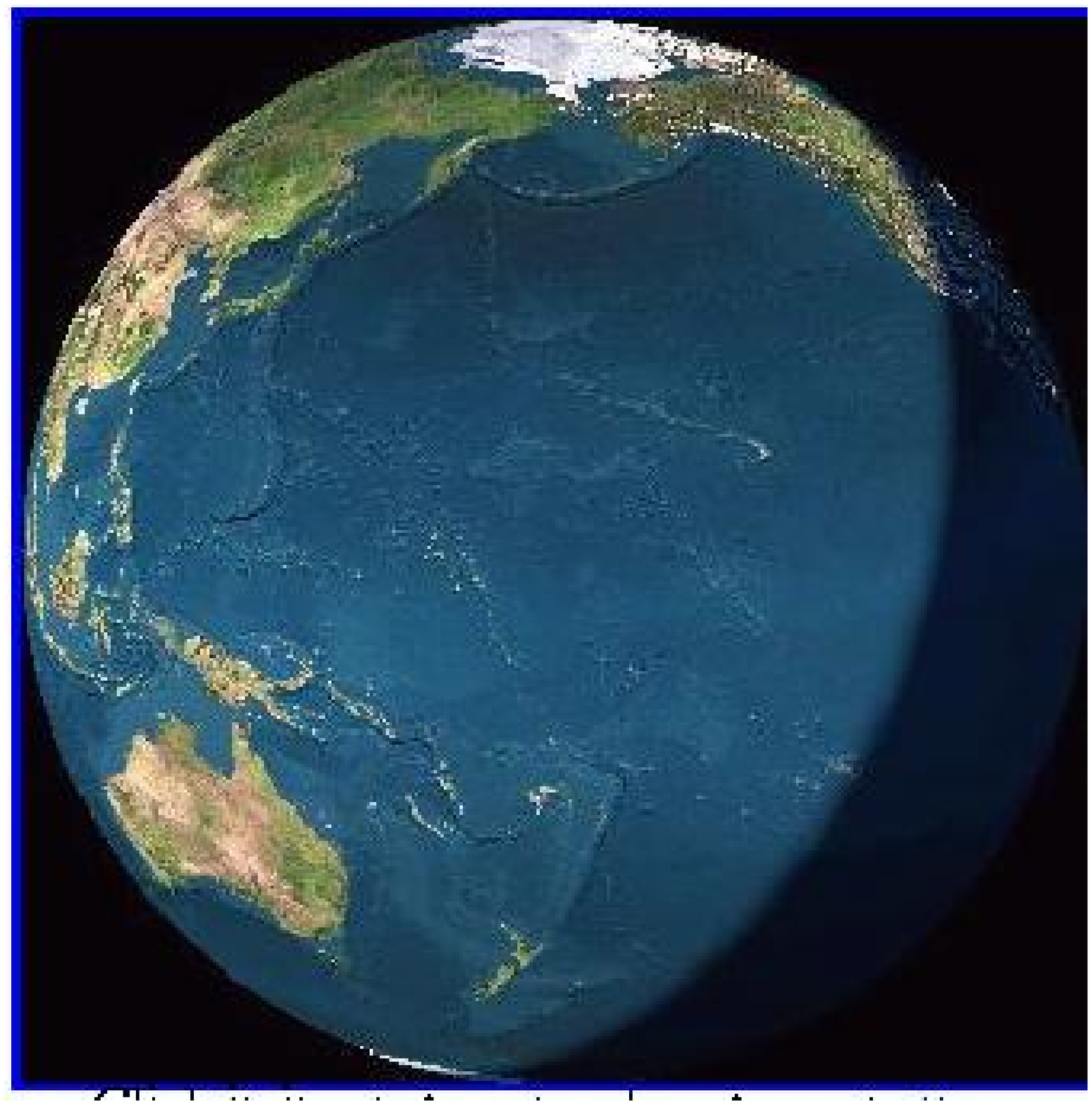} 
 \caption{ { \footnotesize \emph{   Fig. on left: Earth shine  spectrum of the Earth observed from the Steward Observatory of the University of Arizona
 (black plot) compared with the synthetic one produced in almost the same conditions of  illumination and cloud cover (red).
The yellow plot at the bottom of the figure, is the disk-averaged synthetic cloud-free spectrum. Plots in light blue
show the contribution of different cloud-types. 
 Fig. on the right: we show the view of the Earth at the time of the measurement (Woolf et al., 2002),  (Walker, 1994). 
 } } }  \label{fig:validavis}
\end{figure}

Our model produces an acceptable fit with the IR and optical data as we can see in fig. \ref{fig:validair} and \ref{fig:validavis}.
In both these figures we have plotted the recorded  (black plot) and the synthetic spectra (red plot),
and the contribution of the cloud free (yellow plot) and totally cloud covered spectra  as well (light blue plots).

In the IR  (fig. \ref{fig:validair}) the clear-sky spectrum is the uppermost curve  (warmest) since with no clouds we can observe the warm lower layers in the troposphere and the surface. 
The main effect of adding clouds is to decrease the emitted flux (colder temperature  where clouds form).
The reduced thermal contrast between the cloud tops and space also reduces the amplitude of the spectral features: the lowest curve (cirrus clouds) is almost a blackbody at the tropopause temperature. The emission spike
 in the middle of the 15 $\mu$m CO$_{2}$ band and 9.6 $\mu$m O$_{3}$ band is due to the Earth's stratospheric temperature inversion. H$_{2}$O absorbs at wavelengths longer than 15 $\mu$m (rotational bands) and from 5 to 8 $\mu$m (vibrational bands).
CH$_{4}$ and N$_{2}$O are hard to detect at (respectively) 6 and 8 $\mu$m because their absorption
is obscured by the strong water absorption.

In the optical (fig. \ref{fig:validavis}) the contribution of clouds is more complex: clouds affect the atmospheric light paths in several ways compared to the
clear-sky conditions. Because of multiple  scattering inside the clouds, the light path can be either increased
or decreased, and the absorption of the tropospheric trace gases
can be strongly enhanced or reduced. Thin high clouds can  also reduce the absorption paths below the clouds  (Wagner et al., 1998).
As shown in figure  \ref{fig:validavis}, we confirm the presence of O$_{3}$ (Chappuis bands, 0.5-0.7 $\mu$m), H$_{2}$O ($\sim$ 0.7, 0.8, 0.9 $\mu$m),
O$_{2}$ (B band, 0.69 $\mu$m, and  A band, 0.79 $\mu$m) already discussed in (Woolf et al. 2002).
Moreover, we believe the absorption features $\sim$ 0.6 $\mu$m might be explained (C. Miller, private communication) by the oxygen dimer (O$_{2}$)$_{2}$ (absorption bands at 0.577 and 0.63 $\mu$m)
(Wagner et al., 2002). The model is unable to reproduce this feature since (O$_{2}$)$_{2}$ was not initially included in our simulations. 
An extensive discussion about atmospheric species
and spectral signatures in both IR and visible can be found in (Des Marais et al. 2002).

The phase and the viewing geometry for the Woolf et al. (2002) optical observations 
were known and were used in our model to reproduce the observing geometry precisely. Given the observing geometry and the day of the year, we calculated the extreme cases for a completely clear and completely cloudy Earth
(fig. \ref{fig:validavis} yellow plot and light blue plots), which must bound the correct fit. 
However, to model the observed spectrum precisely, we used a realistic spatially resolved cloud cover
for the date of observation  (AIRS Level 2 data and MODIS data slightly adjusted   to obtain a better fit).
In doing that, we have tried to reproduce not only the spectral shape, but also the absolute brightness, neglecting the possible uncertainties
present in the measured spectrum (the data 
were taken through intermittent clouds  and  the method of scattered light subtraction is not precisely known). The best fit was obtained considering $\sim$30\%  of cirrus, $\sim$10\% of Strato-Cumulus and $\sim$10\% of Alto-Stratus
clouds spatially distributed. The oxygen A-band at  0.76 $\mu$m was particularly useful for cloud retrieval (Kuze and Chance, 1994).

\newpage

\section{Results} 
            \subsection*{Effects of Surface types and clouds} 
\begin{figure}[h!]
\begin{center}
\includegraphics[width=5. cm, angle=0]{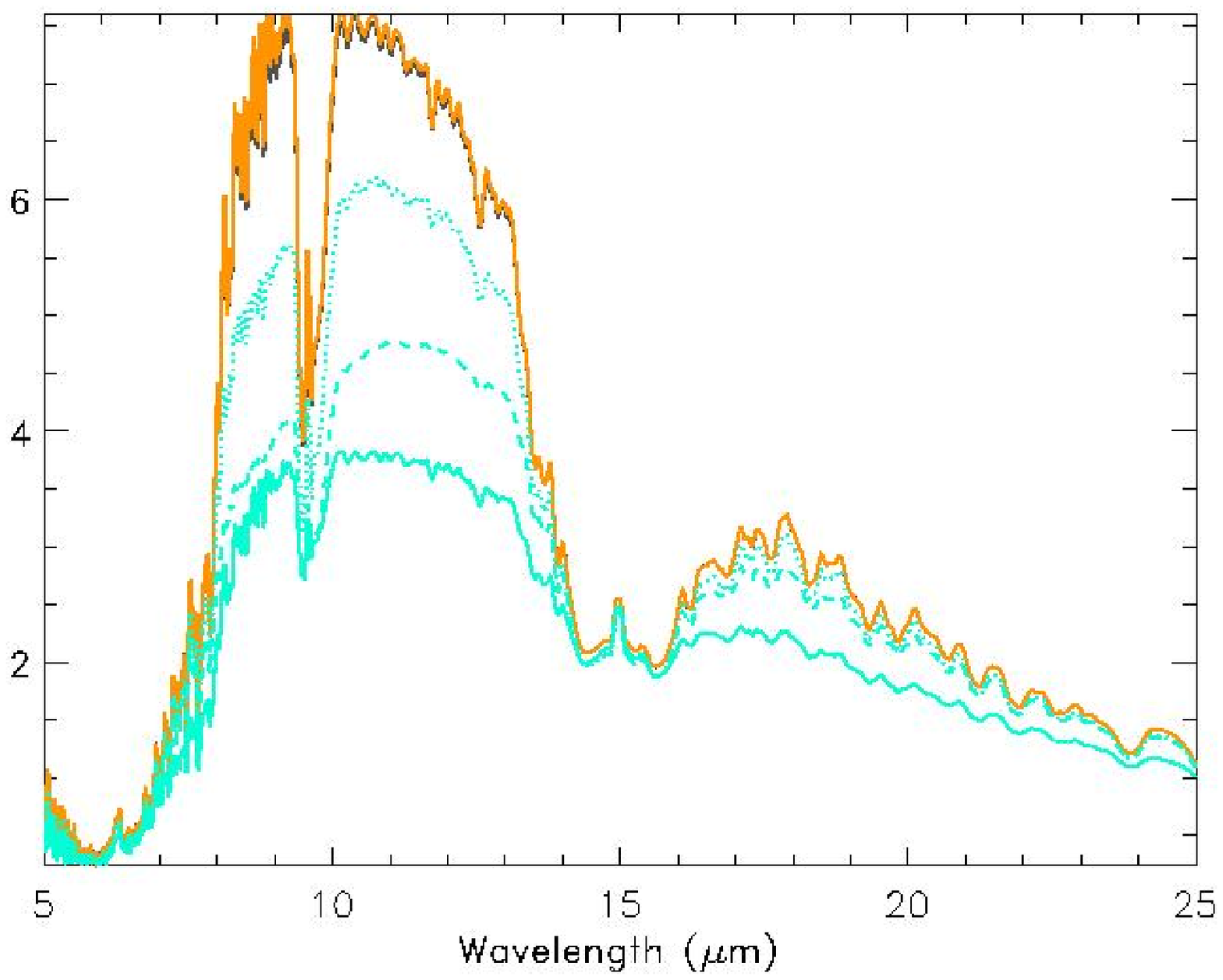}
\includegraphics[width=5. cm, angle=0]{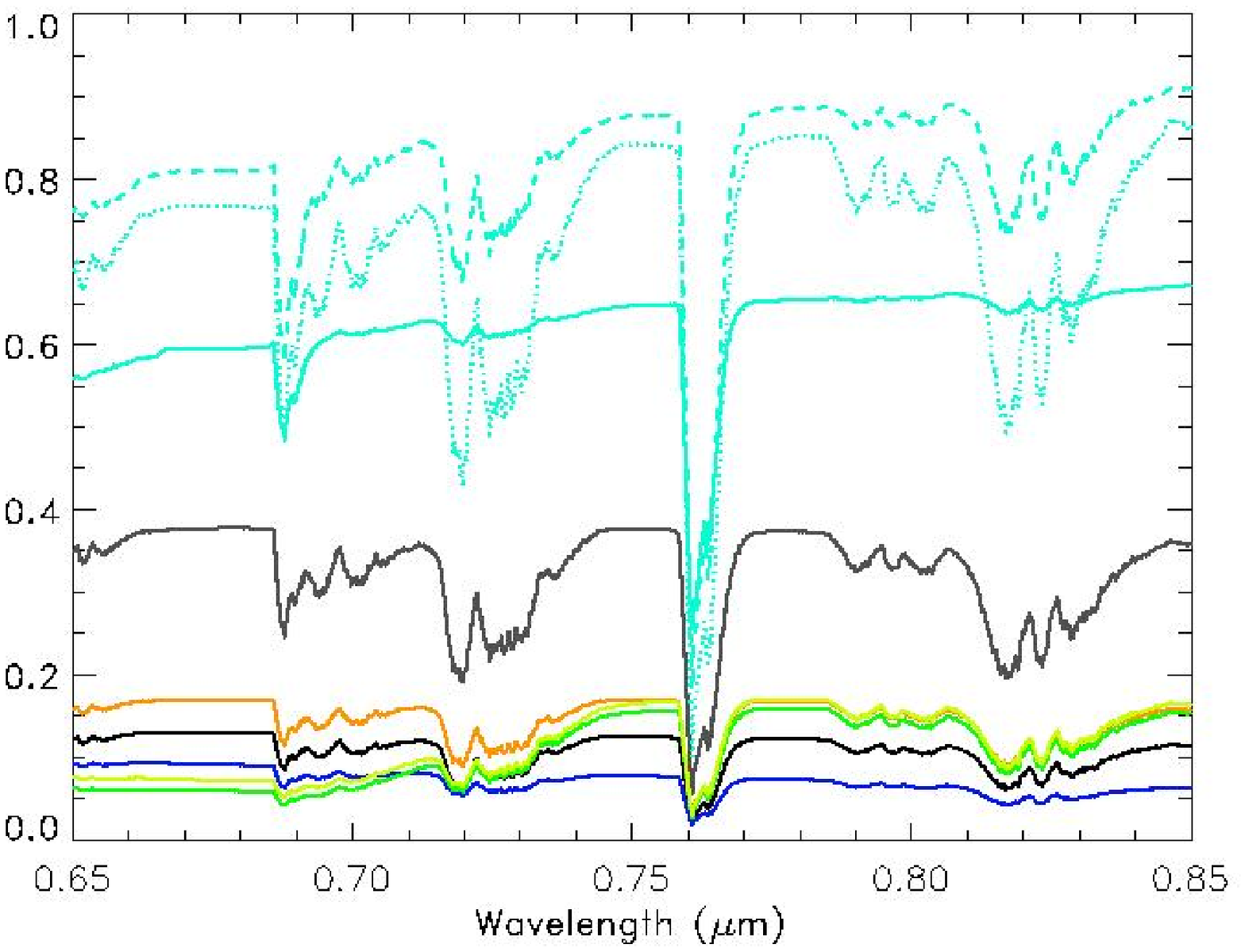} \\
\includegraphics[width=11. cm, angle=0]{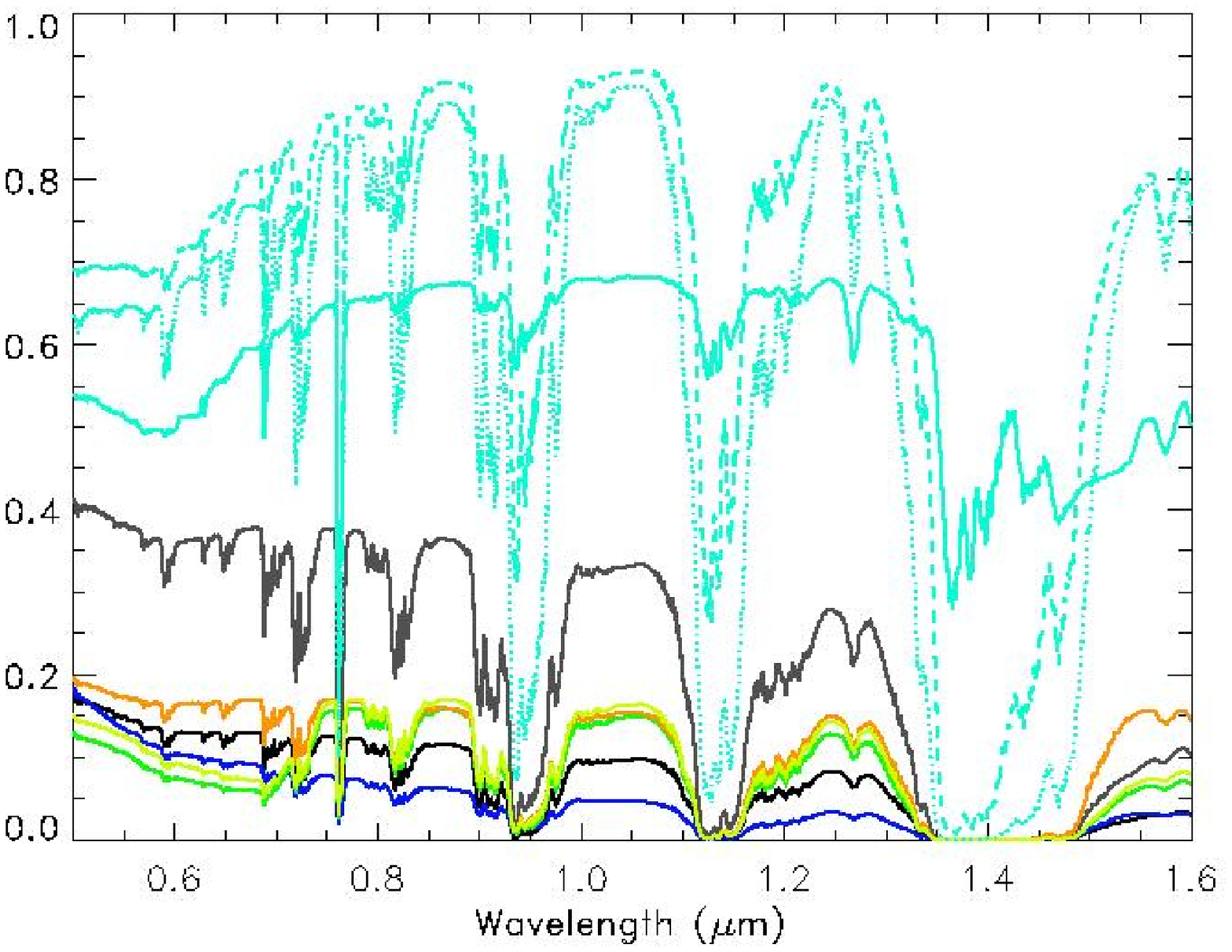} \\
\caption{ { \footnotesize \emph{  Disk-averaged spectra with single surface type coverage (blue: ocean, grey: ice, yellow: desert, green vegetation, black: tundra).
Almost no differentiation in the IR (radiance expressed in W/m$^{2}$/sr). In the visible (radiance divided by the solar radiation at the top of the atmosphere) on the contrary we have pronounced differences in the spectra. The light blue plots in the figures show the contribution of clouds: cirrus (continuous line), alto-stratus (dashed line)
and strato-cumulus (dotted line). 
In the band 0.7-0.8 $\mu$m we can
recognize very clearly the red-edge signal (leafy plants reflect sunlight strongly in this
band, fig. \ref{fig:re}).} } } \label{fig:surf}
	    \end{center} 
\end{figure}

To assess the contributions of the various surface types, we have run some simulations covering the Earth's surface varying one environmental variable at a time.
The plots in fig. \ref{fig:surf} show  disk-averaged spectra of ocean (dark blue curves), ice (grey curves), forest (darker green curves), grass (green curves) and 
tundra (black curves) worlds.
The viewing geometry and illumination are the same in all the plots (view: latitude 0, longitude 0, fully illuminated). 
In the IR it is impossible to distinguish among the different surface types (fig. \ref{fig:surf} top left).
This is due to the fact that
the surface reflectance does not change appreciably in the IR, just the opposite 
in the optical (fig. \ref{fig:albedo}).
The light blue plots in the figures show the contribution of clouds - cirrus (continuous line), alto-stratus (dashed line)
and strato-cumulus (dotted line)-.  In fig.  \ref{fig:surf}, top right, we show a zoom into the 0.65-0.85 $\mu$m band,
where the red-edge signal is detectable. Leafy plants reflect sunlight strongly in this
band. We will focus on the red-edge detectability later on in this section. 

\subsection*{Disk-averaged synthetic spectra of Earth. IR and visible} 
\begin{figure}[h!]
\begin{center}
 \includegraphics[width=9cm,angle=0]{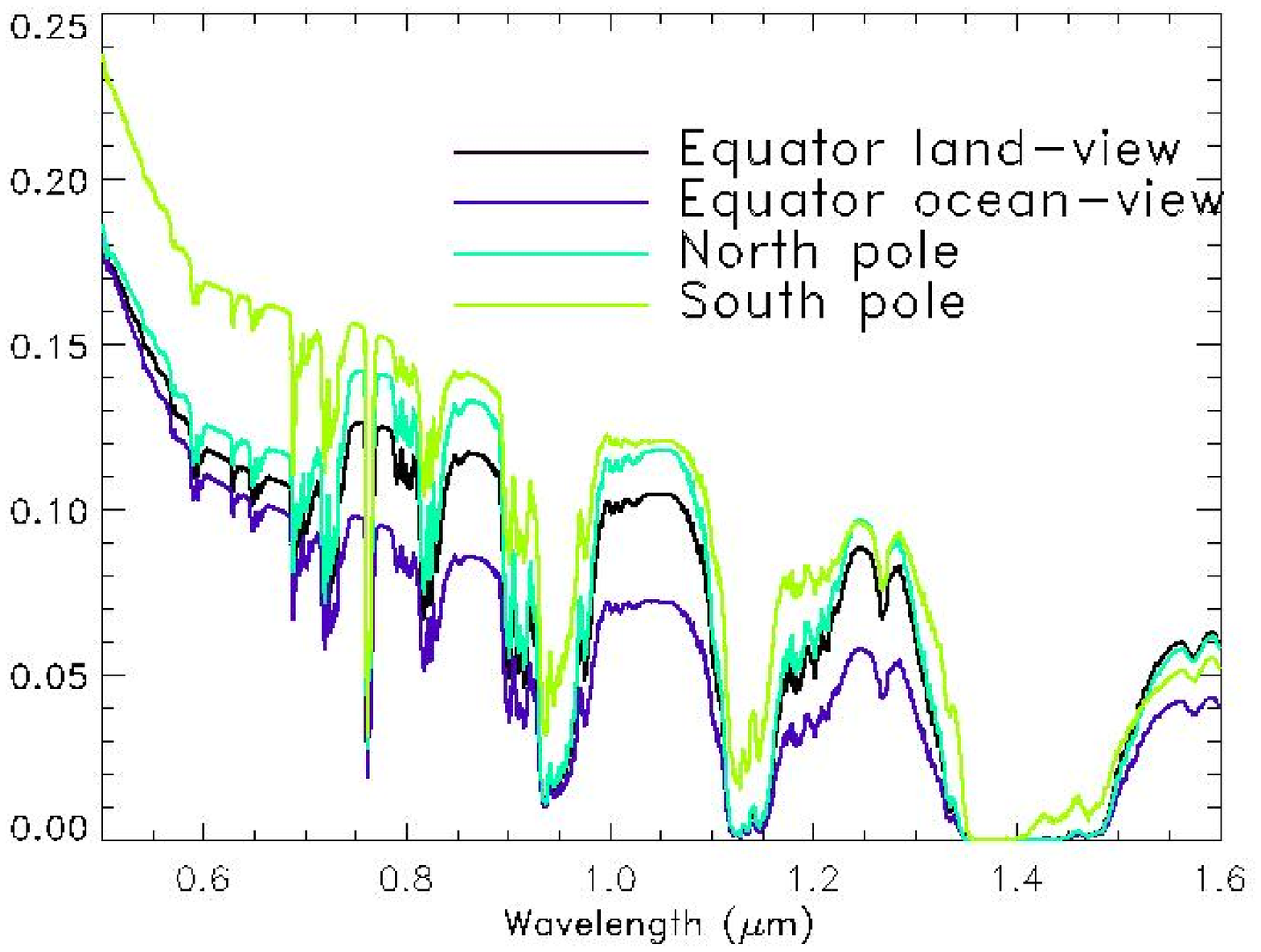}  \\
 \includegraphics[width=9cm,angle=0]{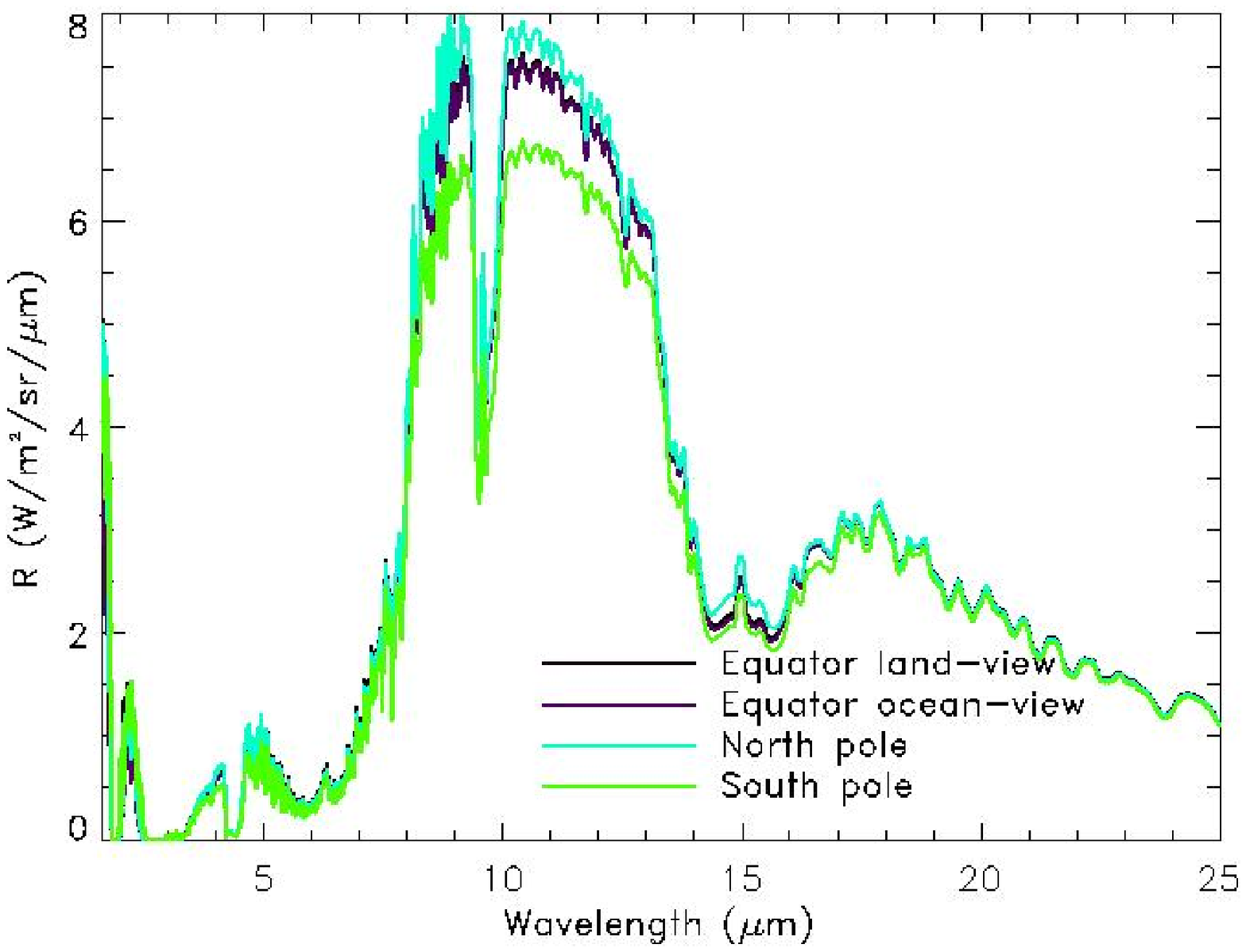}  \\
  \caption{ { \footnotesize \emph{Earth solar (fig. at the top, radiance divided by the solar radiation at the top of the atmosphere) and IR disk-averaged spectra. In this particular simulation there are no clouds and the disks are equally illuminated. We have used AIRS data
corresponding to July 20th 2002 as input for SMART.   
Viewing position: black curve, North Pole; light blue curve, South Pole;
green curve, Equator-longitude 0$^{\circ}$, red curve, equator-longitude
180$^{\circ}$.
  }   }} \label{fig:view}
\end{center}
\end{figure}

Disk-averaged solar and IR spectra of Earth  were generated from several
vantage points (i.e. over the pole, over the equator, etc.) with the same illumination
and with no clouds (fig. \ref{fig:view}).
In the optical, the main differences among the spectra are due to the changed surface composition.
In particular, in  fig.  \ref{fig:view} we show a day in Northern Hemisphere summer where 
the Southern polar cap is more extended (ice has a very high reflectance in the optical, fig. \ref{fig:surf})
and the Northern hemisphere is more forest covered.
There are visible discrepancies between the two equatorial viewing geometries due  to the high presence of land/vegetation in one case  and Ocean on the other.   Both cases simulate  a viewing geometry including a large amount of 
vegetation, and the black and light blue
 plots  show a very distinct red edge signal starting from 0.7 $\mu$m,

In the IR, on the contrary, 
 the observed variations are due to the horizontal temperature gradients. As we expect due to the time of the year, the lowest average 
temperature is 
 the one of the South pole view.
 
\subsection*{Sensitivity to phases}  
\begin{figure}[h!]
\begin{center}
\mbox{\includegraphics[width=1.5 cm]{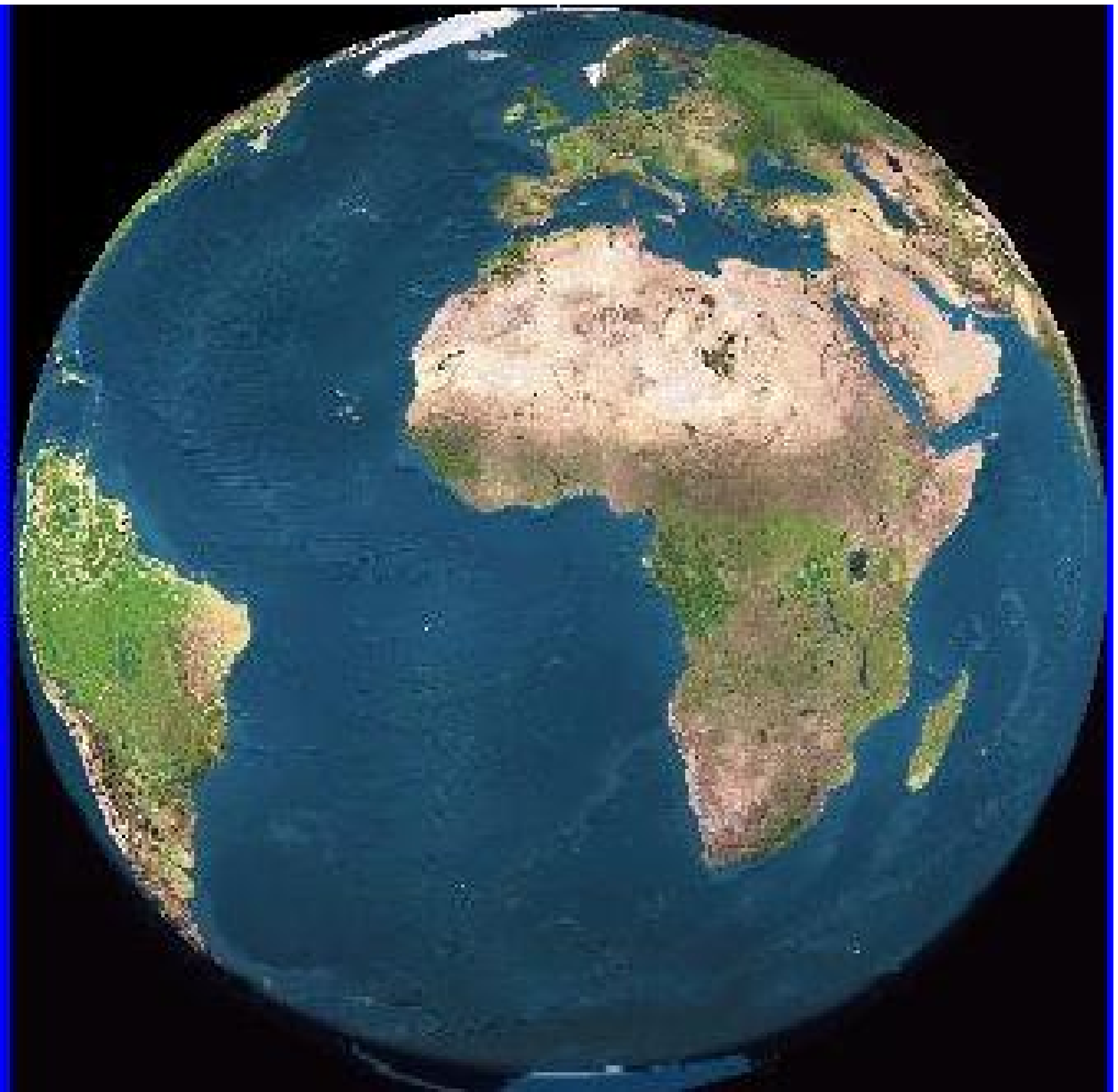} 
\includegraphics[width=7. cm, angle=0]{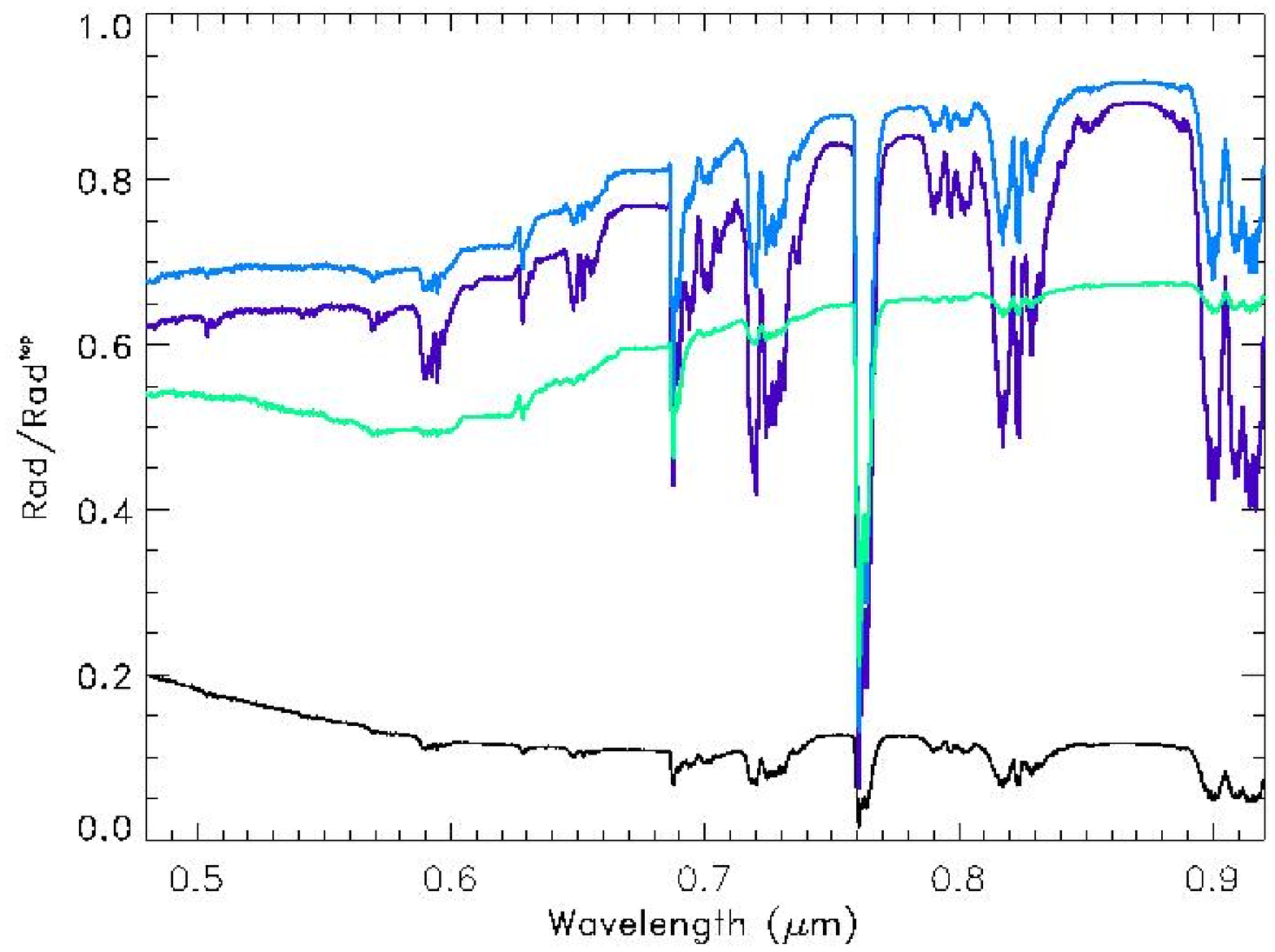} \includegraphics[width=6. cm, angle=0]{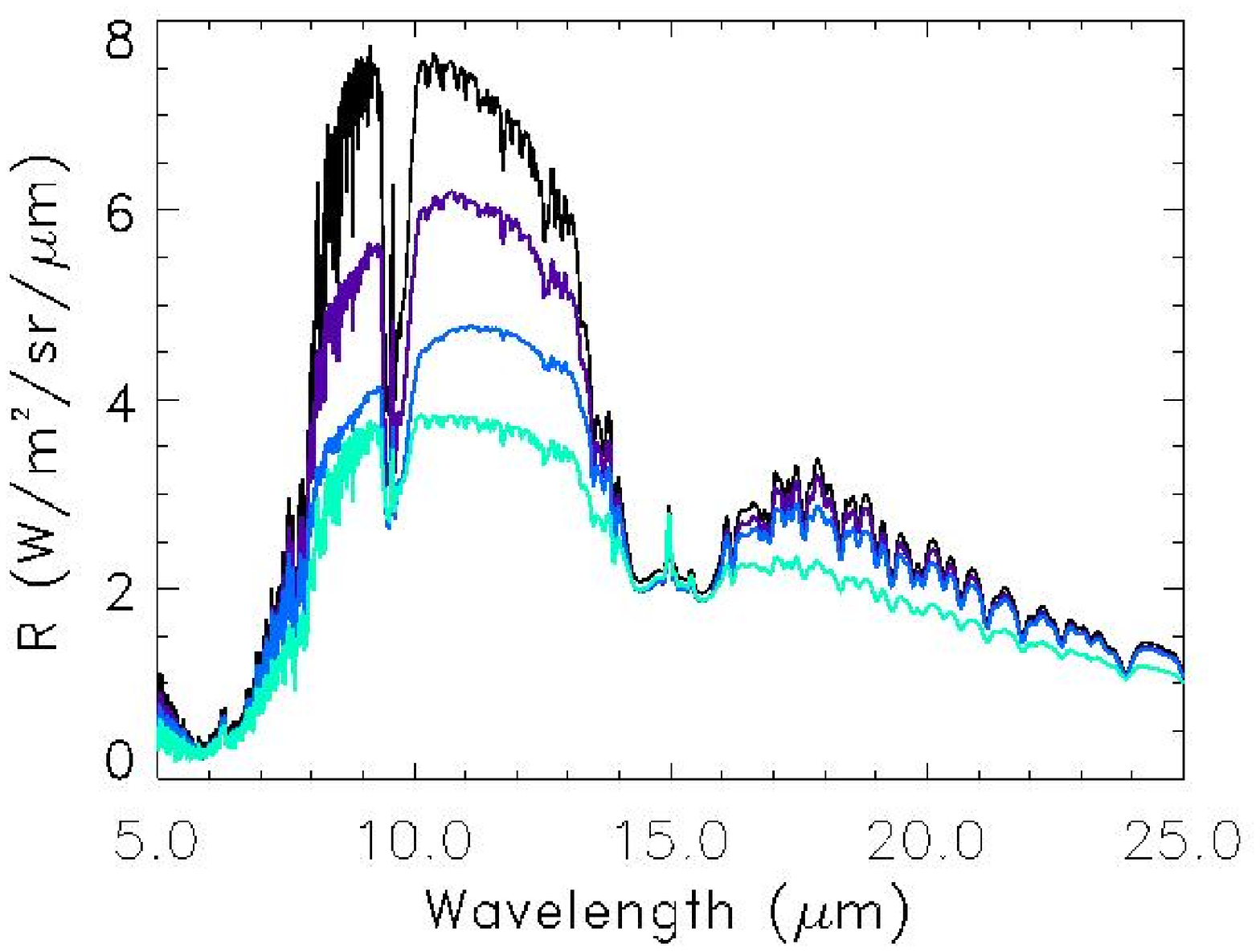} }  \\
\mbox{ \includegraphics[width=1.5 cm]{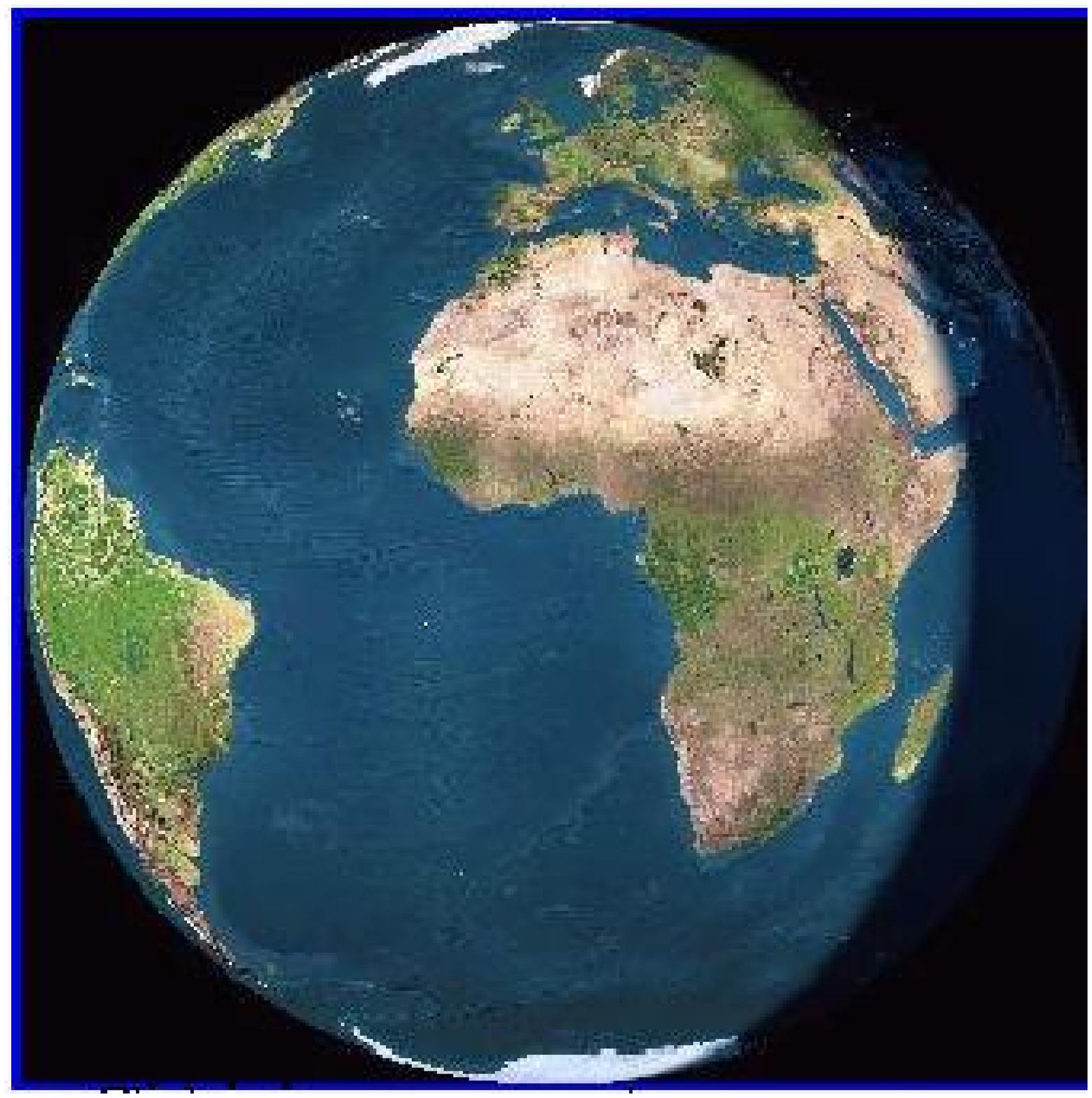} 
\includegraphics[width=7. cm, angle=0]{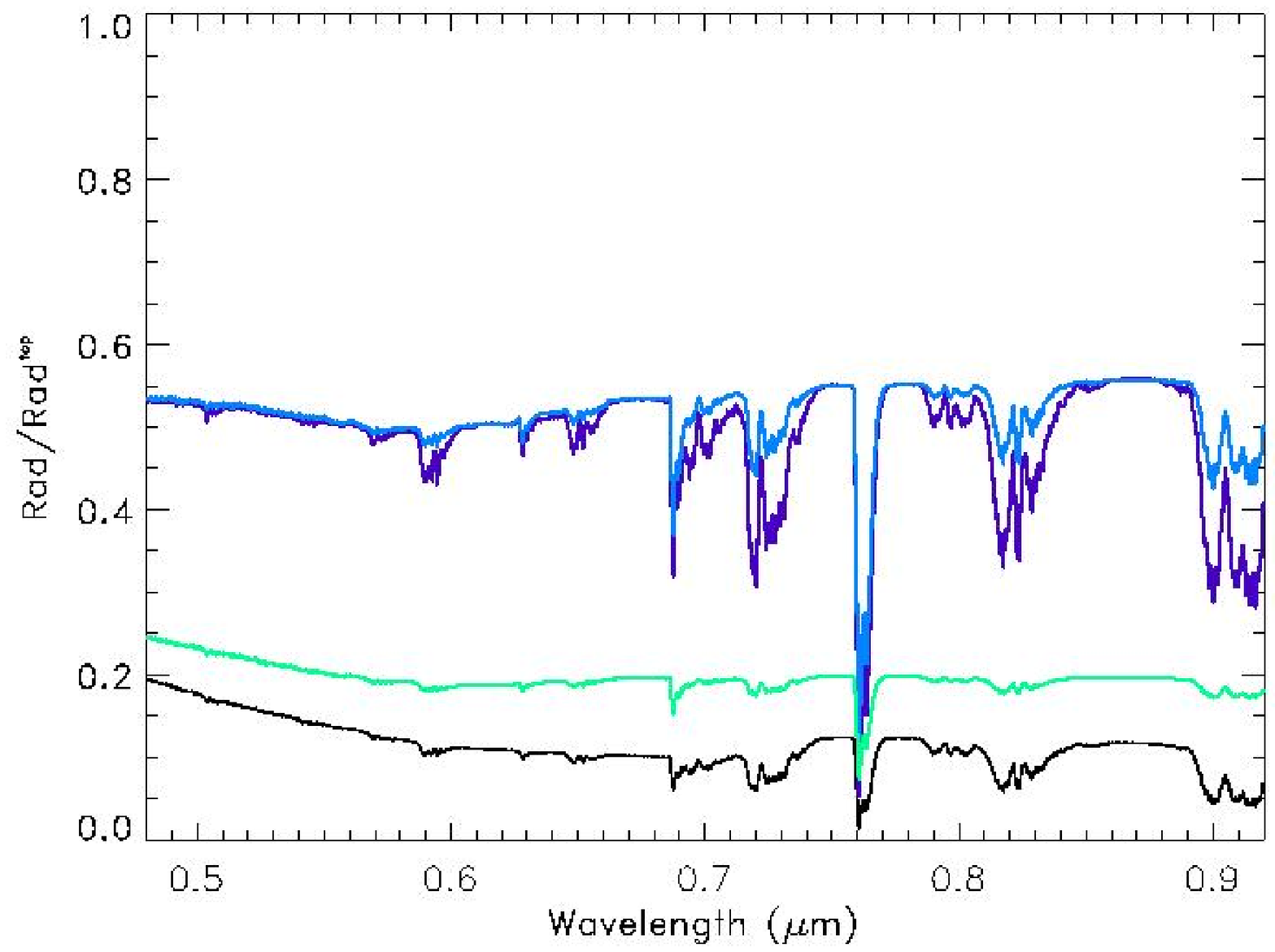}  \includegraphics[width=6. cm, angle=0]{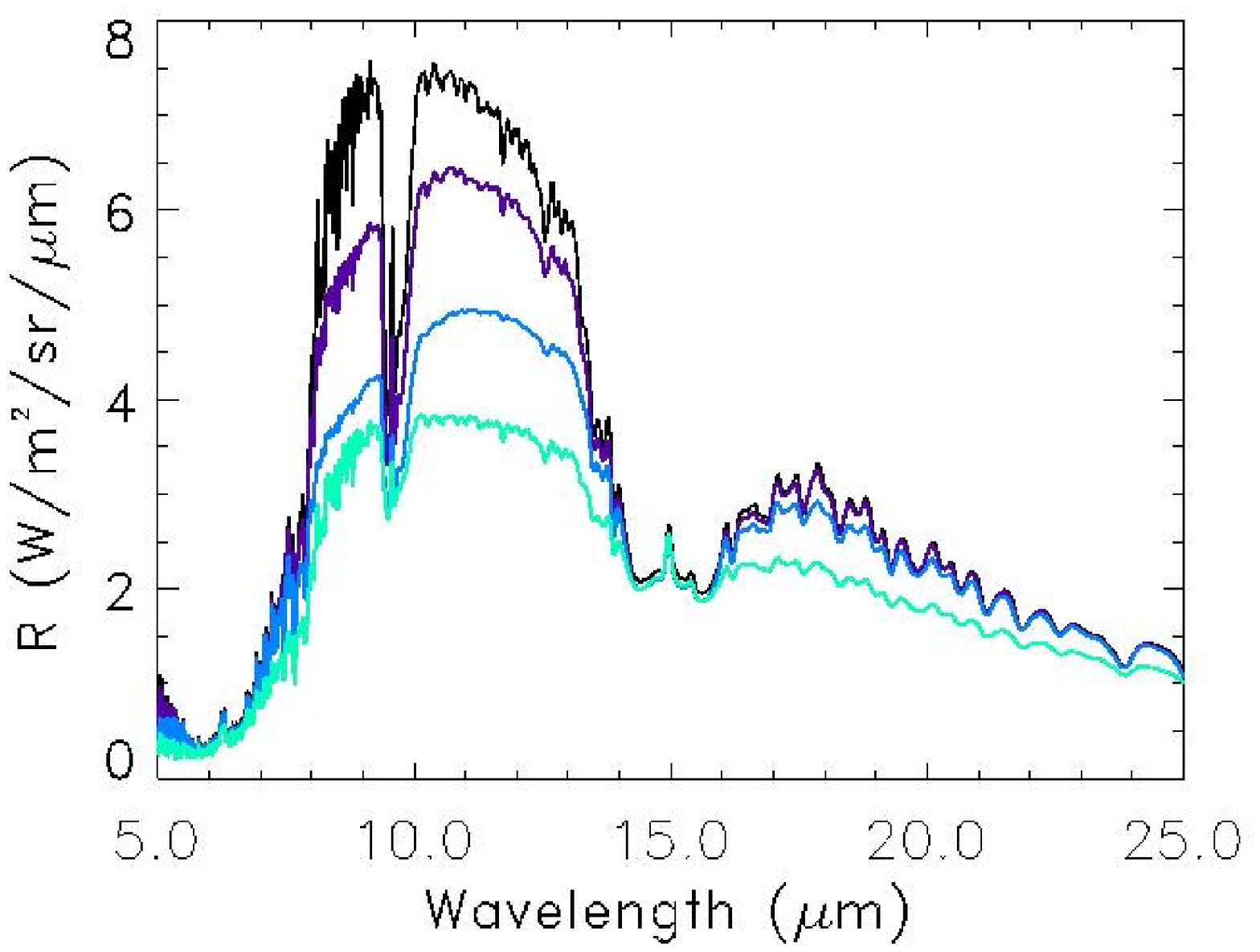} }  \\
\mbox{ \includegraphics[width=1.5 cm]{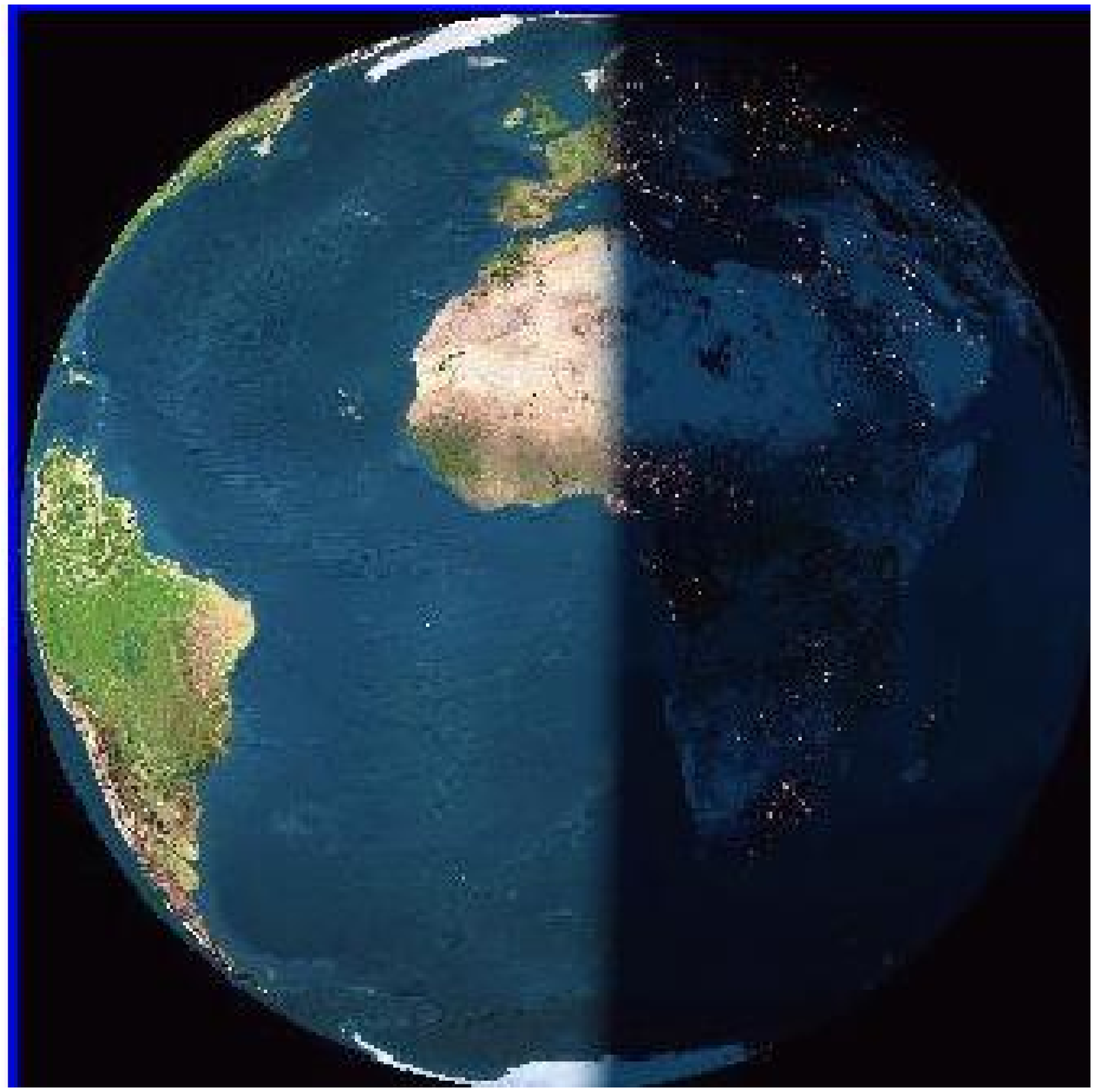}
\includegraphics[width=7. cm, angle=0]{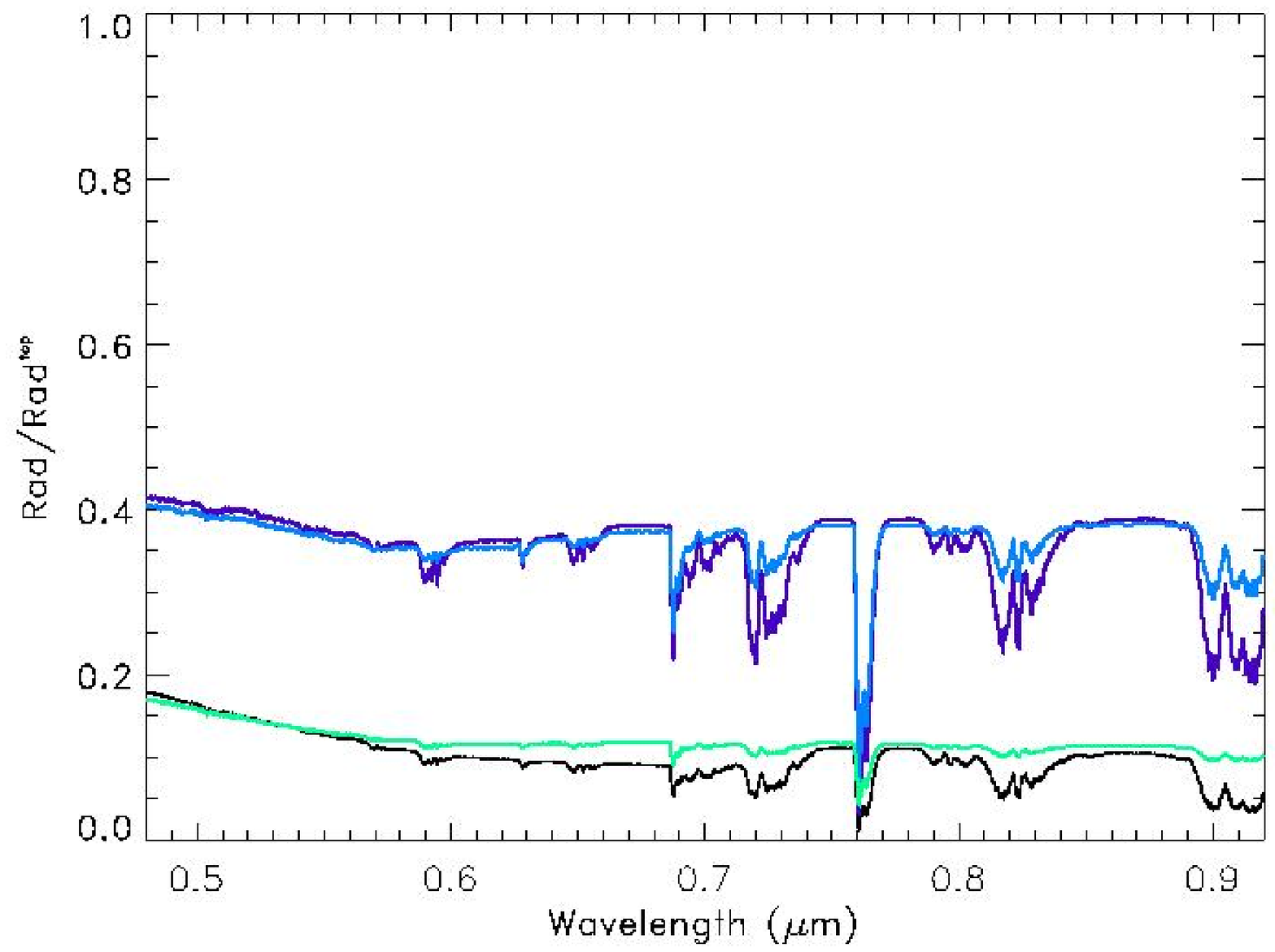}  \includegraphics[width=6. cm, angle=0]{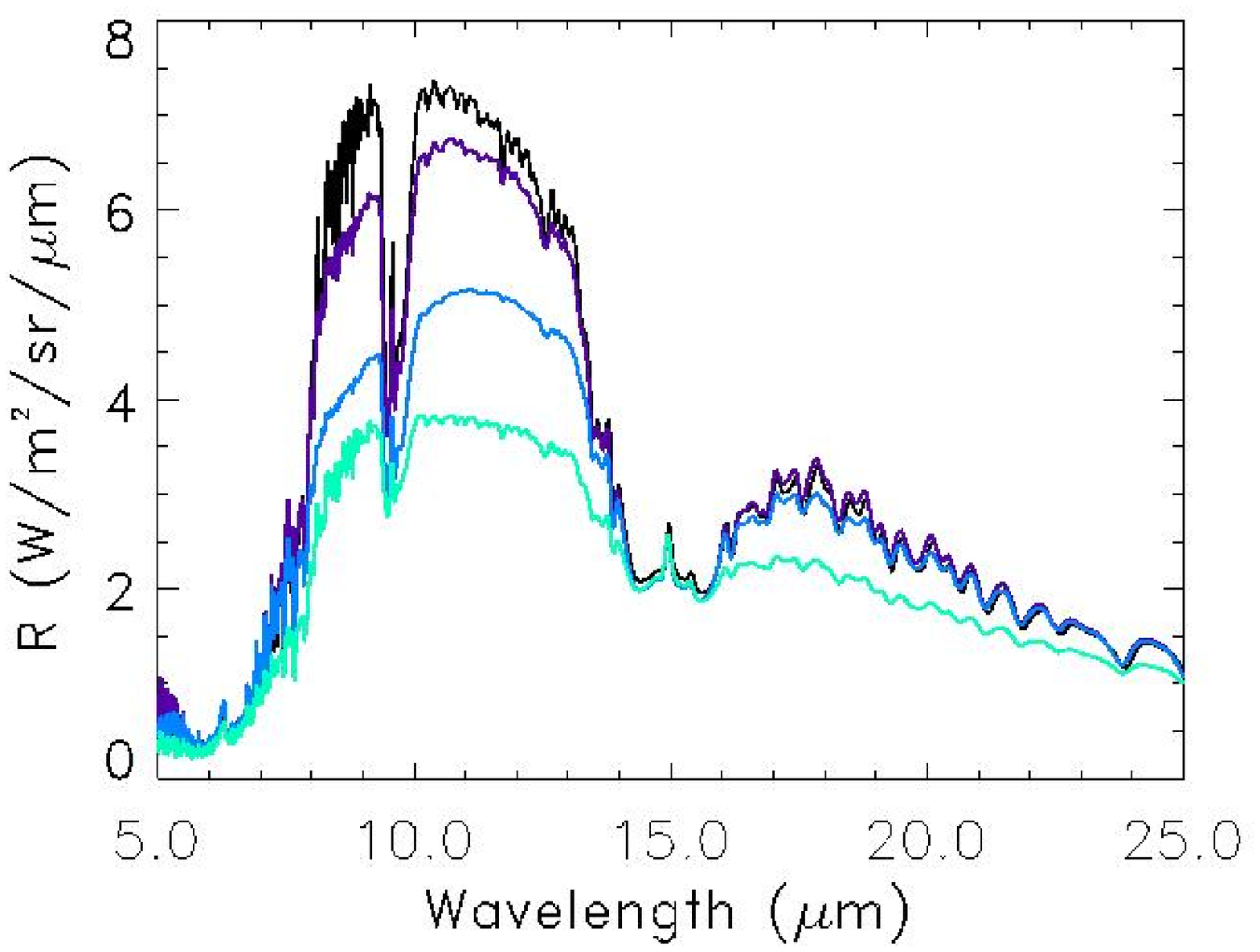} }  \\
\mbox{\includegraphics[width=1.5 cm]{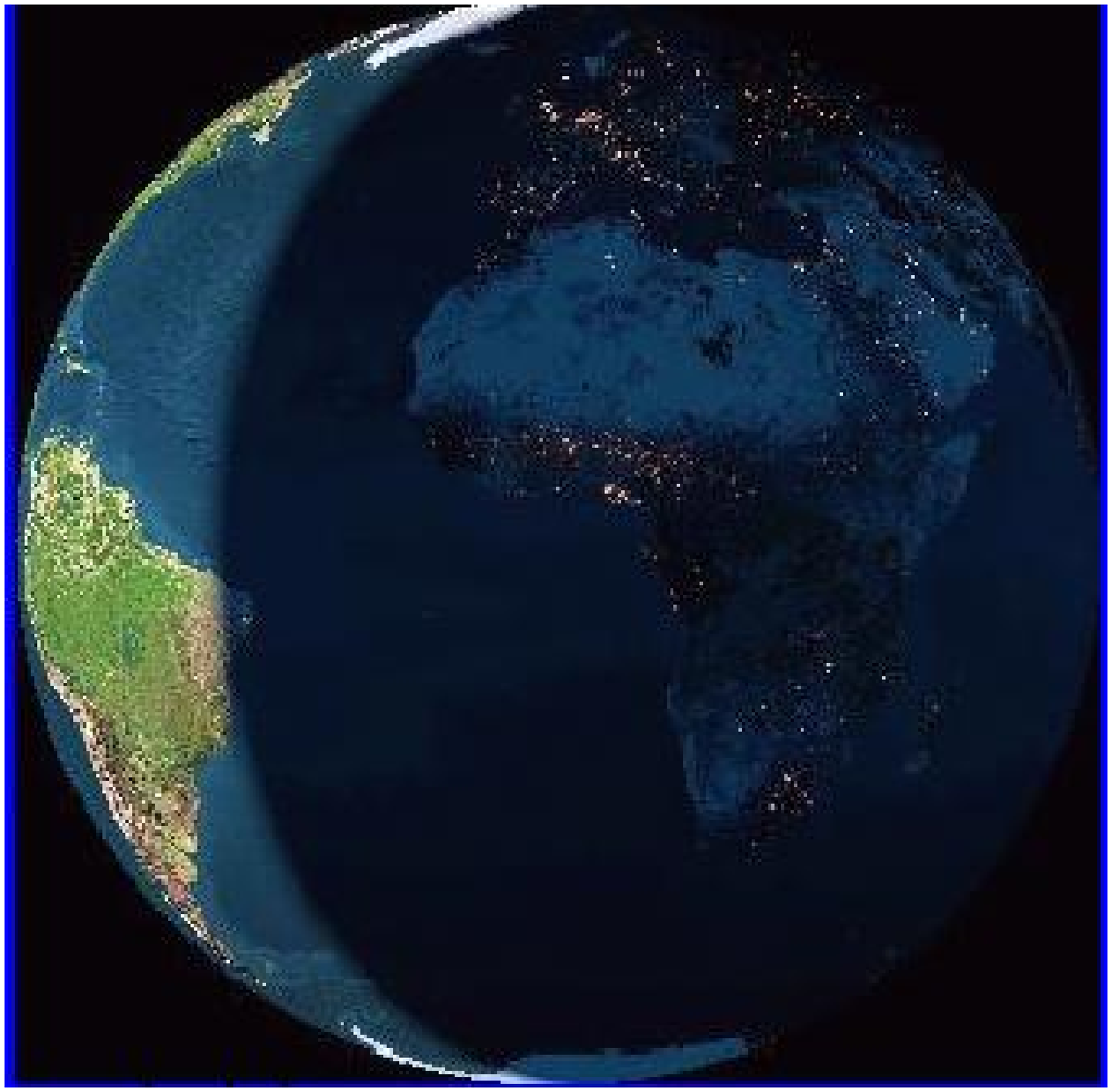}
\includegraphics[width=7. cm, angle=0]{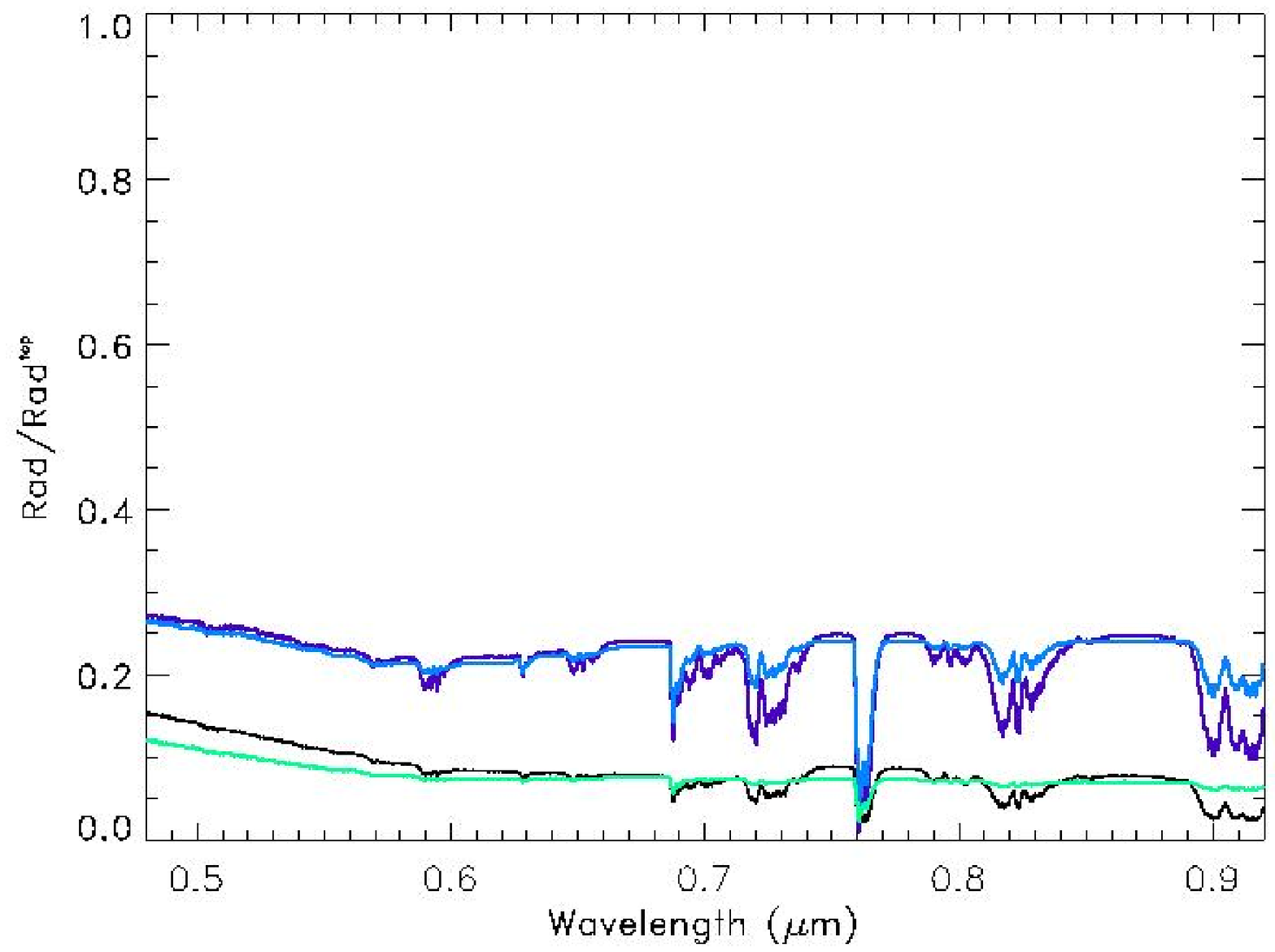} \includegraphics[width=6. cm, angle=0]{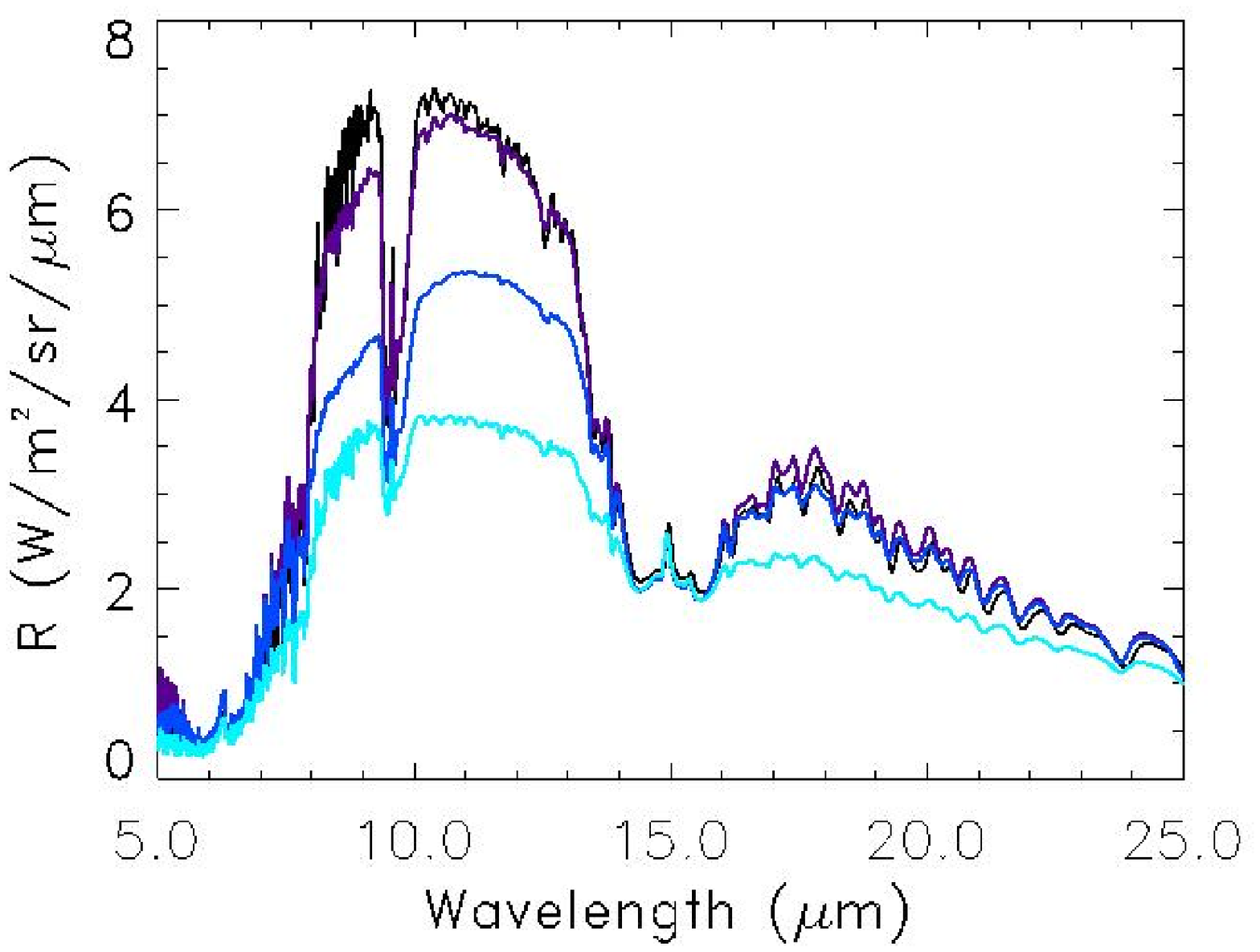} }  \\ 
\caption{ { \footnotesize \emph{ The black plots are the cloud free disk-averaged spectra relative to the four different phases, 
the others represent the fully cloud covered 
disk-averaged spectra (light blue cirrus, blue Alto-Stratus, violet Strato-cumulus). 
    } } } \label{fig:earth_view} \end{center}
\end{figure}

We have run some experiments to explore the sensitivity of disk-averaged spectra to different phases. No pronounced differences are seen
in the IR, except for those caused by  changes in the temperature/cloud distribution
 due to a changed illumination ($\sim$ 6\% for the cloud-free case and $\sim$   12\%
for the low-cloud case, compare the four  
plots on the right in fig. 
\ref{fig:earth_view}).
Figure \ref{fig:earth_view} shows the   results for the optical as well. For the cloud free disk-averaged spectra (black plots) there are only slight discrepancies 
among the four cases. On the contrary, dramatic changes can be seen for the fully cloud covered 
disk-averaged spectra (light blue cirrus, blue Alto-Stratus, violet Strato-cumulus).
The differences  are both in intensities and shapes.
This behavior is mostly due to variations in atmospheric pathlengths above the clouds.
When the visible disk is fully illuminated all the angles from 0 to 90 contribute symmetrically to the final integration,  
when less than only half of the disk is illuminated, only the angles close to 90 contribute to the final result.

\newpage

\subsection*{Light curves}  
\begin{figure}[h!]
\begin{center}
\mbox{\includegraphics[width=6 cm, angle=0]{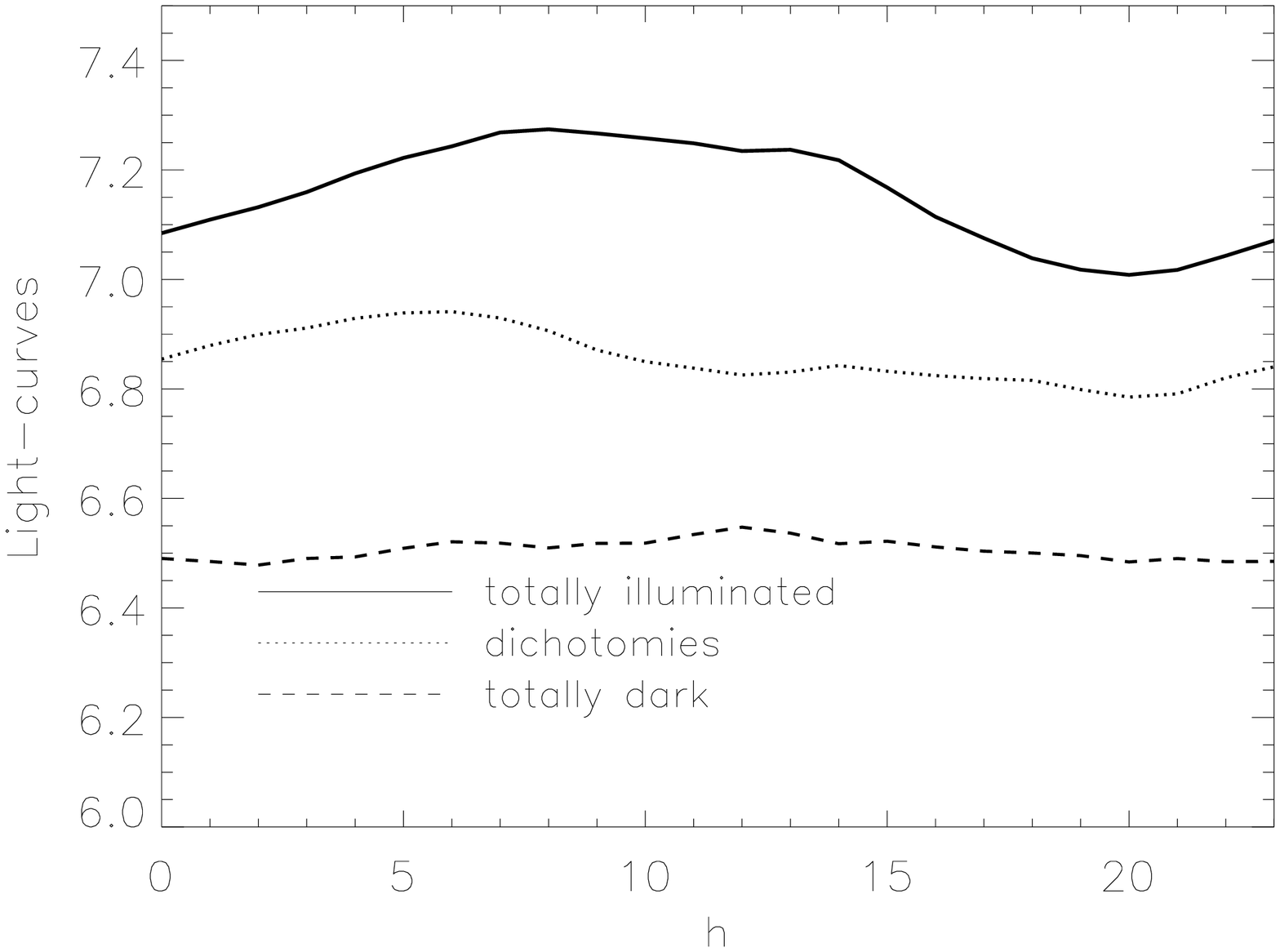}
\includegraphics[width=6 cm, angle=0]{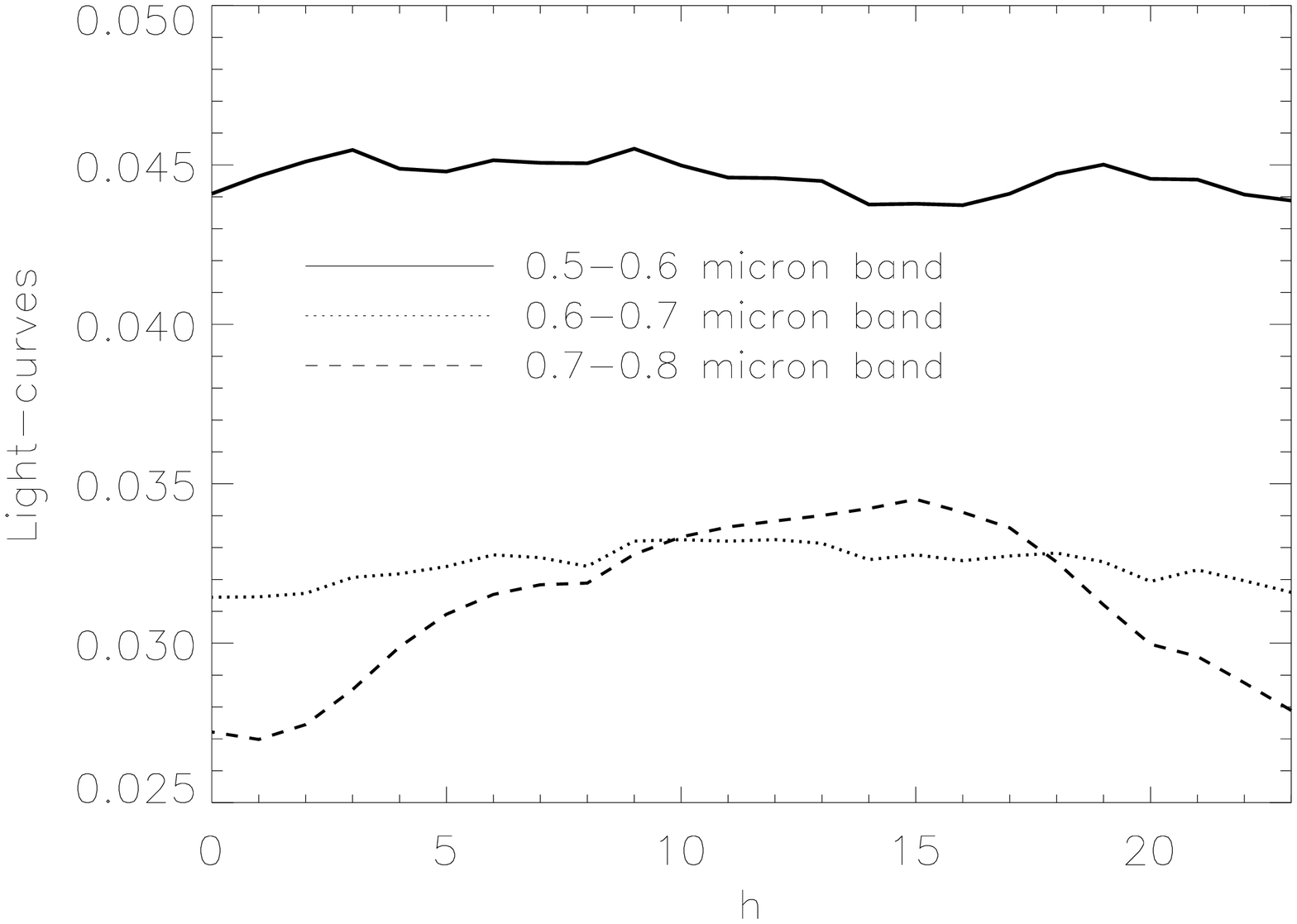} }
\caption{ { \footnotesize \emph{
Fig. on the left: light-curves following the diurnal rotation of the planet  for the spectral interval 8-13 $\mu$m
in the IR for three different phases.
Fig. on the right: light-curves following the diurnal rotation of the planet  for the spectral intervals 0.5-0.6 $\mu$m,
 0.6-0.7 $\mu$m, and  0.7-0.8 $\mu$m
in the visible. We see almost no variability in the 0.5-0.6 and 0.6-0.7 bands, the 0.7-0.8 band on the contrary shows a pattern due to the presence of the red-edge.
    } } } \label{fig:lc} \end{center}
\end{figure}

Time dependent variations in the disk-averaged spectra, or "light curves"
can provide additional information about spatial variations and periodicities. Fig. \ref{fig:lc} left shows
light-curves  for the spectral interval 8-13 $\mu$m following the diurnal rotation of the planet  for three different phases.
The phases considered are
totally illuminated, totally dark  and dichotomies. 
 The quantity plotted is $ \frac{ \int_{\lambda_{1}}^{\lambda_{2}} \, \mathcal{I} \, \text{d} \lambda }{\int_{\lambda_{1}}^{\lambda_{2}} \,  \text{d} \lambda} $, where $\mathcal{I} (\lambda)$ is the disk-averaged radiation,  $\lambda_{1}$ and $\lambda_{2}$ are the extremes of the chosen interval (8-13 $\mu$m). 
Although all the three curves clearly show              the daily periodicity, the rotational variations in this spectral range
 are of the order of magnitude of 3\% for the fully illuminated disk and 0.3\% for the totally dark disk. It is unlikely that this generation of 
extrasolar terrestrial planets missions will be sensitive to those fluctuations.
The relatively small horizontal gradients in the thermal are due to the presence of a quite thick atmosphere
and an Ocean that act as a buffer. On a planet like Mars, where the atmosphere is one thousand time smaller than the one on Earth and where there is not a huge reservoir of liquid water on the surface,
heat is exchanged above all  through conduction and radiation. As a result,
the temperature gradients between day and night and between different positions on the surface are 
extreme (Tinetti et al, 2004). The fluid component on Earth is playing a fundamental role in the global
energy balance, by storing heat through the water cycle (subtraction and release of latent heat) and in the smoothing of horizontal gradients through convection.

  Fig. \ref{fig:lc} right shows light curves in the visible for totally illuminated disks. 
For the chosen bands 0.5-0.6 $\mu$m,
 0.6-0.7 $\mu$m these light curves are almost flat lines.
The band  0.7-0.8 $\mu$m shows on the contrary high variability due to the presence of the red edge signal
(see next paragraph for a more detailed discussion).

\subsection*{Detectability of surface biosignatures}
\paragraph{Vegetation ``Red edge'' and NDVI idex} 
 The grass and forest albedo curves in fig. \ref{fig:albedo} illustrate the red edge, which begins between .680 $\mu$m  and .700 $\mu$m 
 and rises to a plateau at .761 $\mu$m.  The low reflectance in the .450-.520 $\mu$m  (blue) region is due to absorbance by carotenoids and chlorophylls, and that in the .630-.700 $\mu$m  (sometimes extending up to .730 $\mu$m) region is due only to chlorophyll, with the higher reflectance in the green giving the green color of most plants (Tucker, 1978). Because only the chlorophylls are actually involved in reduction of molecules relevant to photosynthesis (other pigments assist in light harvesting), we are interested primarily in the red portion of the visible absorbed by the chlorophylls for detection of photosynthetic activity.  Some regions of the NIR reflectance plateau may vary according to plant physiological status (Pe$\tilde{\text{n}}$uelas and Filella, 1998).  

	To quantify the strength (hence the potential detectability) of the vegetation ``red edge'', we use a commonly used index of vegetation activity, the Normalized Difference Vegetation Index (NDVI), which is a measure of the contrast between the red and NIR reflectances.  A high NDVI indicates high absorbance by chlorophyll and a well-hydrated status. Note that NDVI is only an indicator of photosynthetic activity, but its use in any actual quantification of photosynthesis is only correlative, because many other vegetation features (e.g. canopy structure, leaf surface characteristics) can cause variation in NDVI.  Nonetheless, it is a useful indicator.  
	NDVI is calculated as:
\begin{equation} \label{eq:ndvi}
NDVI = \frac{NIR - R}{NIR + R} 
\end{equation}
where $NIR$ is near-infrared reflectance, and $R$ is red light reflectance.										
 The choice of bands for calculating NDVI varies among different workers depending on available data.  
Landsat Thematic Mapper sensors have bands at .63-.69 $\mu$m and .74-1.10 $\mu$m for the red and NIR, respectively (Tucker, 1978).  NOAA uses .58-.68 $\mu$m  
 and .725-1.000 $\mu$m  for the visible and NIR, respectively.   MODIS sensors, designed to avoid windows of atmospheric absorbance for bands specific to surface remote sensing, have the red band at .649-.679 $\mu$m  and an NIR band at .855-.875 $\mu$m  (Salmonson, 1990).  
These ranges avoid atmospheric absorbance by O$_{2}$ and H$_{2}$O vapor.  

Here we calculate the disk-averaged NDVI of our  model using .65-.68  $\mu$m for the red and .75-.8  $\mu$m
(band I) or
.855-.875 $\mu$m (band II) for the NIR.

If we calculate the NDVI index for a particular surface type (fig. \ref{fig:surf}), we find the following results:

\begin{tabular}[h!]{lc}
\textbf{Surface type} & \textbf{NDVI-band I} \\
Ocean & -0.15  \\
Grass & 0.34  \\
Forest & 0.4  \\
Tundra & -0.08  \\
Desert & -0.05  \\
Ice & -0.05  \\
High (cirrus) & 0.03 \\
Medium (cumulus) & 0.01  \\
Low (stratus) & 8. 10$^{-3}$   \\
& \\
\textbf{Surface type} & \textbf{NDVI-band II} \\
Ocean & -0.2  \\
Grass & 0.4  \\
Forest & 0.44  \\
Tundra & -0.05  \\
Desert & -0.02  \\
Ice & -6. 10$^{-3}$   \\
High (cirrus) & 0.07 \\
Medium (cumulus) & 0.07  \\
Low (stratus) & 0.09   \\
\end{tabular}
\\
{\par \footnotesize{ \textbf{Table 2 -} Values of NDVI indexes (band I and band II) for different surface and cloud types. In the clear sky case, only vegetation has a positive NDVI index.} } \\

In a clear sky case the only surface showing a positive NDVI index is vegetation. 
Clouds may produce a positive NDVI index for certain phases, viewing geometries and distribution patterns. The numbers
we report in table 2 for clouds refer to a pessimistic case -totally illuminated, cloud uniformly distributed-. Even if those numbers are considerably lower than the vegetation case, 
clouds may cover a larger area  of the visible disk  compared to vegetation. If we keep 0.03 -band I- and 0.09 -band II-  as thresholds to discriminate unambiguously the presence of vegetation, every value between 0 and the thresholds might be explained either by clouds or vegetation.
 
If we calculate the NDVI index for the case of earthshine discussed in the validation paragraph (fig. 8),
Woolf et al. data give 0.015 (band I) and 7.$\cdot 10^{-3}$ (band II). Band I seems to indicate that vegetation is present in the disk average, since NDVI values for clouds for that particular phase and distribution
are all negative. Band II is ambigous, since the the NDVI index is positive but below the threshold. 

\paragraph*{Red-edge signal and clouds}

\begin{figure}[h!]
  \begin{center}
\includegraphics[width=12 cm, angle=0]{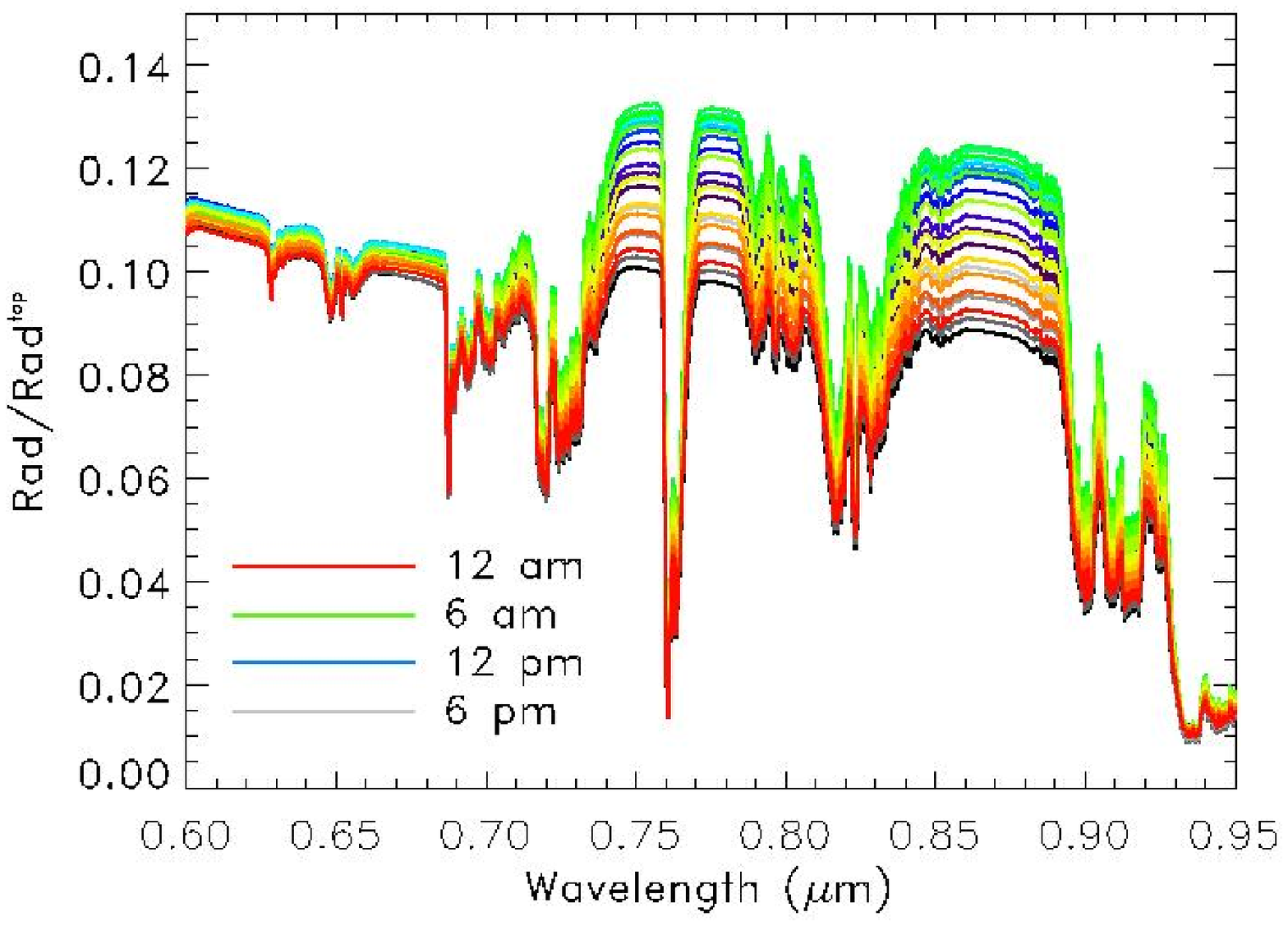} \\
\includegraphics[width=2 cm]{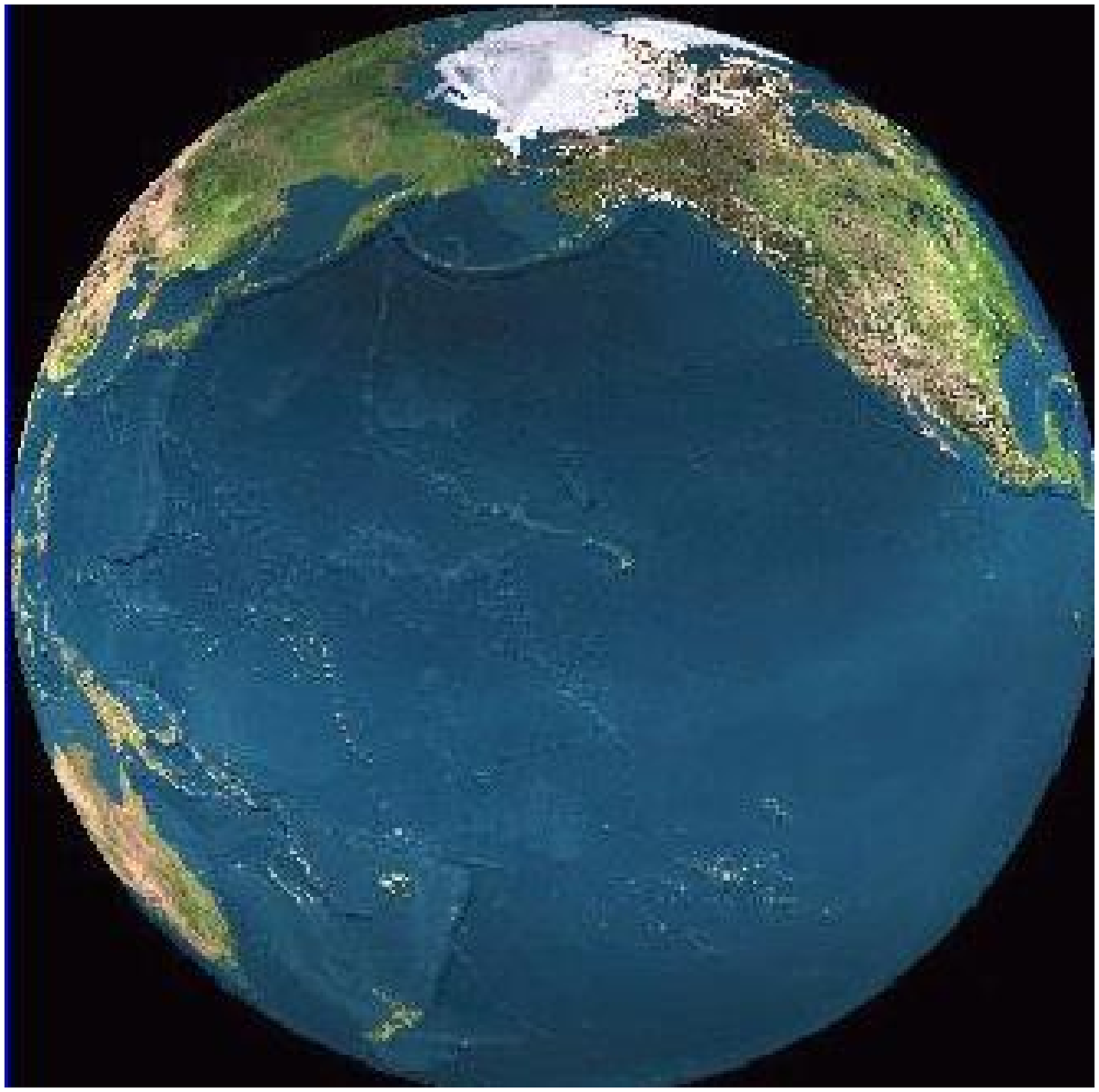} \includegraphics[width=2 cm]{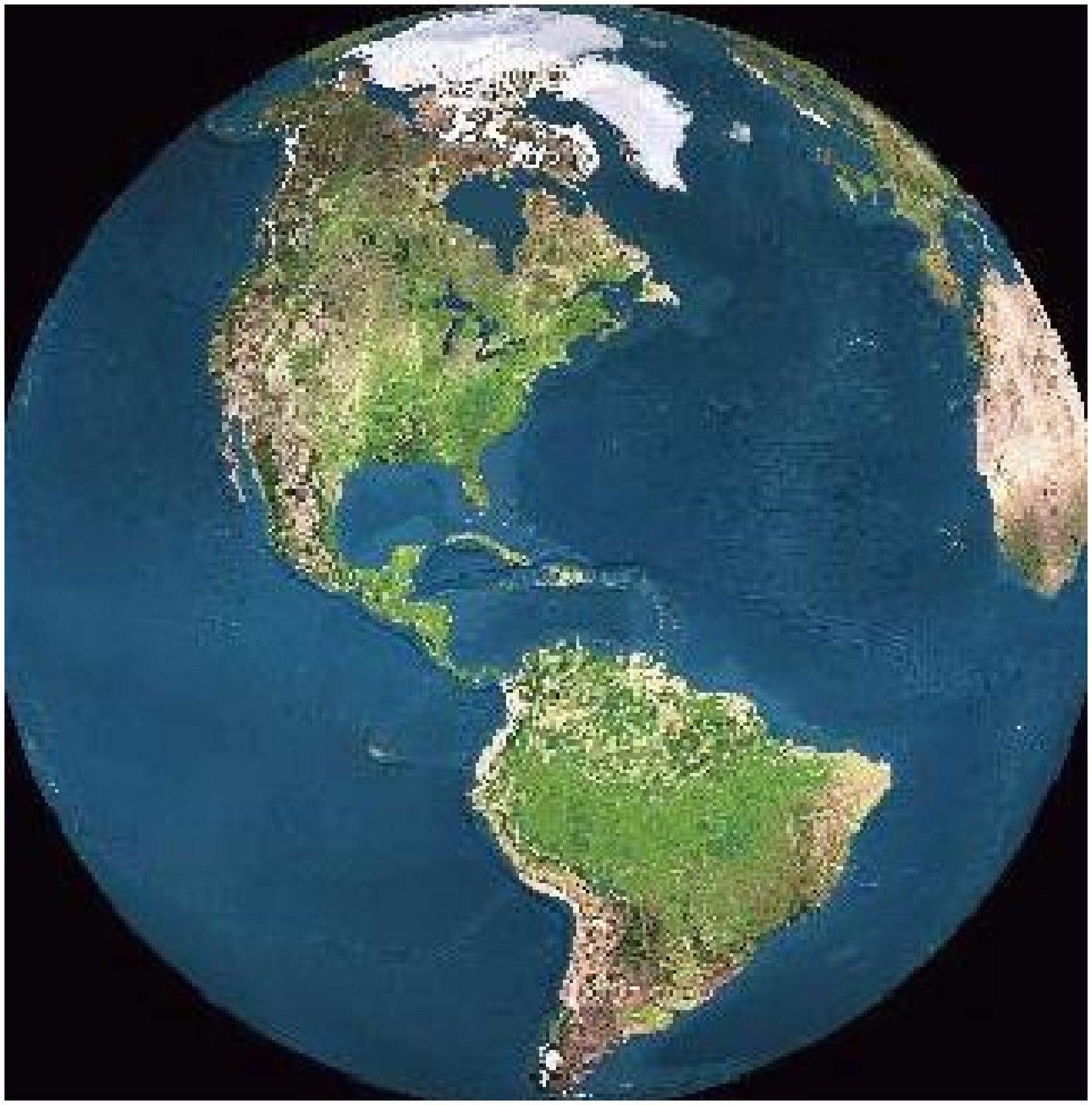}
\includegraphics[width=2 cm]{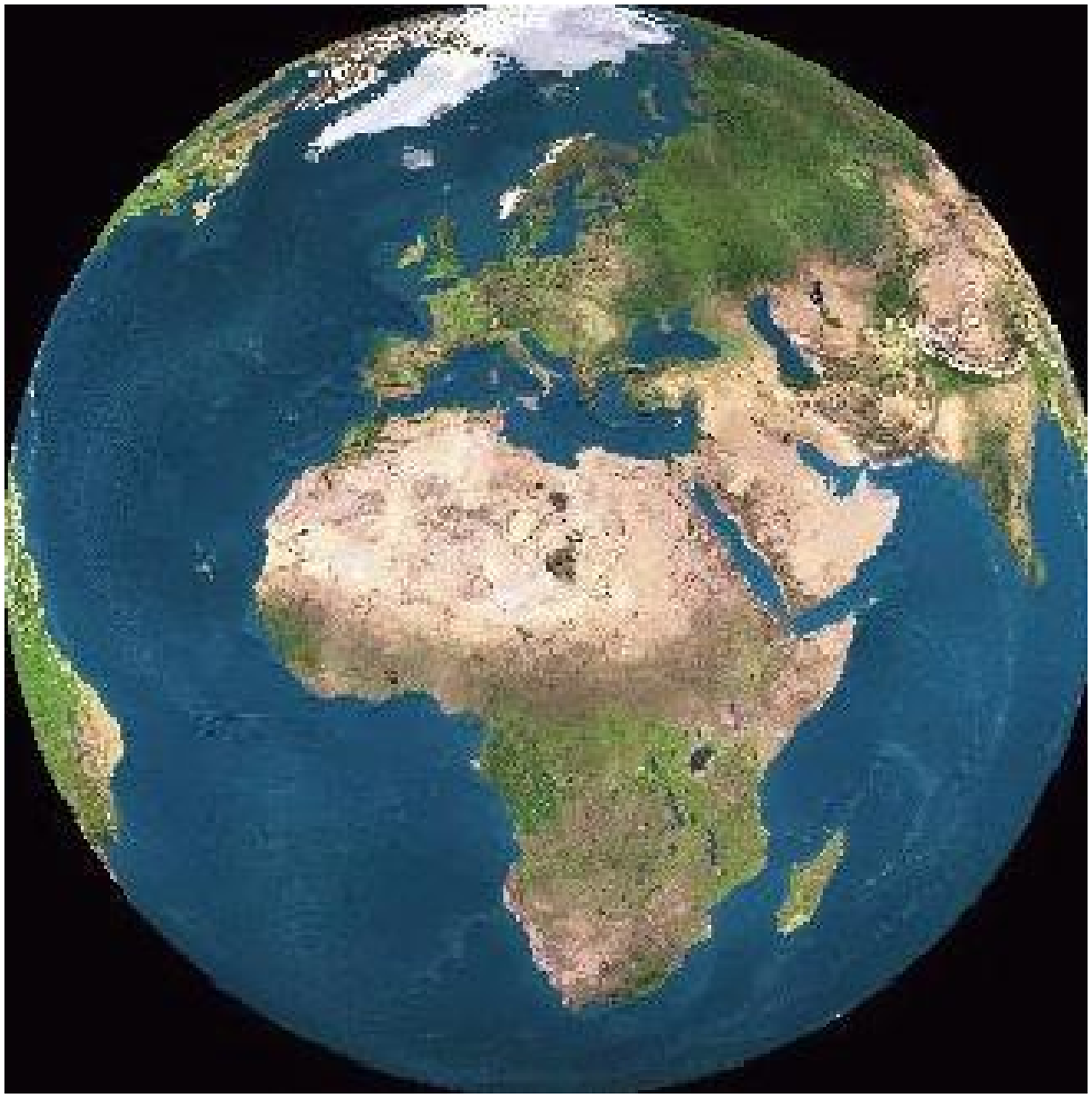} \includegraphics[width=2 cm]{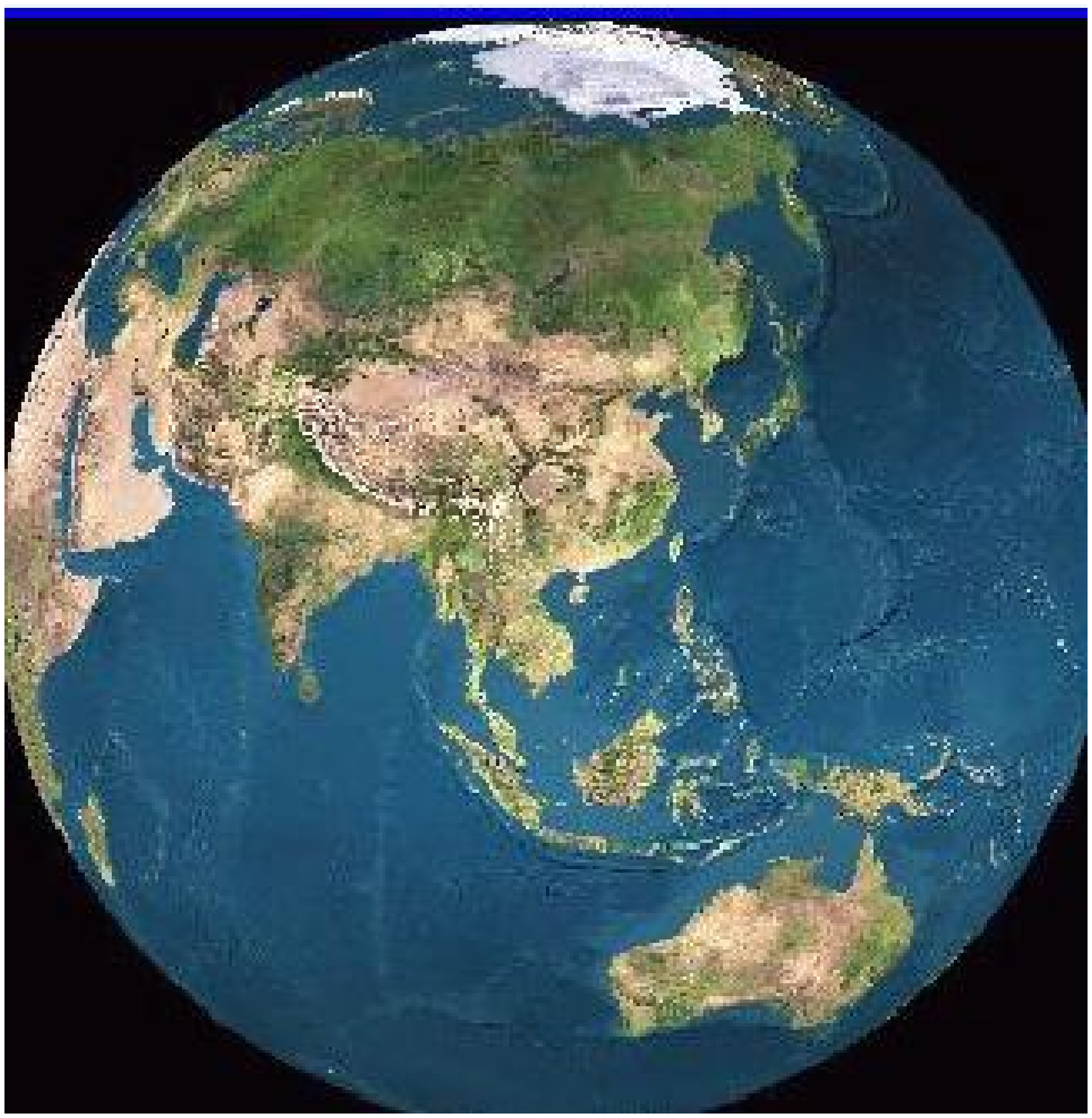} \\
\mbox{\includegraphics[width=7 cm, angle=0]{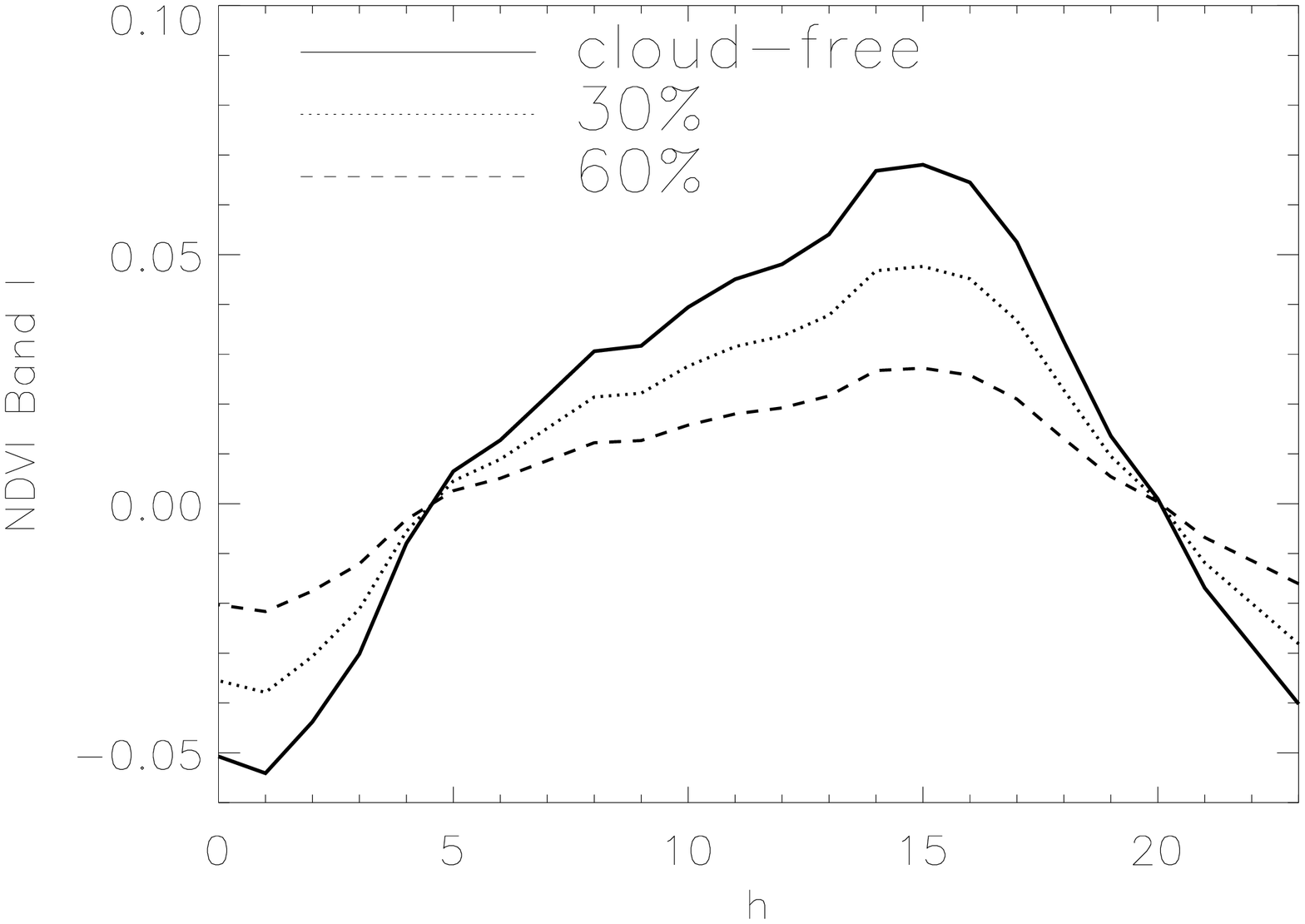}
\includegraphics[width=7 cm, angle=0]{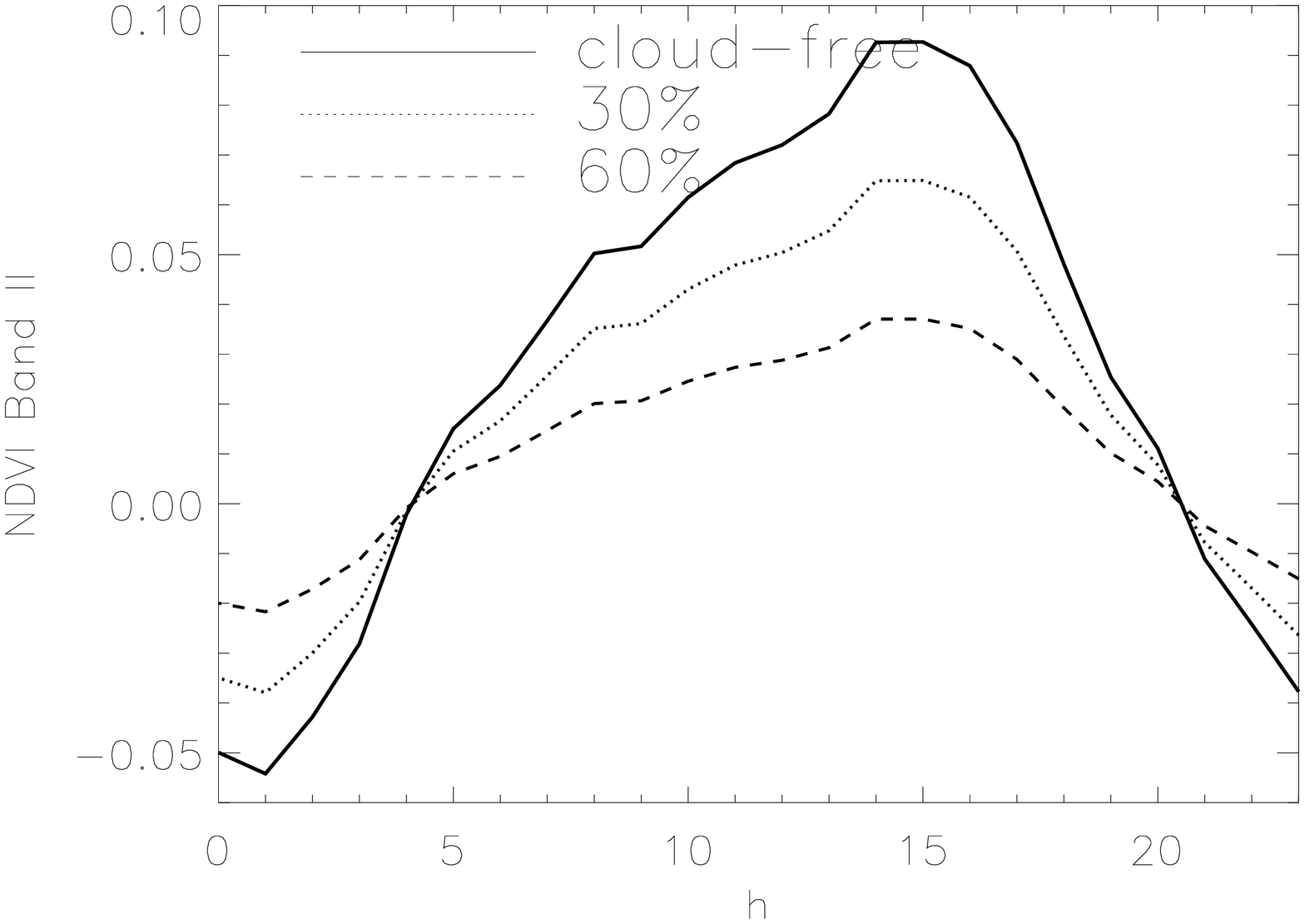} } \\
\caption{ { \footnotesize \em{ Fig. at the top:  disk-averaged spectra in the optical of a cloud-free Earth seen from different views 
following the diurnal rotation
of the planet (ordinates in Rad/Rad$_{top}$).   We show also below examples of the viewing geometries visible during the 
diurnal rotation (12 am, 6 am, 12 pm and 6 pm): as we can see from these pictures the land versus ocean fraction changes considerably
with time. 
 Fig on the bottom: light curves for a cloud-free, 30\% and 60\% cloud cover rotating Earth.
Here we calculate the disk-averaged NDVI of our  model using in eq (\ref{eq:ndvi}) .65-.68  for the red,  .75-.8 for the NIR (band I fig. on the left) and .855-.875 for the NIR (band II fig. on the right). For the cloud free case, when 
the NDVI is less or equal zero, the land  fraction covered by vegetation is not enough to produce a 
potentially detectable signal (less than 20\%).
As we can deduce from the figures the band II produces a stronger signal for the clear sky case, but when clouds are included, the NDVI threshold is considerably higher (table 2): band II appears less sensitive 
to the vegetation signal.
 }} } \label{fig:re}
\end{center}
\end{figure}

Fig. \ref{fig:re} shows modeled disk-averaged spectra over 0.6-.95 microns of different views of a cloud-free Earth during a diurnal cycle.  The NIR reflectance by vegetation is clearly visible, since it is responsible for a variation of 50\% in the signal.  The surface cover of vegetation on Earth
is evidently adequate to produce a strong enough disk-averaged signal to show a red edge feature when there are no clouds. Although the initial rise of the red edge at around 680 nm is obscured by absorbance by atmospheric water vapor, the latter part of the red edge is captured in the rise in the 0.73-0.81 range.  The plateau at .745 microns occurs earlier than the actual vegetation red edge plateau at 0.76 microns due to atmospheric water vapor.
The bottom plot in Figure \ref{fig:re} shows the sensitivity of NDVI, our red edge strength indicator, to different views of the Earth, cloud cover and for two different choices of the NIR band:  0.75-0.8 microns (band I) and
0.855-0.875 microns (band II).  Here,  the red edge can not be discriminated  when the Pacific Ocean dominates the view, while the peak detectability occurs  when Africa and the Eurasian continents  provide a view with the maximum land cover ($\sim$ 40\%). Considering the values in table 2, we can infer that vegetation is clearly present when NDVI index is positive (vegetation cover $\sim$ 20 \%). 

Introducing clouds (uniformly distributed) dampens the variability, as we see the red edge detectability is progressively reduced for cloud cover of 30\% and 60\%. For the clear sky experiment band II seems to be a better choice than band I, but when clouds are taken into account
band II produces a more ambiguous signal compared to band I (Table 2). What level of NDVI should be accepted as a true signal is a subject for the discussion section.

\begin{figure}[h!]
  \begin{center}
\includegraphics[width=9 cm, angle=0]{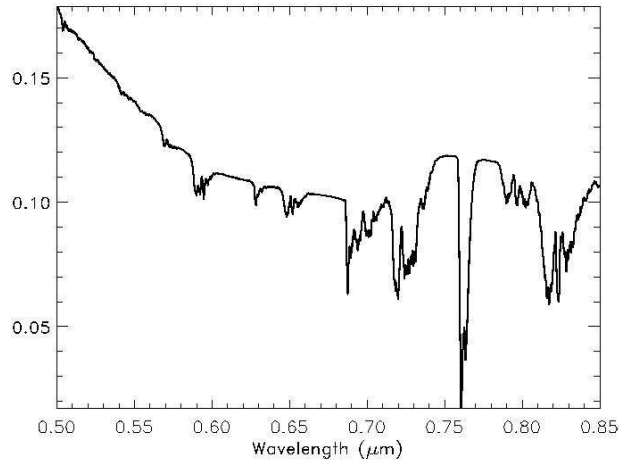}
\caption{ {\em { \footnotesize  Daily average of the sequence shown in the fig. at the top. In this cloud-free case we can still detect the red-edge even if we average over time.
 }} } \label{fig:media}
\end{center}
\end{figure}

The surface cover of vegetation on Earth is evidently adequate to produce a strong enough disk-averaged signal to show the red edge even when we average over the diurnal cycle (fig. \ref{fig:media}). This information can be useful for a mission like TPF-C which most probably will have to deal with integration times of the order of a day.

\begin{figure}[h!]
  \begin{center}
\includegraphics[width=12 cm, angle=0]{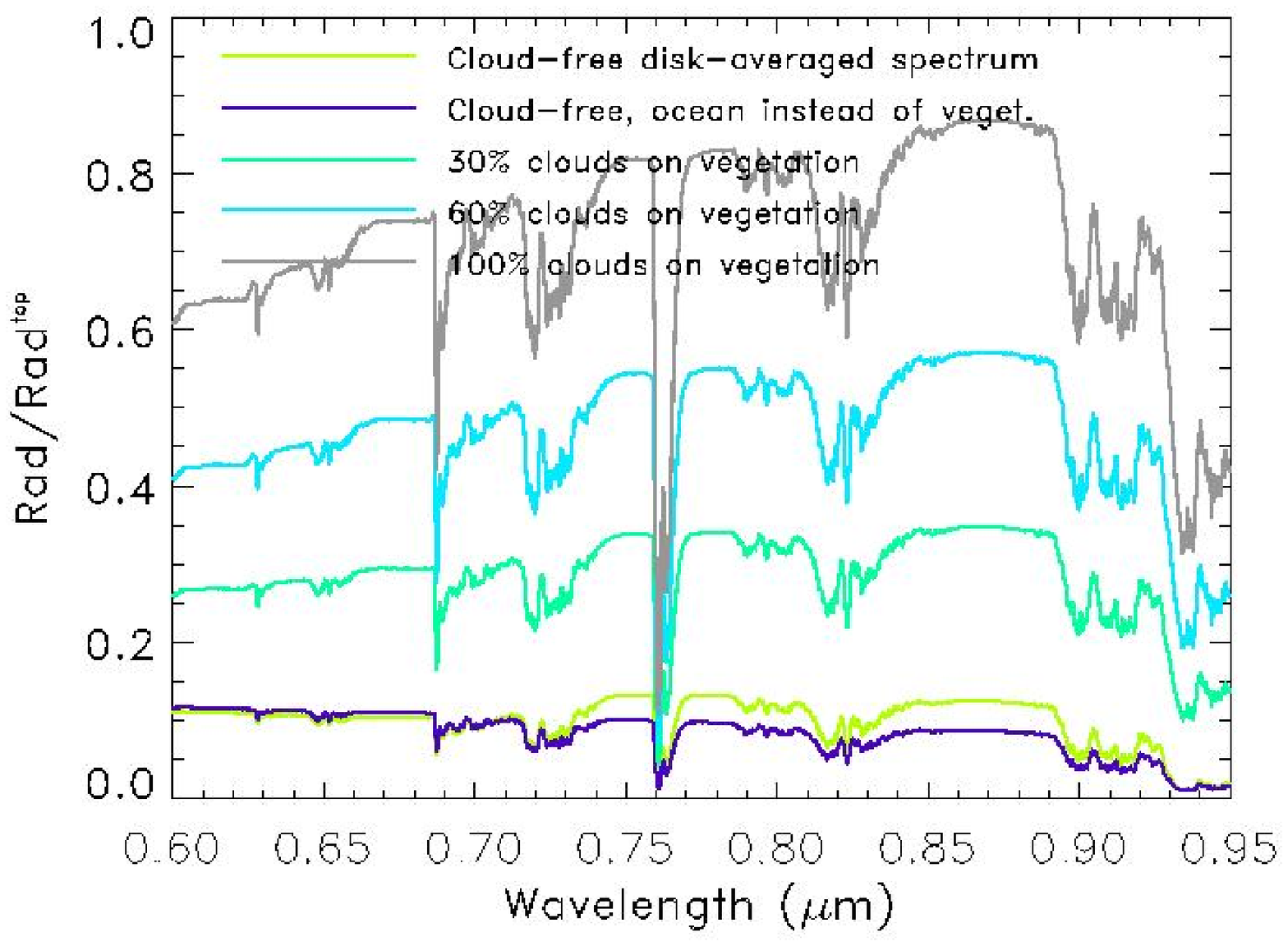} \\
\caption{ { \footnotesize
 \em{ Disk-averaged spectra in the optical of an increasingly cloudy Earth. The clouds are simulated to be present only on the part of the
surface covered by vegetation.  From the vantage point selected 
($\sim$ 40 \% cloud cover) it is easy to detect the red-edge signal (yellow-green curve, cloud-free). As a comparison we
have included the same simulation with ocean replacing  the land vegetation (violet curve).
The grey curve at the top shows the situation when clouds are totally covering the vegetation.
The other plots show intermediate situation (30\% and 60\% cloud cover).    
 }} } \label{fig:cloudre}
\end{center}
\end{figure}

Figure \ref{fig:cloudre}  shows the sensitivity of the disk-averaged and diurnally averaged spectra to percent cloud cover and to land versus ocean cover.  The clouds, in increments from cloud-free to 100\% cloud cover, are simulated to be present only on that part of the surface covered by vegetation.  The violet curve is a simulation with land vegetation completely replaced by cloud-free ocean.  


\paragraph*{Plankton}

Essentially all the light absorption that takes place in natural waters is attributable to four components of the aquatic ecosystem: the water itself, dissolved yellow pigments, the photosynthetic biota (phytoplankton and macrophytes) and inanimate particulate matter (tripton) (J. T. O. Kirk, 1983).

Water absorbs only very weakly in the blue and the green regions of the spectrum,
but its absorption begins to rise as wavelength increases above 550 nm and is quite significant in the red region:
a 1-m thick layer of pure water will absorb about 35\% of incident light  of wavelength 680 nm.

 The general tendency is that as phytoplankton concentration increases, reflectance decreases in the blue (400-515 nm)
and increases in the green (515-600 nm). The increased reflectance in the green may be attributed to the fact that phytoplankton,
being refractive particles, increases scattering at all wavelengths, but in this spectral region absorbs only weakly.

\begin{figure}[h!]
  \begin{center}
\includegraphics[width=12 cm, angle=0]{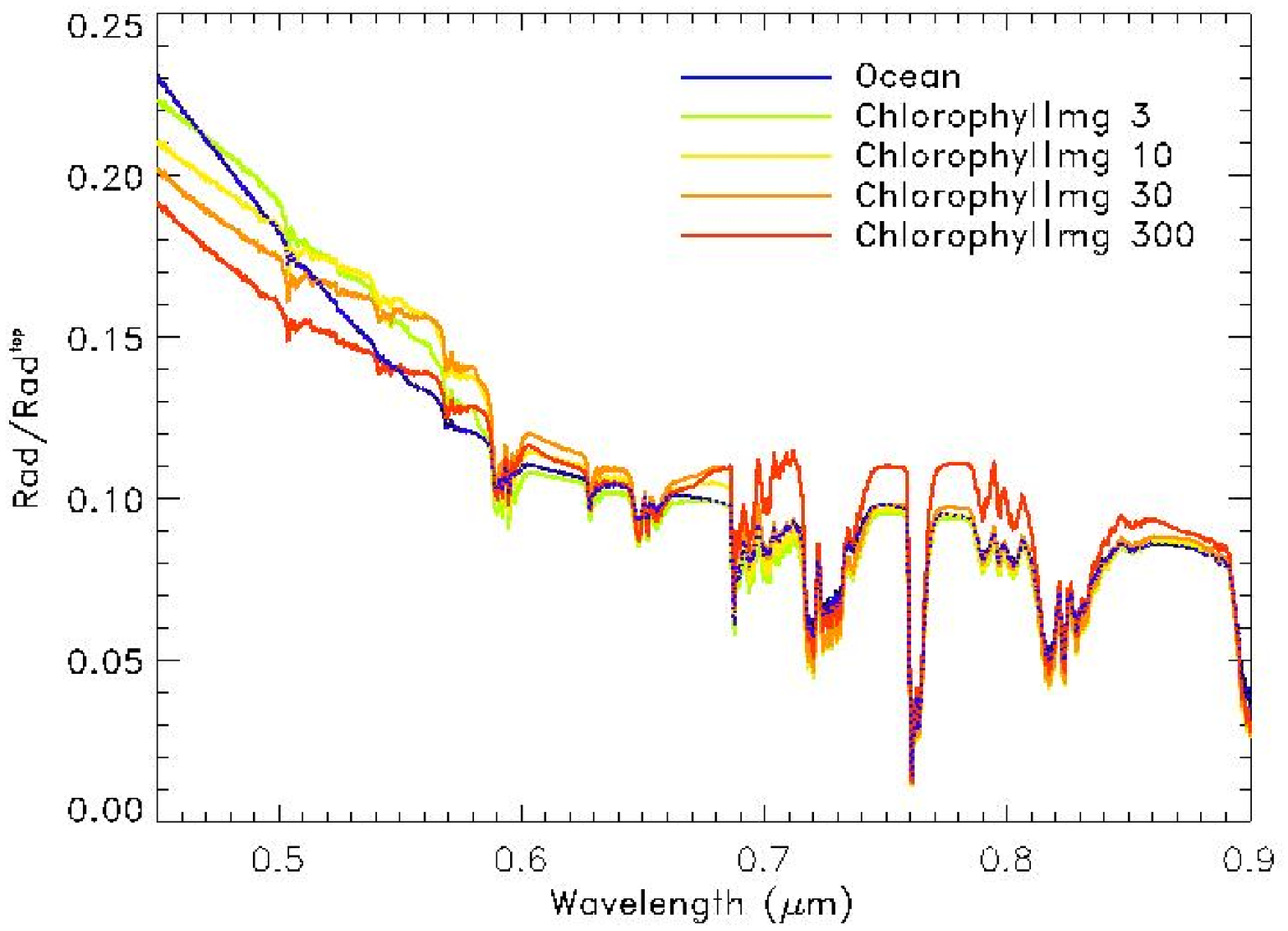} \\
\includegraphics[width=8 cm, angle=0]{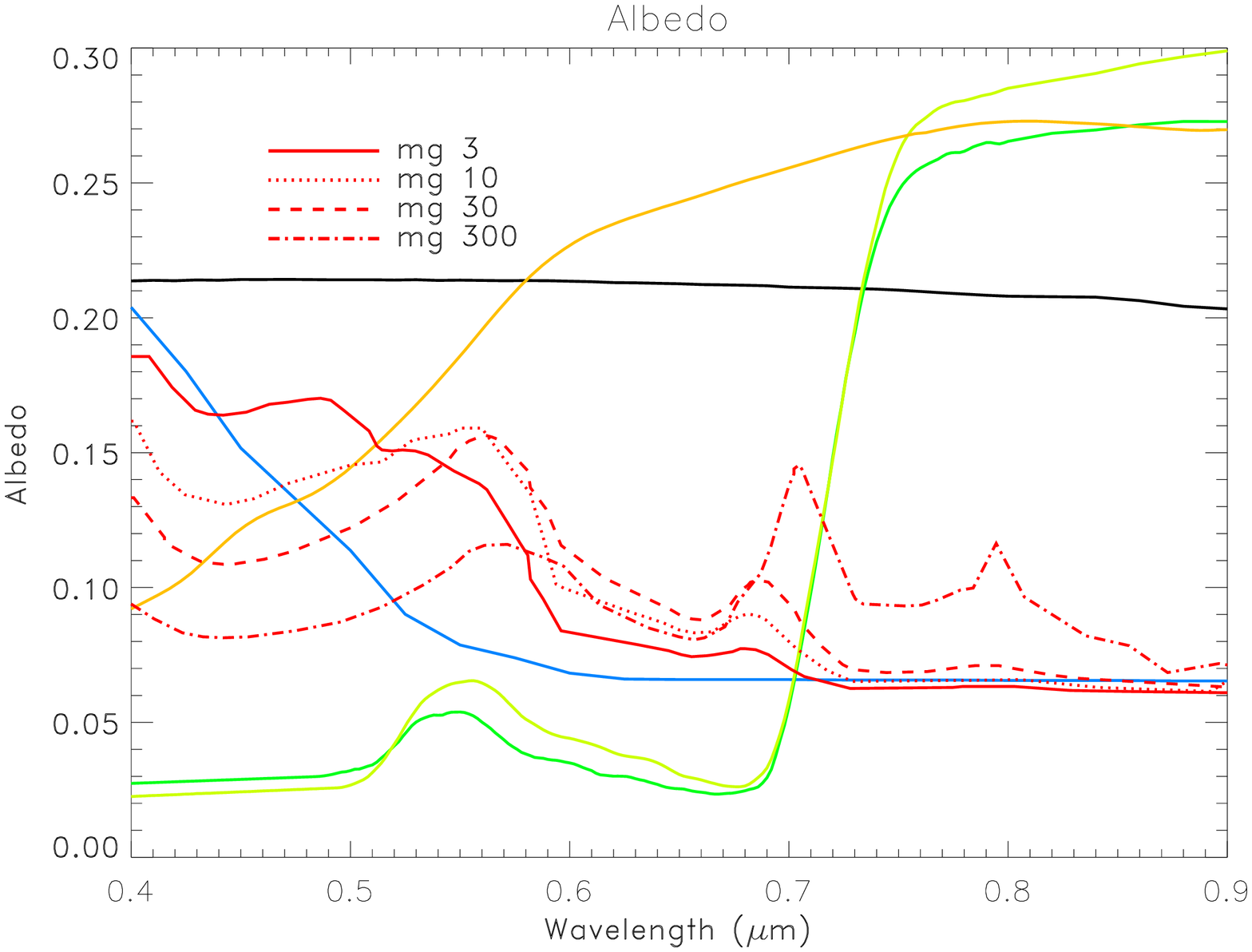} 
\caption{
 {\em { \footnotesize    Disk-averaged synthetic spectra of a  cloud-free, predominantly ocean-view of Earth. The concentration of plankton has been increased from 0 (sterile ocean, blue plot) to 300 mg/m$^{3}$  (extreme concentration, red plot).
For a present day Earth the average concentration floats between 0.1 and 10
  mg/m$^{3}$, with picks up to 30  mg/m$^{3}$. 
 }}} \label{fig:planktonx}
\end{center}
\end{figure}

About 1\% of the light a photosynthesizing cell absorbs is re-emitted as fluorescence, with a peak at about 685 nm (fig. \ref{fig:planktonx}). This fluorescence can show up as a distinct peak 
in the spectral distribution of the upwelling stream or in the curve of apparent reflectance against wavelength.
Calculations indicate that the increased fluorescence associated with an increase in phytoplankton chlorophyll of 1 mg m$^{-3}$
in the water would lead to an additional upward radiance of 0.03 W m$^{-2}$ sr$^{-1}$ $\mu$m$^{-1}$ above the water (Kirk, 1983).

Plankton also features a red-edge, but it is harder to detect than that for land plants.

In fig.  \ref{fig:planktonx} we show disk averaged spectra of an ocean view
of Earth, with an increasing concentration of plankton. This is an
 ``optimistic'' result, since the spectra are cloud-free, and we have chosen 
a particular viewing geometry and a particular season, in which plankton 
can be better detected (NASA-Goddard SeaWiFS website). 
The reflectance spectra used as input for our model were taken from (Gower et al., 2004) and
(Kirk, 1983). 

\begin{figure}[h!]
  \begin{center}
\includegraphics[width=12 cm, angle=0]{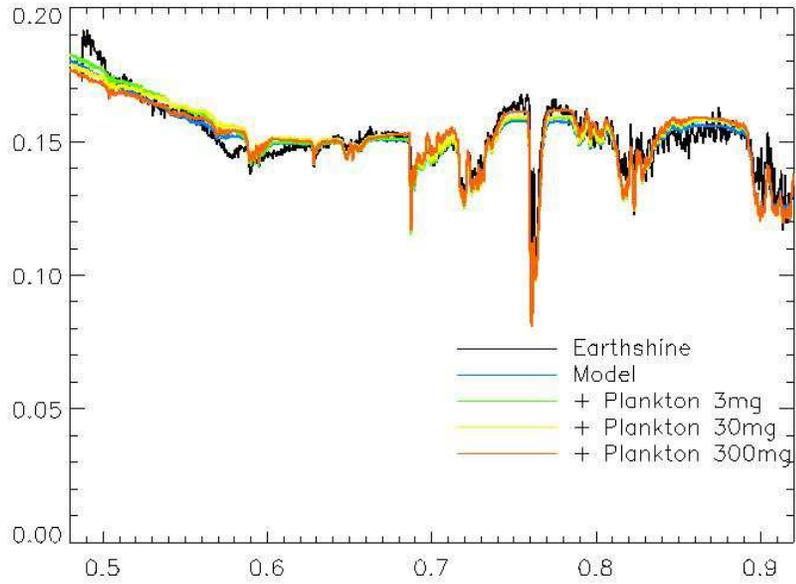} 
\caption{ {\em { \footnotesize  In this figure we show the comparison between the earth-shine spectrum (black plot) and our simulation for a sterile ocean (blue plot). We have then progressively increased the concentration of plankton in the ocean to test 
if its presence can improve our fit.
 } }} \label{fig:planktonxx}
\end{center}
\end{figure}

In fig. \ref{fig:planktonxx} we have increased the concentration of plankton and compared our result with the earthshine spectrum. When clouds are taken into account the presence plankton is less detectable.
In the month of the year when earthshine was recorded (June) we do not expect to have too much plankton. 
Adding the contribution of plankton to our simulation does not  necessarily
improve the fit between the data and the model. For high concentrations there is a better agreement with the recorded spectrum for wavelengths longer than 0.73 $\mu$m, since the red-edge signal is increased, but the fit is worse for shorter wavelengths. For low concentrations the contribution of plankton is negligible.

\subsection*{Simulation of a TPF detection of an Earth-like  planet} 
Our synthetic disk-averaged spectra were run through a TPF observation system 
simulator for different spectral resolutions.  The TPF book design (TPF book, 1999) has been 
assumed for these calculations and the planet was placed around a G star that is 10pc distant.   
TPF will provide only disk-averaged spectra with possible spectral 
resolving power ($\lambda/\delta\lambda$) of $\sim$75 (visible) and $\sim$25 (MIR), depending on the final architecture 
(TPF book), (Beichman and Velusamy, 1999).
\begin{figure}[h!]
\begin{center}
\includegraphics[width=14 cm]{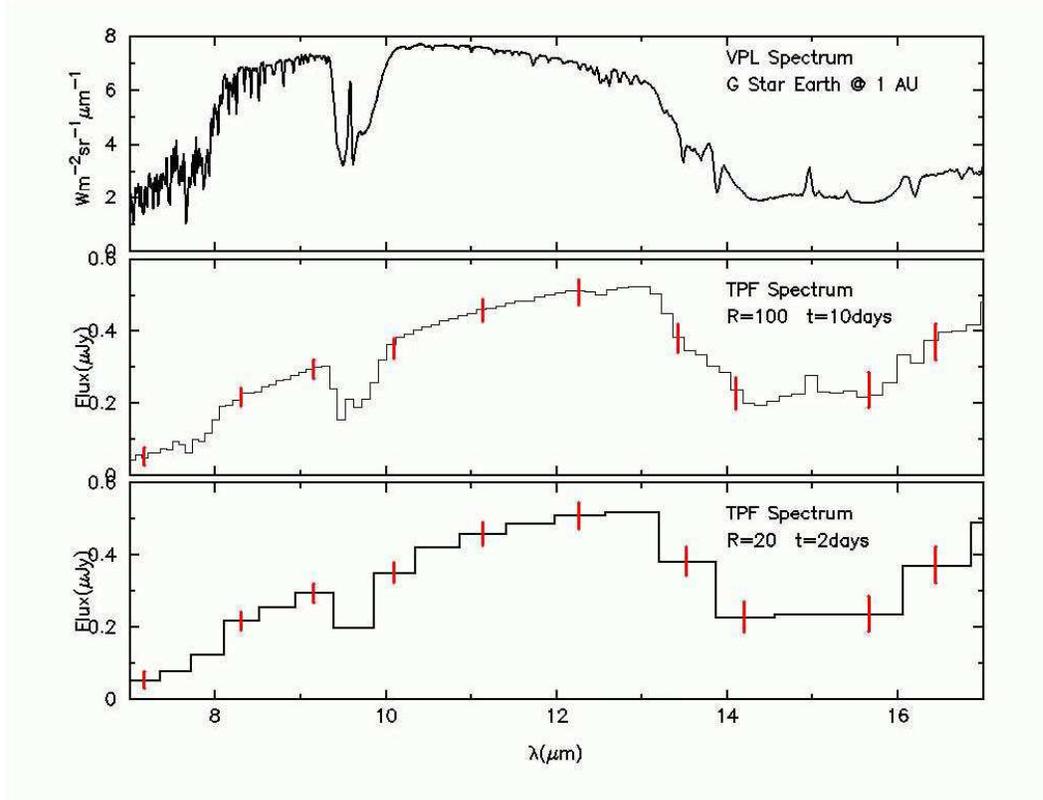} \\
	\caption{ { \footnotesize \emph{     Simulation of TPF-Interferometer detection of a  Earth-like planet orbiting around a 
G star.  The top spectrum in each set is at high-resolution, and the middle and 
lower panels show R$\sim$100 (at 10 days integration) and R$\sim$20 (at 2 days
integration) 
respectively.  1-$\sigma$ error bars are shown in red. 
  We recall that the wavelength range currently proposed and desirable should be 6.-17 microns  for the interferometer, with spectral resolution of 25-50.  } }}   \label{fig:velu1}
	    \end{center} 
\end{figure} 
\begin{figure}[h!]
\begin{center}
\includegraphics[width=14 cm]{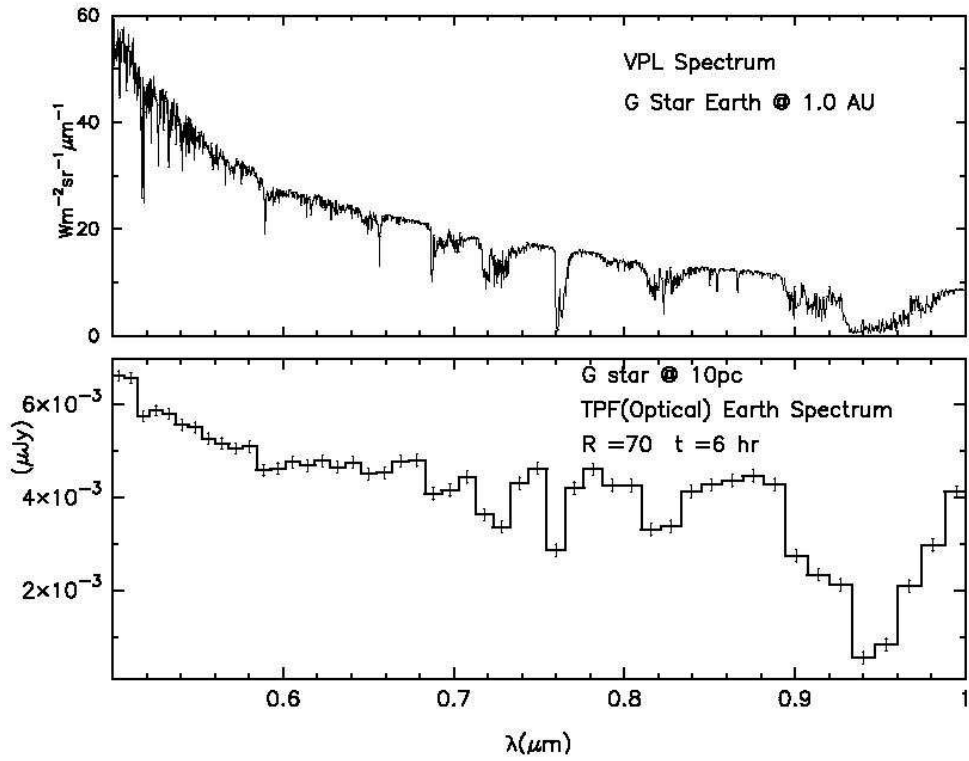} \\
\caption{ { \footnotesize \emph{   Simulation of TPF-Coronograph detection of a  Earth-like planet orbiting around a 
G star. These plots have been produced by processing the synthetic spectra with an observational system simulator of the TPF coronograph, at spectral resolutions R= 70 
with corresponding integration time of 6h.  The error bars 
in the spectra   represent rms noise (1 sigma) in the TPF 
measurement in each spectral channel for the integration.   } } }  \label{fig:velu2}
	    \end{center}
\end{figure} 

\section{Discussion}

Ford et al. (2002), as we already mentioned in the introduction, developed a model to evaluate the photometric variability in some specific bands in the optical (450, 550, 650, and 750 nm) due to daily rotation and seasons.
 In their paper, they predict a daily variation as high as 150\% for their cloud-free Earth model (750 nm band). According to our results
for the clear sky case
(see fig. \ref{fig:re}), the maximum daily variation is about 50\%  (750 nm band). The seasonal variation though, can be considerably higher  due to the changes in ice cover, which can dramatically increase the surface albedo
(fig. \ref{fig:surf} and  \ref{fig:view}). Clouds tend to increase the overall brightness and variability, reducing the fractional variation in the reflected light curve compared to the cloud free case
(fig. \ref{fig:surf}, \ref{fig:re} and \ref{fig:cloudre}). These conclusions although, in the case of an extrasolar terrestrial planet,  should be taken with some caveats, since they depend on the cloud pattern and rate of cloud formation
on that planet. We confirm their hypothesis that the diurnal rotation variation of an Earth-like planet
is lower in the IR  than in the optical (fig. \ref{fig:lc}), since the temperature gradients are not so dramatic
compared with the  surface albedo variations. This is due to the presence of a relatively thick atmosphere and of an ocean that act as buffer to the temperature excursions that otherwise would occur (see for comparison the Mars case (Tinetti et al, 2004)).

Previous workers have utilized Earthshine data to look for a vegetation signature in a disk-averaged spectrum (Woolf et al., 2002;  Arnold, et al., 2002).  In the case of Arnold et al., we cannot directly compare our results with theirs, since we do not know the details of these observations (spectra recorded, viewing geometries, time, illumination, etc.).  We agree, though, at least qualitatively, with their conclusions.

	The Earthshine data show that the atmosphere on Earth is clear enough to allow some of the vegetation signal to show through, but it is weakened.  Our model and disk- and diurnally-averaged NDVI show that the vegetation signal has the potential to be quite strong but also completely hidden (fig. \ref{fig:re}, \ref{fig:media} and \ref{fig:cloudre}).  The problem of clouds obscuring vegetation on land is compounded by the fact that vegetation promote conductance of moisture from the soil to the atmosphere, and vegetation roughness promotes moisture convergence and hence cloud cover at the regional and continental scales (Dickinson, 1989;  Friend and Kiang, accepted).  Cloud cover will tend to favor vegetated areas (and vice versa). The role of transpiration processes in the hydrological cycle on an extrasolar planet, should therefore be considered in narrowing the range of plausible detectable biosignatures.
	Choice of bands for calculation of the red edge detectability must clearly take into account the obscuring of surface signatures not only by clouds but also by other atmospheric gases.  Oxygen is a potential biosignature produced by photosynthesis and water is a prerequisite for known life; therefore, both will be present if a vegetation red edge is present, while both obscure parts of the red edge.  The choice of red and NIR bands for our red edge  index (disk-average NDVI) followed the choices of   .65-.68  for the red and .75-.8
(band I) or .855-.875 (band II) for the NIR  since the O$_{2}$ and H$_{2}$O absorbance lines all fall along the red edge feature.
NIR band I is subject to physiological variability, whereas the band II wavelength range is fairly stable, even across mosses, lichens, and cyanobacteria. Unfortunately NDVI band II loses its sensitivity when we consider
 the impact of level of cloud cover on the strength of the vegetation signature
(fig. \ref{fig:re}).

  With spectral resolutions of $\sim$ 75  in the visible for TPF-C, it will be important that the ranges of the sensor bands be designed both to detect atmospheric gases, but also to exclude them for surface sensing. 
	Given constraints of sensor design and damping effects of clouds and diurnal averaging, we will need to determine how strong an NDVI signal is an actual indicator of photosynthesis.

Therefore, unless we uncover a potentially abiotic source of a red edge, we may be able to say that any statistically significant departure of the NDVI from zero implies the existence of Earth-like land-based photosynthetic activity.  Given potential TPF-C sensor bands at 650-680 nm and 750-800 nm, the red-edge detectability for the Earth requires at least 20\% average diurnal land vegetation cover unobscured by clouds to achieve an index  that is greater than zero.
 So, this means a planet with a high percent of water over the surface could still reveal land surface photosynthetic activity.  When clouds are included, the thresholds for discriminating unambiguously
the red-edge become 0.03 (band I) and 0.08 (band II). These numbers although might be  significantly smaller and even negative depending on phase, viewing geometries and cloud distribution.

Moving from land to marine vegetation we have dedicated a paragraph on the detectability of plankton (fig. \ref{fig:planktonx}, \ref{fig:planktonxx}).
In fig.  \ref{fig:planktonx} we show disk averaged spectra of an ocean view
of Earth, with an increasing concentration of Plankton. As we have already pointed out, this is an
 ``optimistic'' result, since we have supposed to have a clear sky, and we have chosen 
a particular viewing geometry and a particular season, in which Plankton 
can be better detected (NASA-Goddard SeaWiFS website). When clouds are taken into account, a more realistic case,
(fig. \ref{fig:planktonxx}) plankton is very hard to detect, even for high concentration.

The concentration of 300 mg of Chlorophyll is not realistic 
for our present day planet, but it is useful to understand the trend of how
plankton can possibly modify the final spectra. In the case of an early Earth
or an extrasolar terrestrial planet this hypothesis might be plausible.
It is out of the interests of this paper to discuss further  the likelihood of the occurrence of photosynthesis on Earth-like planets outside our solar system on which life has
evolved. Based on knowledge of photosynthesis on Earth and of stellar evolution Wolstencroft and Raven in (Wolstencraft and Raven, 2002) conclude that it is likely that photosynthesis would have evolved on Earth-like planets in response to the same
evolutionary factors as have been involved on Earth. The chemical intermediates and catalysts would probably be different, but the substrates and end-products would have been the same.
Modeling of photosynthesis on Earth-like planets orbiting stars of different spectral types shows that cooler stars, with maximum radiation output at longer wavelengths, may require more than the
two light reactions used in oxygen-evolving photosynthesis on Earth; such photosynthesis would be limited by the attenuation of radiation by water. Detection of photosynthesis will be based on
spectroscopy using most probably photosynthetic pigments which could have very different absorption
properties from those on Earth.

\section{Conclusions}
\begin{itemize}
\item Our approach is feasible. The comparison with the experiment
confirms: the model works (see fig. \ref{fig:validair}, \ref{fig:validavis})!
\item We can distinguish among different surface types looking at the Earth's spectra in the visible (0.7-0.8 $\mu$m band, red-edge), there are almost no differences
in the IR (fig. \ref{fig:surf}).
\item The contribution of clouds is dramatic. We have to include them in order to have a realistic model 
 (see fig. \ref{fig:validair}, \ref{fig:validavis}, \ref{fig:cloudre}, \ref{fig:earth_view})
\item Our disk-averaged spectra are sensitive to phases above all in the optical and in particular
when clouds are present (fig. \ref{fig:earth_view} ).
\item We have generated disk-averaged spectra from different vantage points. In the optical the main differences are due to a changed surface composition. In the IR on the contrary the fluctuations are due to 
the temperature gradients.
\item
 The lack of high spectral 
resolution smears out the spectral features. However some of the strongest 
features  are  detectable with TPF (fig. \ref{fig:velu1}, \ref{fig:velu2}). 
\item Time dependent variations in the disk-averaged spectra, or "light-curve"
can provide additional information about spatial variations (fig.  \ref{fig:lc}, \ref{fig:re}, \ref{fig:cloudre}).
\item The red-edge biosignature could be potentially detectable (depending on the instrumentation capabilities) when we consider a cloud-free Earth. If clouds are randomly distributed and in high \%, it is very hard to detect this signal (fig.  \ref{fig:re}).
\item Plankton could be potentially discriminated in disk-averaged spectra, depending on the concentration, season of the year, viewing geometry and above all cloud cover (fig. \ref{fig:planktonx}, and  \ref{fig:planktonxx}).
 
\end{itemize}

\section{Acknowledgments}

\textbf{\emph{This work has been supported by NASA Astrobiology Institute and National Academies. }}

\section*{References}

\footnotesize{  \textbf{(Arnold et al., 2002)} Arnold L., Gillet S., Lardière O., Riaud P., Schneider J.: 2002, ``A test for the search for life on extrasolar planets: Looking for the terrestrial vegetation signature in the Earthshine spectrum''
Astronomy and Astrophysics, 392, 231 } \\ \\
\footnotesize{ \textbf{(Beichman and Velusamy, 1999)}
 C.A. Beichman, and T. Velusamy,  "Sensitivity of TPF Interferometer 
for Planet Detection", {\it Optical and IR Interferometry 
from Ground and Space}. S.C.Unwin, and R.V. Stachnick, eds. ASP Conference 
Series, 
Vol.194 (San Francisco: ASP), 405, 1999. } \\
\\
\footnotesize{    \textbf{(Ceccato et al., 2002)} Ceccato, P., N. Gobron, S. Flasse, B. Pinty, and S. Tarantola, \emph{Designing a Spectral Index to Estimate Vegetation Water Content from Remote Sensing Data, Part I: Theoretical Approach}, Remote Sensing of Environment, 82, 188-197, 2002 }
\\ 
\\
{\footnotesize \textbf{(Christensen, Pearl, 1997)}  \emph{Initial data from the Mars Global Surveyor thermal emission spectrometer experiment: Observations of the Earth}, P. R. Christensen,
J. C. Pearl, Journal of Geophysical research Vol. 102, May 25, 1997} \\
\\
{ \footnotesize  \textbf{(Crisp, 1997)} Crisp D., 
\emph{ Absorption of sunlight by water vapor in cloudy conditions: A
partial explanation for the cloud absorption anomaly. },
Geophys. Res.
Lett.,  24, 571-574, 1997. } \\
\\
\footnotesize{    \textbf{(Darwin website)}  http://www.esa.int/esaSC/120382\_index\_0\_m.html } \\
\\
\footnotesize{    \textbf{(Defries and Townshend, 1994)}
 DeFries, R. S. and J. R. G. Townshend, \emph{ NDVI-derived land cover classification at global scales}. International Journal of Remote Sensing, 15(17):3567-3586. Special Issue on Global Data Sets, 1994.
}
\\ \\
\footnotesize{    \textbf{(Dickinson, 1989)}
Dickinson, R. E., 1989: 
\emph{Modeling the effects of Amazonian deforestation on regional surface  climate:  a review.} 
Agricultural and Forest Meteorology 47: 339-347
}
\\
\\
\footnotesize{  \textbf{(Fishbein et al., 2003)}
E. Fishbein, C.B. Farmer, S.L. Granger, D.T. Gregorich, M.R. Gunson, S.E. Hannon, M.D. Hofstadter, S.-Y. Lee, S.S Leroy;
\emph{Formulation and Validation of Simulated Data for the Atmospheric Infrared Sounder (AIRS)}, 
IEEE TRANSACTIONS ON GEOSCIENCE \& REMOTE SENSING, Vol. 41, N. 2, Feb 2003 1. 
} \\
\\
\footnotesize{  \textbf{(Ford et al., 2001)} Ford, E. B., S. Seager, and E. L. Turner. 2001. Characterization of extrasolar terrestrial planets from diurnal photometric variability, Nature 412, 885-887. } \\
 \\
\footnotesize{    \textbf{(Friend and Kiang, accepted) } Friend, A. D. and N. Y. Kiang (2005). 
\emph{Land Surface Model Development for the GISS GCM:  Effects of Improved Canopy Physiology on Simulated Climate.} Journal of Climate: accepted }
\\ 
\\
\footnotesize{    \textbf{(Giorgini et al., 1997)}
       Giorgini JD, Yeomans DK, Chamberlin AB, Chodas PW, Jacobson RA, 
       Keesey MS, Lieske JH, Ostro SJ, Standish EM, Wimberly RN;
    BULLETIN OF THE AMERICAN ASTRONOMICAL SOCIETY (BAAS),v28,No.3,p.1158, 1996. \\
\emph{JPL's On-Line Solar System Data Service}, 
 http://ssd.jpl.nasa.gov/ } \\
\\
{\footnotesize  \textbf{(Gorski et al., 1998)} Healpix (Hierarchical Equal Area and iso-Latitude Pixelisation) is a curvilinear partition of the sphere into exactly equal area
quadrilaterals of varying shape. 
Healpix was originally designed for the ESA Planck and NASA WMAP (Wilkinson Microwave Anisotropy Probe) missions by Krzysztof M. G\'orski, Eric Hivon, Benjamin D. Wandelt, (1998). \newline
 http://www.eso.org/science/healpix/index.html} \\
\\
{\footnotesize  \textbf{(Goode et al., 2001)}, Goode  P.R., J. Qiu, V. Yurchyshyn, J. Hickey, M.C. Chu, E. Kolbe, C. T. Brown, S. E. Koonin, \emph{Earthshine observations of the earth's reflectance}, 
Geophysical Research Letters, 28, 1671-1674, 2001. } \\
\\
{\footnotesize  \textbf{(Goode et al., 2003)}, Goode P.R., E. Pall\'e, J. Qiu, V. Yurchyshyn, J. Hickey, P. Monta\`n\'es Rodriguez, M.C. Chu, E. Kolbe, C.T. Brown, and S.E. Koonin, \emph{Earthshine and the earth s albedo II: Observations and simulations over three years}, submitted JGR, 2003. } \\
\\
\footnotesize{    \textbf{(Goody, Yung, 1996)}  
R. M. Goody, Y. L. Yung, \emph{Atmospheric Radiation: Theoretical Basis}, Oxford Univers
ity Press, 1996 } \\
\\
\footnotesize{  \textbf{(Gower et al., 2004)} J. Gower, S. King, W. Yan, G. Borstad and L. Brown,
\emph{Use of the 709 nm band of merits to detect intense plankton blooms and other conditions in 
coastal waters}. proc. MERIS User Workshop, Frascati Italy, 10-13 Nov. 2003 (ESA SP-549, May 2004) } \\
\\
\\
\footnotesize{  \textbf{(Hanel et al., 1992)} R. A. Hanel, B. J. Conrath, D. E. Jennings, R. E. Samuelson,
\emph{Exploration of the Solar System by Infrared Remote Sensing},
Cambridge Planetary Science Series 7, 1992 } \\
\\
\footnotesize{    \textbf{(Harrison et al., 1990)}  Harrison, E. F., P. Minnis, B. R. Barkstrom, V. Ramanathan, R. D. Cess, and G. G. Gibson, 
\emph{Seasonal variation of cloud radiative forcing derived from the Earth Radiation Budget Experiment},  J. Geophys. Res., 95, 18687-18703, 1990.
}
\\ \\
\footnotesize{    \textbf{(Hartmann, 1994) }} Hartmann, D.L., \emph{ Global Physical Climatology}. Academic Press, 1994
\\
\\
\footnotesize{    \textbf{(Kanamitsu et al., 1991)}
M.~Kanamitsu, J.~C. Alpert, K.~A. Campana, P.~M. Caplan, D.~G. Deaven,
  M.~Iredell, B.~Katz, H.~L. Pan, J.~Sela, and G.~H. White,
 \emph{Recent changes implemented into the global forecast system at NCEP},  Weather and Forecasting, vol. 6, pp. 1--12, 1991.
\\ \\
\footnotesize{    \textbf{(Kanamitsu, 1989)}
M. Kanamitsu,
\emph{Description of the NCEP global data assimilation and forecast
  system,}
Weather and Forecasting, vol. 4, pp. 334--342, 1989. }
Logan1998}
\\ 
\\
\footnotesize{  \textbf{(Kirk, 1983)} J. T. O. Kirk, \emph{Light and Photosynthesis in Aquatic Ecosystems}, 
Cambridge University Press 1983 } \\
\\
\footnotesize{    \textbf{(Kurucz, 1995)} Kurucz, RL, \emph{The solar spectrum: atlases and line identifications},  Laboratory and Astronomical 
     High Resolution Spectra, Astron. Soc. of the Pacific Conf. Series 81, 
     (eds. A.J. Sauval, R. Blomme, and N. Grevesse) San Francisco: Astron. 
     Soc. of the Pacific, pp. 17-31.   } \\
\\
\footnotesize{    \textbf{(Kuze and Chance, 1994) } Kuze A., and K. V. Chance, \emph{Analysis of cloud top height and cloud coverage from satellites using the O$_{2}$
A and B bands},
J. Geophys. Res. 99, 14,481-14,491, 1994 }
\\
\\
\footnotesize{    \textbf{(Liou, 2002)}  K. N. Liou, International Geophysics Series, vol. 84.
\emph{An Introduction to Atmospheric Radiation}, Elsevier Science, 2002. } \\
\\
\footnotesize{    \textbf{(Logan, 1988)} Logan, J. A.,
\emph{An analysis of ozonesonde data for the troposphere: Recommendations
  for testing 3-d models, and development of a gridded climatology for
  tropospheric ozone,}
 Journal of Geophysical Research, vol. 104, pp. 16,115--16,149,
  1998. }
\\ \\
\footnotesize{    \textbf{(Lovel and Belward,1997)}
T.~R. Loveland and A.~S. Belward,
 \emph{The international geosphere biosphere programme data and
  information system global land cover data set DISCover},
 \em Acta Astronautica, vol. 41, no. 4-10, pp. 681--689, 1997 }.
\\
\\
\footnotesize{    \textbf{(Manabe and Strickler, 1964)}   
Manabe, S., and R.F. Strickler, \emph{Thermal equilibrium of the atmosphere with a convective adjustment}. J. Atmos. Sci., 21, 361-185, 1964 
} 
\\
\\
{ \footnotesize \textbf{(Meadows and Crisp, 1996)} V. S. Meadows and D. Crisp, 
\emph{Ground-based near-infrared observations of the Venus nightside: The thermal structure and water abundance near the surface},
(Journal of Geophysical Research, vol. 101, 4595-4622, Feb. 1996) } \\
\\
\footnotesize{    \textbf{(MODIS website)}  
 http://modis.gsfc.nasa.gov/ }\\
\\
\footnotesize{    \textbf{(Muinonen et al., 1989) } Muinonen K., K. Lumme, J. peltoniemi, and M. I. William, \emph{Light Scattering by randomly oriented crystals},
Appl. Opt., 28, 3051-3060, 1989}
 \\
 \\
\footnotesize{    \textbf{(NCEP Development Division Staff, 1988)},
 \emph{Documentation of the ncep global model},
 Tech. {R}ep., {NOAA/NCEP} Development Division, Washington DC 20233,  1988. } 
\\ 
\\
 \footnotesize{    \textbf{(Oort and Peixoto, 1992) }
Peixoto, J.P. \& Oort, A.H., \emph{ Physics of Climate}. American Institute of Physics, 1992
}
\\  
\\
{\footnotesize  \textbf{(Pall\'e et al., 2003)}, Pall\'e, E., P. Monta\`n\'es Rodriguez, P.R. Goode, J. Qiu, V. Yurchyshyn, J. Hickey, M.C. Chu, E. Kolbe, C.T. Brown, and S.E. Koonin,
\emph{ The Earthshine Project: Update on photometric and spectrometric measurements}, submitted, 2003. }
\\ \\
{\footnotesize  \textbf{(Pall\'e, 2003)}, Pall\'e, E., P.R. Goode, J. Qiu, V. Yurchyshyn, J. Hickey, P. Monta\`n\'es Rodriguez, M.C. Chu, E. Kolbe, C.T. Brown, and S.E. Koonin, \emph{Earthshine and the earth's albedo III: Long-term variations in reflectance}, submitted JGR, 2003. } \\
\\
\footnotesize{    \textbf{(Penuelas and Filella, 1998)}, Penuelas, J. and I. Filella,
\emph{Visible and near-infrared reflectance techniques for diagnosing plant physiological status.} 
Trends in Plant Science 3(4): 151-156.
\\ 
\\
\footnotesize{    \textbf{(Potter and Klooster, 1999)} Potter, C. S. and S. A. Klooster, \emph{ Dynamic global vegetation modeling (DGVM) for prediction of plant functional types and biogenic trace gas fluxes}. Global Ecology and Biogeography Letters. 8(6): 473-488, 1999.
} \\
\\
{\footnotesize  \textbf{(Qiu, J., 2003)}, Qiu, J., P.R. Goode, V. Yurchyshyn, J. Hickey, E. Pall\'e, P. Monta\`n\'es Rodriguez, M.C. Chu, E. Kolbe, C.T. Brown, and S.E. Koonin, \emph{Earthshine and the earth s albedo I: Precise and large-scale nightly measurements}, submitted JGR, 2003. }
\\ \\
\footnotesize{    \textbf{(Rothman et al., 2003)}
L.S. Rothman; , A. Barbe, D. Chris Benner, L.R. Brown, C. Camy-Peyret, M.R. Carleer , K. Chance, C. Clerbaux, V. Da
na, V.M. Devi, A. Fayt, J.-M. Flaud, R.R. Gamache, A. Goldman, D. Jacquemart, K.W. Jucks, W.J. Lafferty, J.-Y. Mand
in, S.T. Massie, V. Nemtchinov, D.A. Newnham, A. Perrin, C.P. Rinsland, J. Schroeder, K.M. Smith, M.A.H. Smith, K. 
Tang, R.A. Toth, J. Vander Auwera , P. Varanasi, K. Yoshino, \\
\emph{The HITRAN molecular spectroscopic database: edition of 2000 including updates through 2001}, \\
\mbox{Journal of Quantitative Spectroscopy \& Radiative Transfer \S2 (2003) 5-44} } \\
\\
\footnotesize{    \textbf{(Salomonson, 1990)}, Salomonson, V. V.,  \emph{Execution phase (C/D) spectral band characteristics of the EOS Moderate Resolution Imaging Spectrometer-NADIR (MODIS-N) facility instrument}. The Hague, The Netherlands, XXVIII COSPAR 1990, Symposium No. S.1, Session 6: 12. }
\\
\\
\footnotesize{  \textbf{(SeaWiFS website)} SeaWiFS Project, NASA Goddard Space Flight Center, \newline
http://seawifs.gsfc.nasa.gov/SEAWIFS.html } \\
 \\
\footnotesize{    \textbf{(Segelstein, 1981) } Segelstein, D., \emph{ The complex refractive index of water"}, M.S. Thesis, University of Missouri-Kansas City,
1981
}
\\
\\
\footnotesize{    \textbf{(Stamnes et al., 1988) } Stamnes, K., SC
Tsay, W. Wiscombe, and K. Jayaweera,
 \emph{Numerically stable algorithm for discrete-ordinate-method radiative transfer in multiple scattering and emitting layered media}, 
Appl. Opt., 27, 2502-2509, 1988. }
\\ 
\\
\footnotesize{  \textbf{(TPF book)} C.A. Beichman, N. J. Woolf,  and C.A. Lindensmith, Eds. {\it The 
Terrestrial Planet 	Finder (TPF)}, (JPL: Pasadena), JPL 99-3,1999. } \\ \\
\footnotesize{    \textbf{(TPF website)} http://planetquest.jpl.nasa.gov/TPF/tpf\_index.html } \\
\\
\footnotesize{  \textbf{(Tinetti et al.)} G. Tinetti, V. S. Meadows, D. Crisp,
W. Fong, T. Velusamy, H. Snively; \emph{Disk-averaged synthetic spectra of Mars}, submitted to Astrobiology, 2004. \newline
http://arxiv.org/abs/astro-ph/0408372    } \\
\\
\footnotesize{\textbf{(Tucker,  1978)}, Tucker, C. J., 
\emph{A comparison of satellite sensor bands for vegetation monitoring.}
Photogrammetric Engineering and Remote Sensing 44(11): 1369-1380. } \\
\\ 
\footnotesize{    \textbf{(Turnbull, 2004) } M. Turnbull, PhD thesis, University of Arizona }
\\
\\
\footnotesize{  \textbf{(Walker, 1994)} Earth and Moon Viewer, first implemented by
John walker in Dec. 1994. \newline http://www.fourmilab.ch/cgi-bin/uncgi/Earth } \\
\\
\footnotesize{    \textbf{(Wagner et al., 1998) } Wagner T., F. Erle, L. Marquard, C. Otten, K. Pfeilsticker, T. Senne, J. Stutz,
and U. Platt, \emph{Cloudy sky optical paths as derived from differential optical absorption spectroscopy observations},
J. Geophys. Res. 103, 25, 307-25, 321, 1998 }
\\ 
 \\
\footnotesize{    \textbf{(Wagner et al., 2002) } T. Wagner, C. von Friedeburg, M. Wenig, C. Otten, and U. Platt, \emph{UV-visible observations of atmospheric O$_{4}$ absorptions using direct moonlight and zenith-scattered sunlight for clear-sky and cloudy sky conditions}, Journal of Geophysical Research, Vol. 107, N. D20, 4424, 2002}
\\ 
  \\
\footnotesize{    \textbf{(Warren, 1984) }
Warren, S.,  \emph{Optical constants of ice from the ultraviolet to the microwave}, Appl. Opt., 23, 1206-1225, 1984.}
\\  
\\
\footnotesize{    \textbf{(Wolstencroft and Raven, 2002)}  Wolstencroft R D \& Raven J A (2002)
\emph{ Photosynthesis: likelihood of occurrence and possibility of detection on Earth-like planets. } Icarus 157:535-548. } \\
\\
\footnotesize{  \textbf{(Woolf et al., 2002)} N. J. Woolf and P.S Smith, and W. A. Traub and K. W. Jucks,
``The Astrophysical Journal'', 547:430-433, 2002, July 20. } \\

\end{document}